\documentclass[11pt,cite,epsfig,psfrag]{article}
\usepackage[utf8]{inputenc}
\usepackage{placeins}
\usepackage{geometry}
 \usepackage{graphicx}
 \usepackage{subfig}
  \usepackage{amsmath} 
  \graphicspath{{images33/}}
  \usepackage[usenames, dvipsnames]{color}
\usepackage{wrapfig}
\usepackage{boxedminipage}

\usepackage{fullpage}
\usepackage{amsmath}
\usepackage{amssymb}
\usepackage{graphicx}
\usepackage{mathrsfs}
\usepackage{wrapfig}
\usepackage{boxedminipage}
\usepackage{epsfig}
\usepackage{setspace}
\usepackage{float}
\usepackage{hyperref}
\usepackage{enumerate}
\usepackage{cite}
\usepackage[font=footnotesize,labelfont=rm]{caption}

\def\be{\begin{equation}}
\def\ee{\end{equation}}
\def\bea{\begin{eqnarray}}
\def\eea{\end{eqnarray}}

\numberwithin{equation}{section}

 \newcommand{\RN}[1]{%
   \textup{\uppercase\expandafter{\romannumeral#1}}%
 }
\begin{document}

\thispagestyle{empty}

\vskip 2cm 


\begin{center}
{\Large \bf Heat Engines for Dilatonic Born-Infeld Black Holes}
\end{center}

\vskip .2cm

\vskip 1.2cm

\centerline{ \bf Chandrasekhar Bhamidipati\footnote{chandrasekhar@iitbbs.ac.in} and 
 Pavan Kumar Yerra \footnote{pk11@iitbbs.ac.in}
}

\vskip 7mm 
\begin{center}{ School of Basic Sciences\\ 
Indian Institute of Technology Bhubaneswar \\ Bhubaneswar 751013, India}
\end{center}

\vskip 1.2cm
\vskip 1.2cm
\centerline{\bf Abstract}
\noindent
In the context of  dilaton coupled Einstein gravity with a negative cosmological constant and a Born-Infeld field,
we study heat engines where charged black hole is the working substance. Using the existence of a notion of thermodynamic mass and volume (which depend on the dilaton coupling),  the mechanical work takes place via the $pdV$ terms present in the first law of extended gravitational thermodynamics. Efficiency is analyzed as a function of dilaton and Born-Infeld couplings, and results are compared with analogous computations in the related conformal solutions in the Brans-Dicke Born-Infeld theory and black holes in Anti de Sitter space-time.


\newpage
\setcounter{footnote}{0}
\noindent

\baselineskip 15pt

\section{Introduction}
Recent interest in treating the cosmological constant $\Lambda $ as a dynamical 
parameter \cite{Caldarelli:1999xj}-\cite{Henneaux:1989zc} have led to important extensions of the classical thermodynamic properties of a black hole\cite{Bekenstein:1973ur,Bekenstein:1974ax,Hawking:1974sw,Hawking:1976de}, which relates the mass $M$, surface gravity $\kappa$, and outer horizon area $A$ of a black hole solution to the  energy, temperature, and entropy  ($U$, $T$, and  $S$, resp.) according to (in geometrical units where $G,c,\hbar, k_{\rm B}$ are set to unity): 
\begin{eqnarray} \label{oldTD}
M=U \ , \ T=\frac{\kappa}{2 \pi} \ ,\ S=\frac{A}{4} \ .
\end{eqnarray}
Now, the cosmological constant treated as pressure  $ p=-\Lambda/8\pi$, has a conjugate variable, the thermodynamic volume $V$ associated with the black hole. In this extended thermodynamics, temperature and entropy continue to be related to
surface gravity and area as usual, while, mass, however, turns out to be related to enthalpy $H$\cite{Kastor:2009wy}: $M=H\equiv U+pV$. The First Law now reads:
\begin{eqnarray} \label{newFirstLaw}
dM=TdS+Vdp .
\end{eqnarray}
The black holes may have other parameters such as gauge charges, angular momentum, coupling constants (Gauss-Bonnet, Born-Infeld etc.,)  which enter additively with their conjugates in the First Law (\ref{newFirstLaw}) in the usual way. For static black holes, thermodynamic volume $V$ is just the geometric volume (defined in terms of the horizon radius $r_+$) of the black hole in question~\cite{Parikh:2005qs}, but, in general, the two volumes differ leading to novel physics such as in rotating black holes, AdS-Taub-nut geometry, and black holes with dilaton fields(see for instance \cite{BID,Cvetic:2010jb,Johnson:2014xza,Hawking:1998jf}). Extended thermodynamical phase space treatment leads to an exact identification of small to large
black hole phase transition in charged AdS and related black holes to Van der Waals liquid-gas phase transition \cite{6}, including
an exact map of critical exponents. Furthermore, the phase transitions occur in the $p-T$ plane as 
opposed to $Q-T$ plane and hence identical parameters are now being compared on both sides\cite{8}. 

The possibility of extracting mechanical work from heat energy via the $pdV$ term present in eqn.  (\ref{newFirstLaw}) has led to the proposal of a holographic heat engine in \cite{Johnson:2014yja}, where, the working substance is a black hole solution of the gravity system. Several holographic engines have since been studied\cite{Belhaj:2015hha,Sadeghi:2015ksa,Caceres:2015vsa,Setare:2015yra}. The black hole in particular, provides an equation of state. Work can be extracted from such an engine by defining 
a cycle in state space where there is a net input heat flow $Q_H$, a net output heat flow $Q_C$, and a net output work W, such that $Q_H = W + Q_C$. 

\begin{figure}[h]
\begin{center}
{\centering
\includegraphics[width=2.2in]{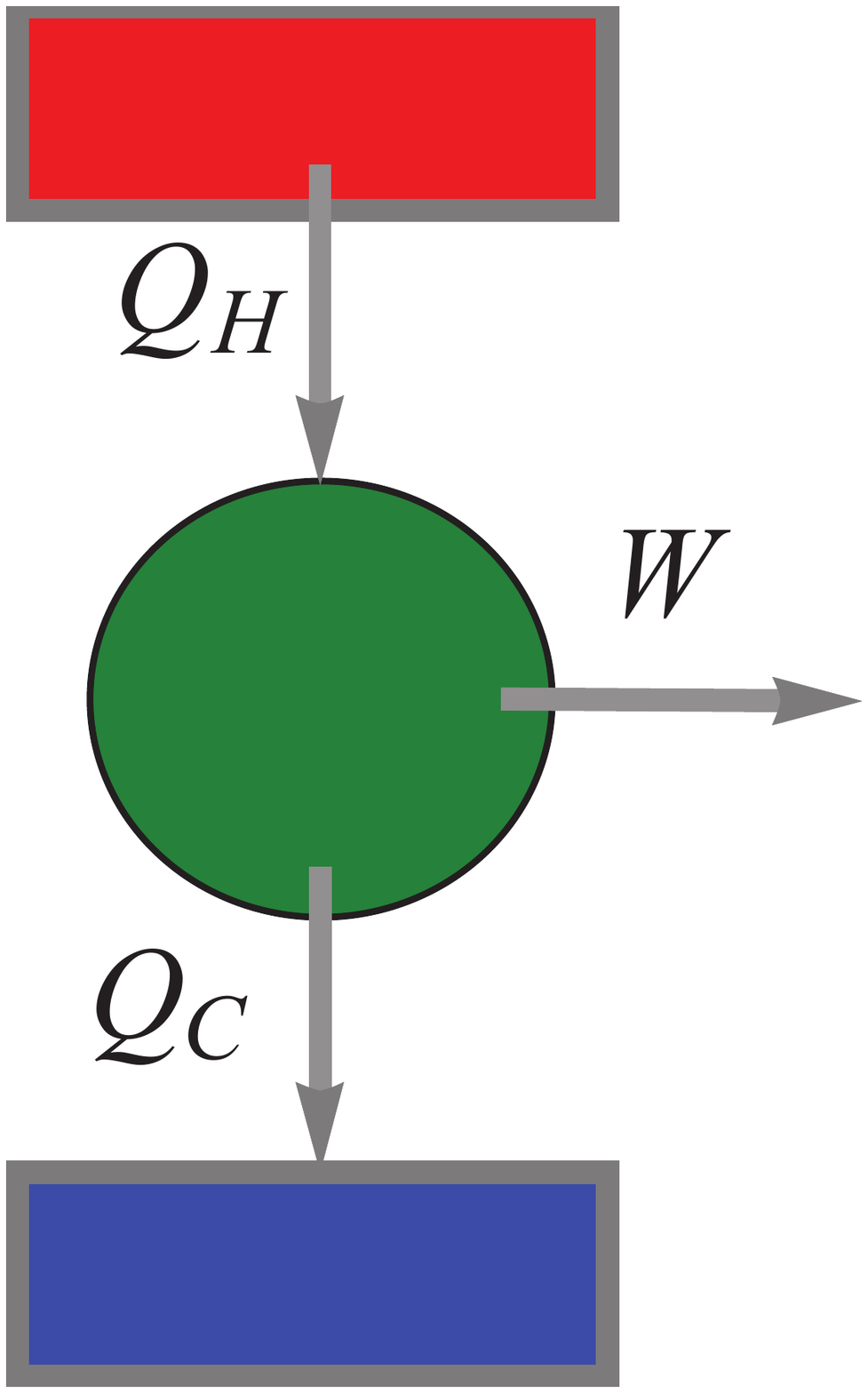} 
\includegraphics[width=2.2in]{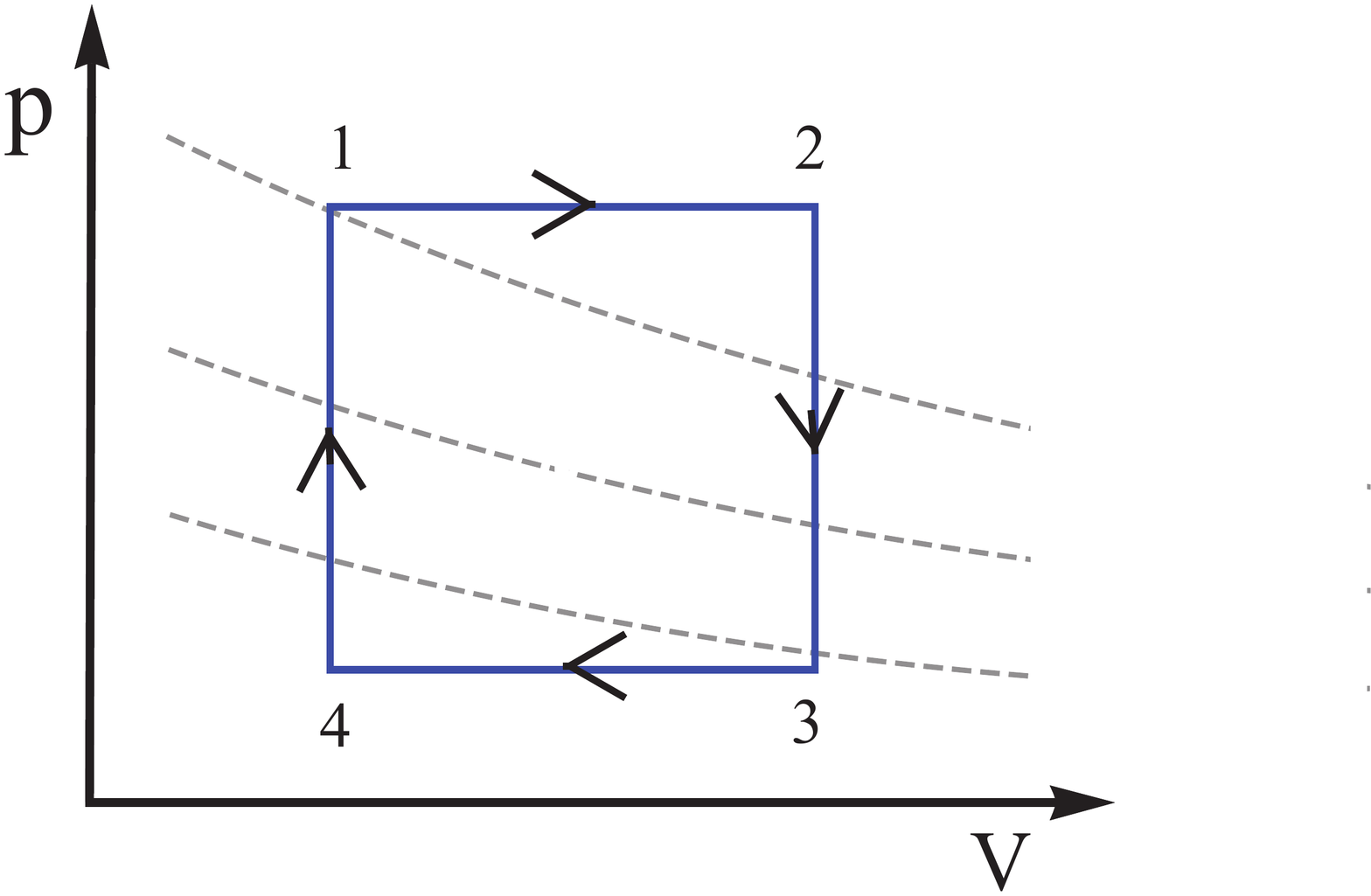} 
\caption{Heat Engine and a rectangular cycle with flows.}   \label{fig:cycle}
}
 \end{center}
\end{figure}

The efficiency of such heat engines can be written in the usual way for heat engines as $\eta=W/Q_H=1-Q_C/Q_H$. Its value depends crucially on the equation of state provided by the  black hole and the choice of cycle in state space. Considering the thermodynamical cycle in figure (\ref{fig:cycle}) advocated 
in \cite{Johnson:2014yja,Johnson:2015ekr,Johnson:2015fva} for static black holes, the entropy and the volume turn out to be dependent on each through $r_+$. This means that isochores are adiabats and hence, the heat flows in the cycle in figure (\ref{fig:cycle}) occur only along top and 
bottom lines. Formal computation of efficiency proceeds via the evaluation of $\int C_p dT$ along those isobars, where~$C_p$ is the specific heat at constant pressure. This in general being difficult, efficiency was evaluated in various limits in \cite{Johnson:2015ekr,Johnson:2015fva}. An exact formula for efficiency 
\begin{eqnarray} \label{eq:efficiency-prototype}
\eta=1-\frac{M_{3}-M_{4}}{M_{2}-M_{1}}\ ,
\end{eqnarray}
was later obtained in\cite{Johnson:2016pfa}, using the fact that mass of the black is just the enthalpy and total heat flow along an isobar is change in enthalpy. For static black holes, $M$ can be written as a function of $p$ and $V$. 
Let us note that defining heat engines via cycles in state space (with dynamical cosmological constant) represents a journey through a family of holographically dual field theories \cite{Maldacena:1997re,Witten:1998qj,Gubser:1998bc,Witten:1998zw,Aharony:1999ti} (at large $N_c$). The exact holographic dictionary corresponding to heat engines and their efficiency needs further study. Nevertheless, we restrict to applications to black holes in Einstein gravity with higher derivative corrections, which are interesting on their own right. Einstein gravity is considered to be an effective description of the underlying quantum gravity, such as, string theory
in the low energy limit. Thus, it is interesting to study the effect of stringy corrections on heat engines and their
efficiency. The effect of Gauss-Bonnet and Born-Infeld higher curvature corrections on efficiency was analyzed in~\cite{Johnson:2015ekr,Johnson:2015fva}. 

\vskip 0.2cm

\noindent
In this paper, we study the efficiency of heat engines in the presence of another stringy effect, namely, the dilaton field (Einstein gravity non-minimally coupled to dilaton is present in the low energy limit of string theory~\cite{lowstring}). 
In particular,  we consider the Einstein-Maxwell dilaton system in the presence of two Liouville type potentials and also dilatonic black holes coupled to nonlinear Born-Infeld theory (Dilatonic-BI) in an extended phase space, in fixed charge ensemble~\cite{Dehghani:2016wmw}. We also consider its conformally consistent counter part, the Brans-Dicke Born-Infeld theory (BD-BI) studied in~\cite{BDED}. BD (Brans-Dicke) theory has been significant in the explanation of the cosmic inflation~\cite{hyp_inflation}, and  consistent with  Dirac's large number hypothesis and Mach's principle~\cite{weinberg}.  Thermodynamics of charged black holes in Brans-Dicke theory have been studied in~\cite{BDth1,BDth2,BDth3,BDth4}. This theory  produces the solar system experimental observations with a specific domain of BD parameter $\omega$~\cite{solar}. 

Moreover, the presence of the dilaton field in Einstein-Maxwell theory changes the causal structure of the space-time and modifies the thermodynamic properties of the black holes in a non trivial way. Rich structure and $p-V$ criticality in black holes with higher derivative couplings, Born-Infeld and dilaton black holes have been studied earlier\footnote{see for instance~\cite{CDB1,CDB2,All,Zhao:2013oza,BI,BID,sheykhi}}. In the case, when there are Liouville type potentials for the scalar fields, the solution is asymptotically neither flat nor AdS. 
Furthermore, an exponential or Liouville potential represents the higher-dimensional cosmological constant present in non-critical string theories \cite{nonsusy,moudas}, non-trivial curved adS backgrounds \cite{CLSZ}, or leading $g_s$ corrections to critical string theories in a flat background.  Starting from standard AdS/CFT duality in higher dimensions, holography of models with Liouville type scalar potentials are generated upon dimensional reduction procedure \cite{d-reduction}. In~\cite{Ahar}, it has been conjectured that the linear dilaton spacetimes, which arise as near-horizon limits of dilatonic black holes, might exhibit holography. \\

Non-asymptotically flat/AdS black hole spacetimes have been actively studied for possible extensions of holography\cite{Ahar,MW,PW,CHM,Cai,Clem,Mitra,SDRP,yaz,DHSR}. The usual notions of thermodynamic mass, thought of as enthalpy of a space time, akin to an
AdS black hole go through for more general backgrounds.
For instance, in the context of black holes in Liftshitz space-times (which are asymptotically neither AdS nor flat) \cite{mass_in_lifshtz, ref42_of_mann}, it has been argued that introducing pressure\footnote{Recently,  there have also been much more general proposals, that pressure should be introduced not just for black holes,  but for {\it all} space-times, resulting in a generalized notion of thermodynamic volume\cite{Johnson:2014xza}. } together with thermodynamic volume and studying extended thermodynamic phase structure (in spite of the fact that all thermodynamic quantities now depend on the dilaton coupling constant) is physically and holographically sensible, with applications to condensed matter systems and quantum criticality. A holographic interpretation for the Van der Waals transition was proposed in  \cite{Johnson:2014yja},  within the extended phase thermodynamics, where, varying the cosmological constant in the bulk corresponds to perturbing the dual CFT, triggering a renormalization group  flow. The transition is then interpreted not as a thermodynamical transition but, instead, as a transition in the space of field theories. Having scalar fields in the bulk (such as the charged Dilaton system in the present manuscript and other examples\cite{Caceres:2015vsa,mass_in_lifshtz}) might turn on certain operators in the boundary theory triggering a non-trivial RG flow.  In particular, there might be solutions of dilatonic theories with Liouville type potentials connecting the IR dynamics to AdS asymptotics in UV\cite{Erdmenger:2016wyp}. \\

Also, since pressure for asymptotically non-flat/ads black holes with Liouville type potentaials has been introduced and the corresponding PV criticality studied in good detail in\cite{sheykhi} and extended to conformally coupled scalars, i.e., the Brans-Dicke-Born-Infeld solutions~\cite{BDED}. Following these works, and the existence of an extended first law with pressure and volume, including the presence of PV criticality allows us to naturally define a heat engine, exactly as in the examples considered for AdS, leading to extension of the results of\cite{Dehghani:2016wmw,BDED,sheykhi,Johnson:2014yja}. The working substance is still the charged black hole, however, the efficiency of heat engines will now depend on the coupling constants provided by dilaton and Born-Infeld theories. A feature of our heat engines is that the volume depends on the coupling constant of dilaton ($\alpha$) and electromagnetic fields ($\beta$) and is not same as the geometrical volume. Existence of an exact formula allows us to study efficiency as a function of both couplings, i.e., $\eta = \eta(\alpha,\beta)$ and take various limits where we keep $\alpha$ fixed while tuning $\beta$ and vice-versa. In particular, in the limit $\alpha \rightarrow 0$ and in the high temperature limit, our exact results agree with the effect of Born-Infeld field on efficiency, captured in\cite{Johnson:2015fva}.

\section{Heat Engines from charged black holes in Dilatonic and Brans-Dicke Born-Infeld theories}
Following the discussion of last section, where a cycle in state space was presented for heat engines from charged black holes, we continue with the computation of efficiency of such engines. We first study the Dilatonic Born-Infeld model and later study the corrections to efficiency of heat engines in Brans-Dicke Born-Infeld model.

\subsection{Dilatonic Born-Infeld Model}

For the purpose of computing efficiency, we start from the relevant expression for mass of the Born-Infeld dilaton black hole~\cite{Dehghani:2016wmw}(details of black hole solutions are reproduced in Appendix A for reference),
\begin{equation}
M= \frac{b^{(n-1)\gamma}(n-1)\omega_{n-1}}{16\pi(\alpha^2+1)}m \, ,
\end{equation}
where, $b$ is an arbitrary positive constant, $\alpha$ is the dilaton coupling constant, $n$ represents the number of spatial dimensions (we restrict to $n >3)$ and $\omega_{n-1}$ is the volume of the constant curvature hypersurface characterizing the horizon . Using the expression for $m$ (see Appendix A), mass can be expressed in terms of other thermodynamic parameters as
\begin{eqnarray} \label{Mass720}
M(r_+, p) &=& \frac{b^{(n-1)\gamma}(\alpha^2+1)\omega_{n-1}}{16\pi}r_+^{n(1-\gamma)-\gamma}\Bigg\{\frac{k(n-1)(n-2)b^{-2\gamma}}{(1-\alpha^2)(n+\alpha^2-2)}r_+^{4\gamma-2}+\frac{16\pi p}{(n+\alpha^2)}r_+^{2\gamma} \nonumber \\ 
 &&-\frac{4\beta^2b^{2\gamma}}{(\alpha^2-n)}\times \Bigg(1-\text{}_{2}F_{1}\left( \left[ -\frac{1}{2},\frac{\alpha ^{2}-n%
}{2n-2}\right] ,\left[ \frac{\alpha ^{2}+n-2}{2n-2}\right]
,-\eta_+\right) \Bigg)\Bigg\} \, .
\end{eqnarray}
Few comments are in order. Here, $p$ is the pressure and $\beta$ is the Born-Infeld parameter, where, $\beta \rightarrow \infty$ corresponds to the Maxwell limit. $\gamma =\alpha ^{2}/(\alpha^{2}+1)$ and $\eta_+ = \eta (r=r_+)$  with $\eta  = \frac{q^{2}b^{2\gamma (1-n)}}{\beta ^{2}r^{2(n-1)(1-\gamma
)}} $ and $r_+$ is the radius of the horizon. $k (> 0)$ is constant characterizing the $(n-1)$ dimensional hypersurface. Temperature 
expressed as a function of other thermodynamic parameters is,
\begin{eqnarray}
T=&\frac{(\alpha ^2+1)}{4\pi
	(n-1)}\left(
\frac{k(n-2)(n-1)b^{-2\gamma}}{(1-\alpha^2)}r_{+}^{2\gamma-1}
+16\pi p\frac{(n-\alpha^2)}{(n+\alpha^2)}r_+ +\frac{4\beta^2b^{2\gamma}}{r_+^{2\gamma-1}}\Big(1-\sqrt{1+\eta_{+}}\Big)\right)
.\end{eqnarray}
 The thermodynamic volume V  is different from the geometrical volume due to dependence on $\gamma$~\cite{Dehghani:2016wmw}
	\begin{equation} \label{eq:2.4}
	V =\frac{b^{(n-1)\gamma}r_+^{n-\gamma(n-1)}}{n-\gamma(n-1)}\omega_{n-1}.
	\end{equation}
	Now  the equation of state $p(V,T)$ for our working substance in the $p-r_+$ plane, or equivalently the $p-V$ plane (using Eq. (\ref{eq:2.4})) is~\cite{Dehghani:2016wmw},
	\begin{eqnarray}
	p &=& \frac{\Gamma T}{r_+}-\frac{k(n-2)(1+\alpha^2)\Gamma}{4\pi(1-\alpha^2)b^{2\gamma}r_+^{2-2\gamma}}+\frac{\beta^2(n+\alpha^2)b^{2\gamma}}{4\pi(n-\alpha^2)r_+^{2\gamma}}\big(\sqrt{1+\eta_+}-1\big) 
	\end{eqnarray}
	where $\Gamma = \frac{(n-1)(n+\alpha^2)}{4(n-\alpha^2)(\alpha^2+1)}$.
A possible scheme for our heat engine (based on the cycle\footnote{(see ref.\cite{Johnson:2014yja},  for reasons on this choice for static black holes)} given in figure~\ref{fig:cycle}) involves specifying values of
temperatures $(T_2,T_4)$ (which  in turn fixes $(T_H,T_C)$ ) and volumes $(V_2,V_4)$. The pressures 
$p_1=p_2$ and $p_4=p_3$ have to be computed from the equation of state and depend on couplings $\alpha$ and $\beta$.
Since, the radii $r_1, r_3$  can be obtained analogously and the mass $M$, written as a function of $r_+$ and $p$ is as in equation (\ref{Mass720}), the efficiency of the engine can now be studied as a function of  couplings $\alpha$ and $\beta$. Considering the cycle given in figure~\ref{fig:cycle},
efficiency of heat engines can be defined entirely in terms of the black hole mass evaluated at the corners as given in (\ref{eq:efficiency-prototype}).
Notice that in the present scheme the Carnot efficiency  $\eta^{\phantom{C}}_{\rm C} = 1-\frac{T_C}{T_H}$,  the upper bound to our engine efficiency  working between  highest and lowest temperatures $T_H$ and $T_C$ respectively, is fixed for all $\alpha, \beta$. Another useful quantity to compare is the efficiency in the Einstein-Maxwell case $\eta_0.$\footnote{(i.e., the limit $\alpha \rightarrow 0$ and $\beta \rightarrow \infty,$ and  we also rescaled the charge $q\rightarrow \sqrt{\frac{(n-1)(n-2)}{2}}q$  to get an exact Reissner-Nordstrom-anti-deSitter black hole)}\\

We first check the case where the dilaton coupling $\alpha$ is set to zero, in which case, we have a pure Born-Infeld black hole.
Efficiency of the heat engine in this case was studied in~\cite{Johnson:2015fva}, in the high temperature limit.  For $n = 4,$ efficiency (equation (\ref{eq:efficiency-prototype})) takes the form as 
\begin{eqnarray}
	&&\eta \Big|_{\alpha = 0} = \frac{\Big(1-\frac{p^{\phantom{4}}_{\rm 4}}{p^{\phantom{1}}_{\rm 1}}\Big)}{\Bigg\{1+\frac{3\sqrt{2}}{8p^{\phantom{1}}_{\rm 1}(\sqrt{V_3}+\sqrt{V_4})}+ \frac{\beta^2}{4\pi p^{\phantom{1}}_{\rm 1}}\Bigg[1-\frac{\Big( \text{}_{2}F_{1}\Big[-\frac{2}{3}, -\frac{1}{2}, \frac{1}{3}, \frac{-\pi^3q^2}{2\sqrt{2}\beta^2V_3^{3/2}}\Big]V_3- \text{}_{2}F_{1}\Big[-\frac{2}{3}, -\frac{1}{2}, \frac{1}{3}, \frac{-\pi^3q^2}{2\sqrt{2}\beta^2V_4^{3/2}} \Big]V_4 \Big)}{(V_3-V_4)} \Bigg] \Bigg\}} \ \ \ \ \ \ \ \ \\ && \text{when $\beta \rightarrow \infty,$ this becomes} \nonumber \\ &&	\eta \Big|_{(\alpha = 0, \ \beta \rightarrow \infty)} = \frac{\Big(1-\frac{p^{\phantom{4}}_{\rm 4}}{p^{\phantom{1}}_{\rm 1}}\Big)}{\Bigg\{1 + \frac{3\sqrt{2}}{8p^{\phantom{1}}_{\rm 1}(\sqrt{V_3}+\sqrt{V_4})} \ - \ \frac{\sqrt{2}  \pi^2 q^2}{16 p^{\phantom{1}}_{\rm 1}  \sqrt{V_4}(V_3+\sqrt{V_3V_4})}  \Bigg\}}  \end{eqnarray}
 \\	\text{For large  $p_1,$ one obtains,} \\
	\begin{equation}	
	\begin{split}	
	\eta \Big|_{\alpha = 0} = \ & \Big(1-\frac{p^{\phantom{4}}_{\rm 4}}{p^{\phantom{1}}_{\rm 1}}\Big)\Bigg\{1 - \frac{1}{p_1} \Bigg(\frac{3\sqrt{2}}{8(\sqrt{V_3}+\sqrt{V_4})} +  \frac{\beta^2}{4\pi }\Bigg[1 \\ & -\frac{\Big( \text{}_{2}F_{1}\Big[-\frac{2}{3}, -\frac{1}{2}, \frac{1}{3}, \frac{-\pi^3q^2}{2\sqrt{2}\beta^2V_3^{3/2}}\Big]V_3- \text{}_{2}F_{1}\Big[-\frac{2}{3}, -\frac{1}{2}, \frac{1}{3}, \frac{-\pi^3q^2}{2\sqrt{2}\beta^2V_4^{3/2}} \Big]V_4 \Big)}{(V_3-V_4)} \Bigg]\Bigg)  + \
	O\Big(\frac{1}{p_1^2}\Big) \Bigg\}, 
	\end{split} \ \ \ \ \ \ \ \ \
	\end{equation}
	 
\begin{figure}[h]
	\begin{center}
		{\centering
		\subfloat[]{	\includegraphics[width=2.8in]{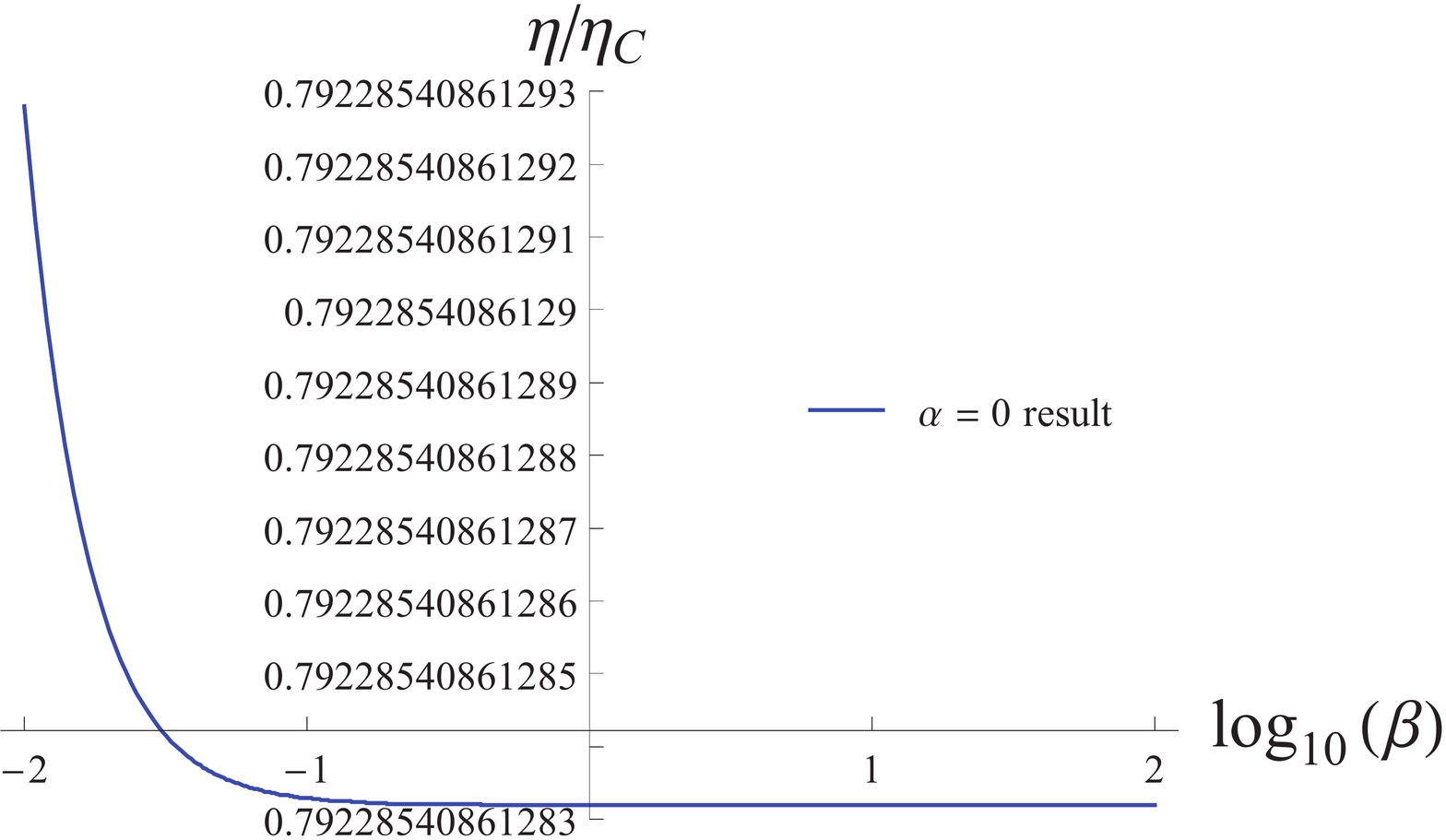} }
		\subfloat[]{	\includegraphics[width=2.8in]{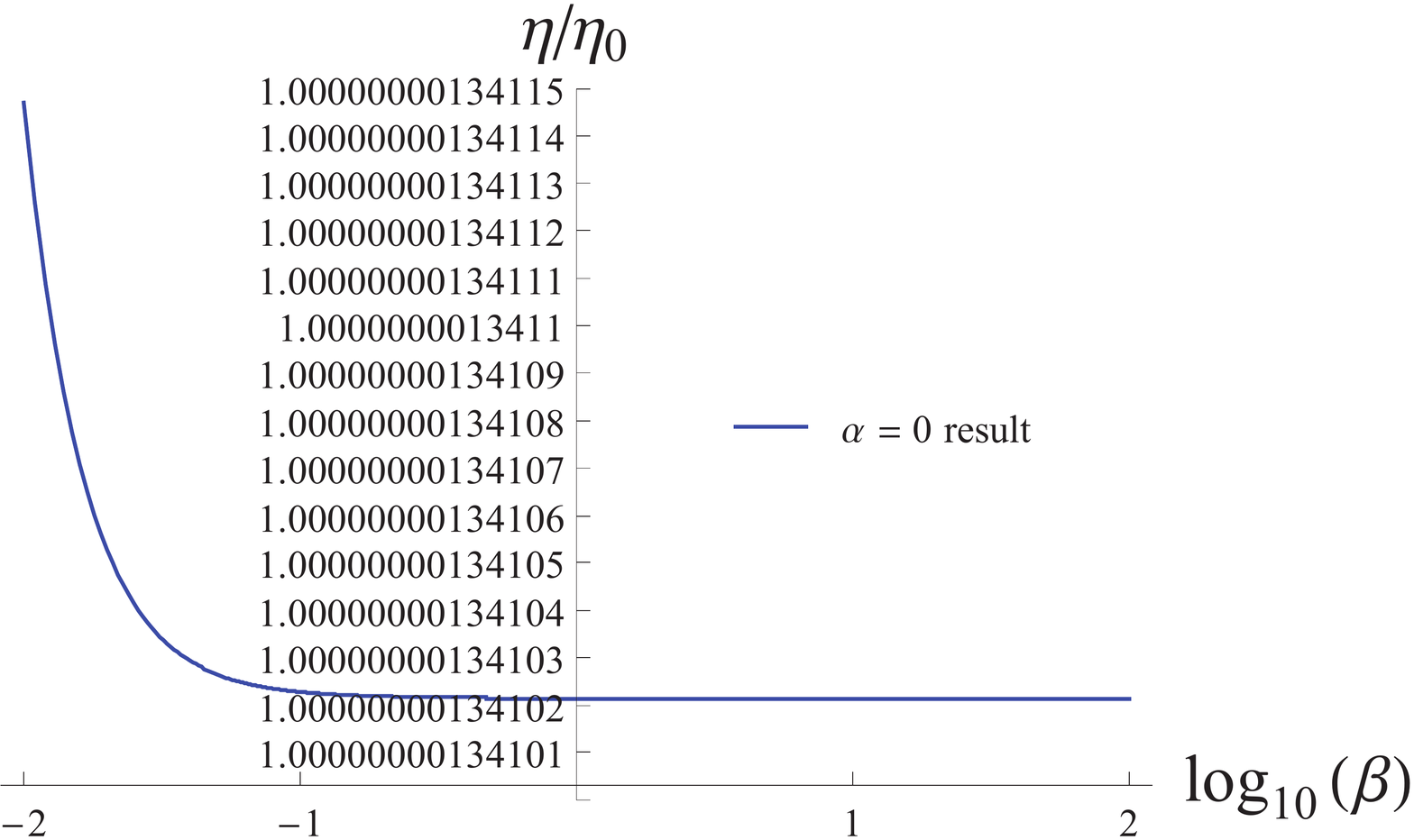} }
			\caption{ For the case $\alpha=0$, (a) The ratio $\eta/ \eta_C$ vs $\log_{10}(\beta).$ (b) The ratio $\eta/ \eta_0$ vs $\log_{10}(\beta).$ (Here, we have chosen the values $n=4,$ $q=0.1,$ $b=1,$ $T_4 \equiv T_C =30,$ $T_2 \equiv T_H =60,$ $V_2 =33000,$ and $V_4 = 15500.$)}   \label{fig:BI}
		}
	\end{center}
\end{figure}

\begin{figure}[h!]
	\begin{center}
		{\centering
			\subfloat[]{	\includegraphics[width=1.5in]{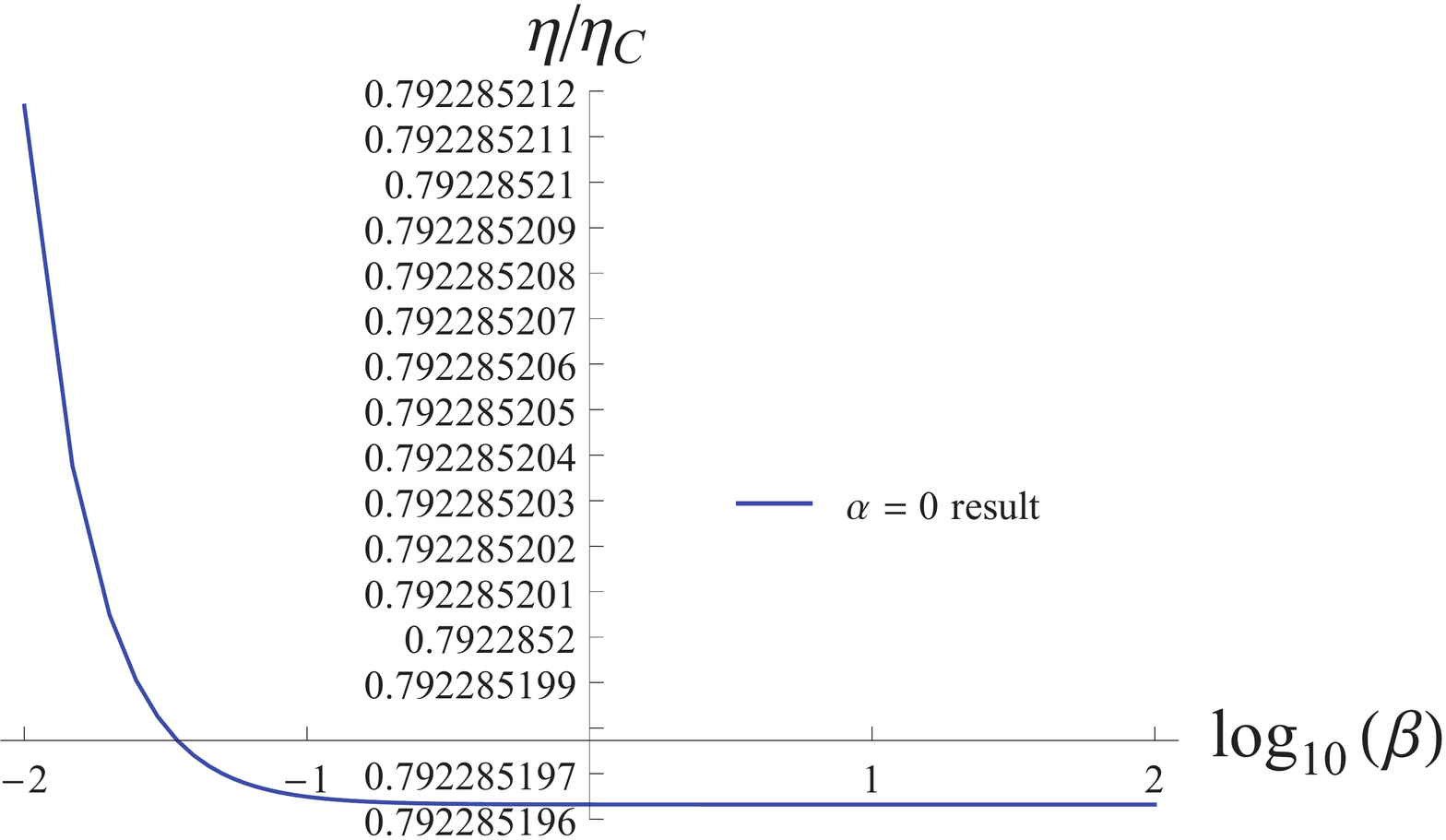} }
			\subfloat[]{	\includegraphics[width=1.5in]{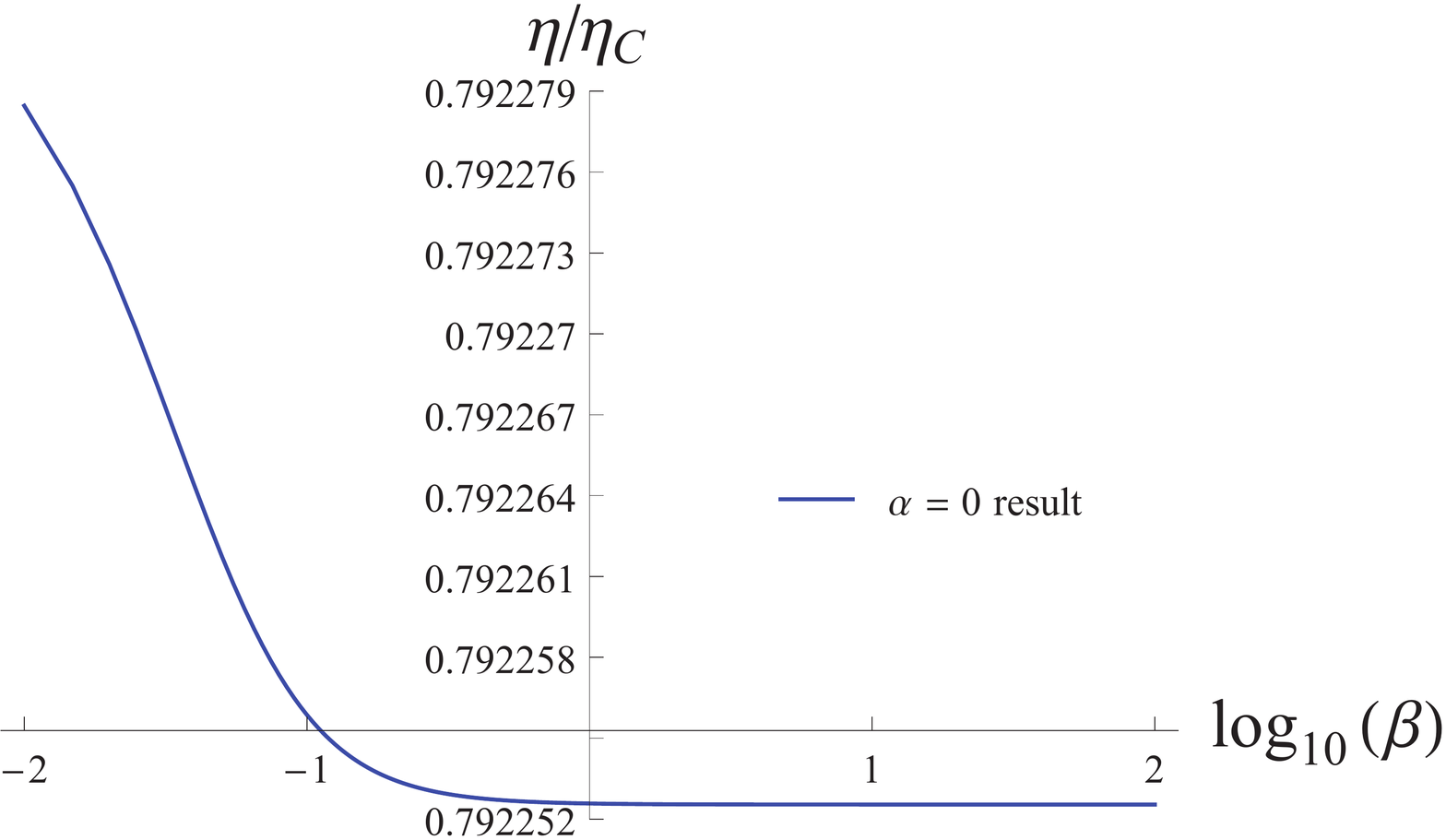} }
			\subfloat[]{	\includegraphics[width=1.5in]{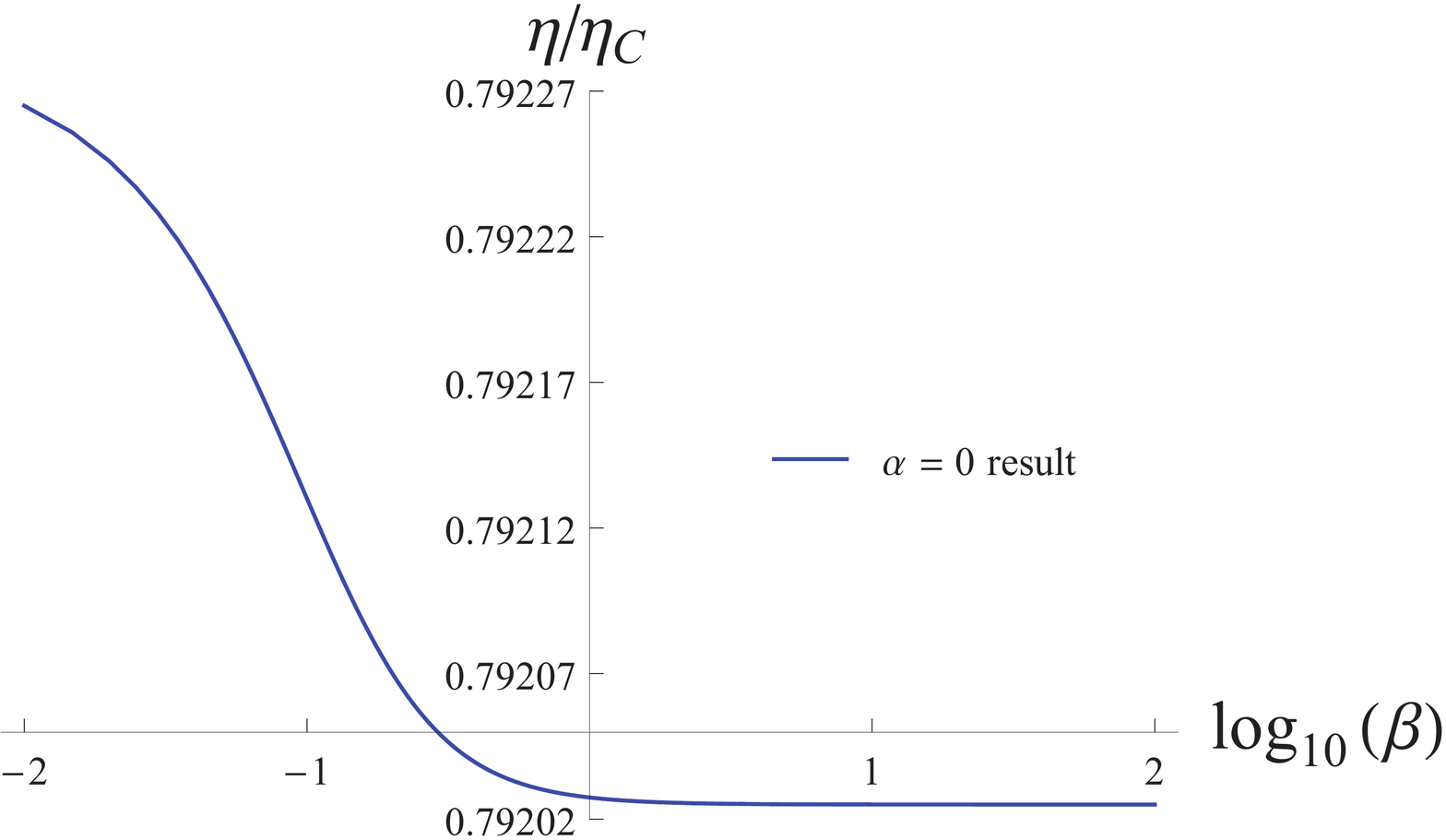} }
			\subfloat[]{	\includegraphics[width=1.5in]{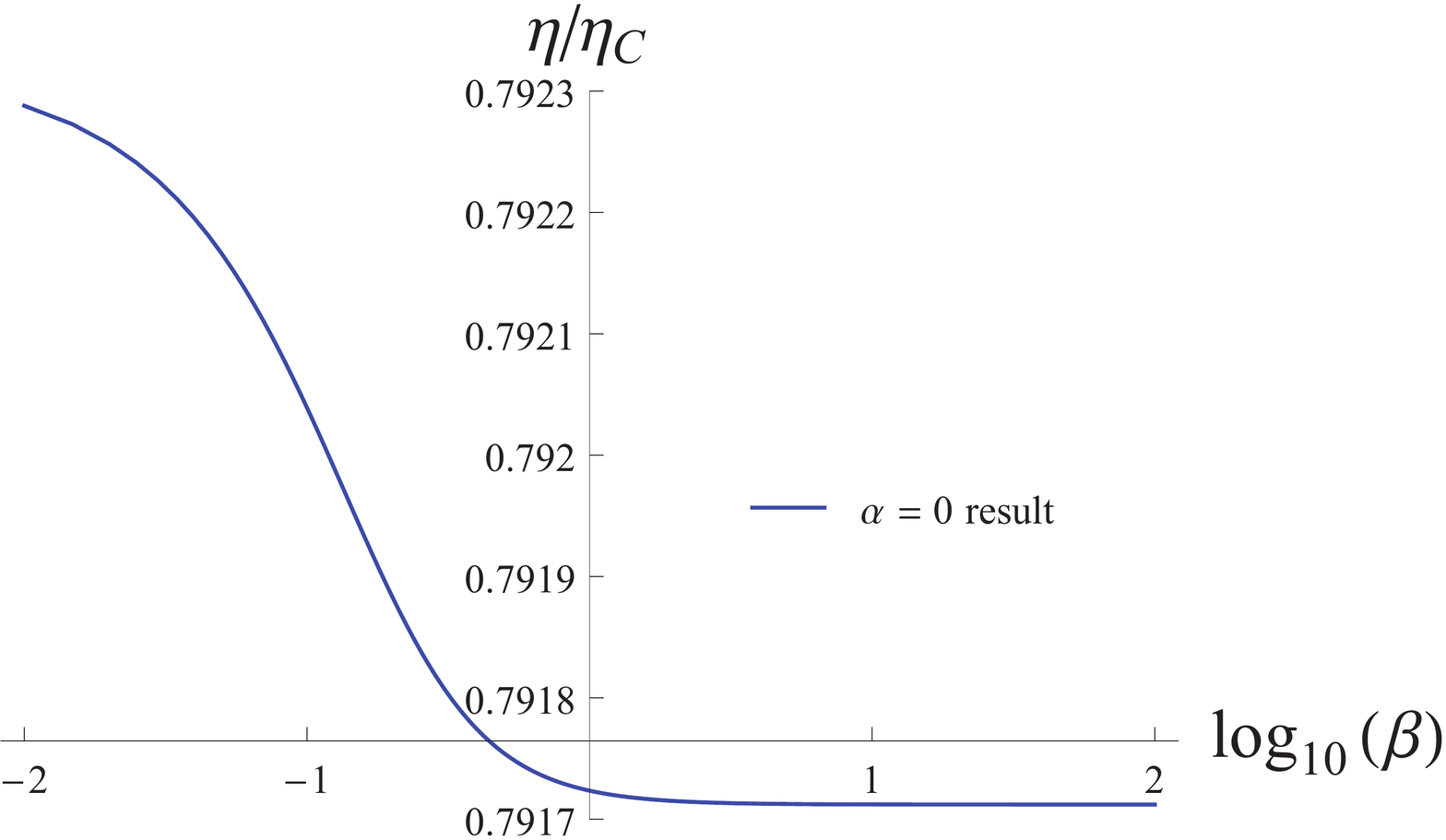} }
			\caption {Effect of q on  efficiency, (a) for $q=2$, (b) for $q=25$, (c) for $q=70$ and (d) for $q=100$,  when other parameters (see the caption of figure 2 for parameter values) are fixed.} \label{different q in beta}  
			
		}
	\end{center}
\end{figure}

\begin{figure}[h!]
	\begin{center}
		{\centering
			\subfloat[]{	\includegraphics[width=2in]{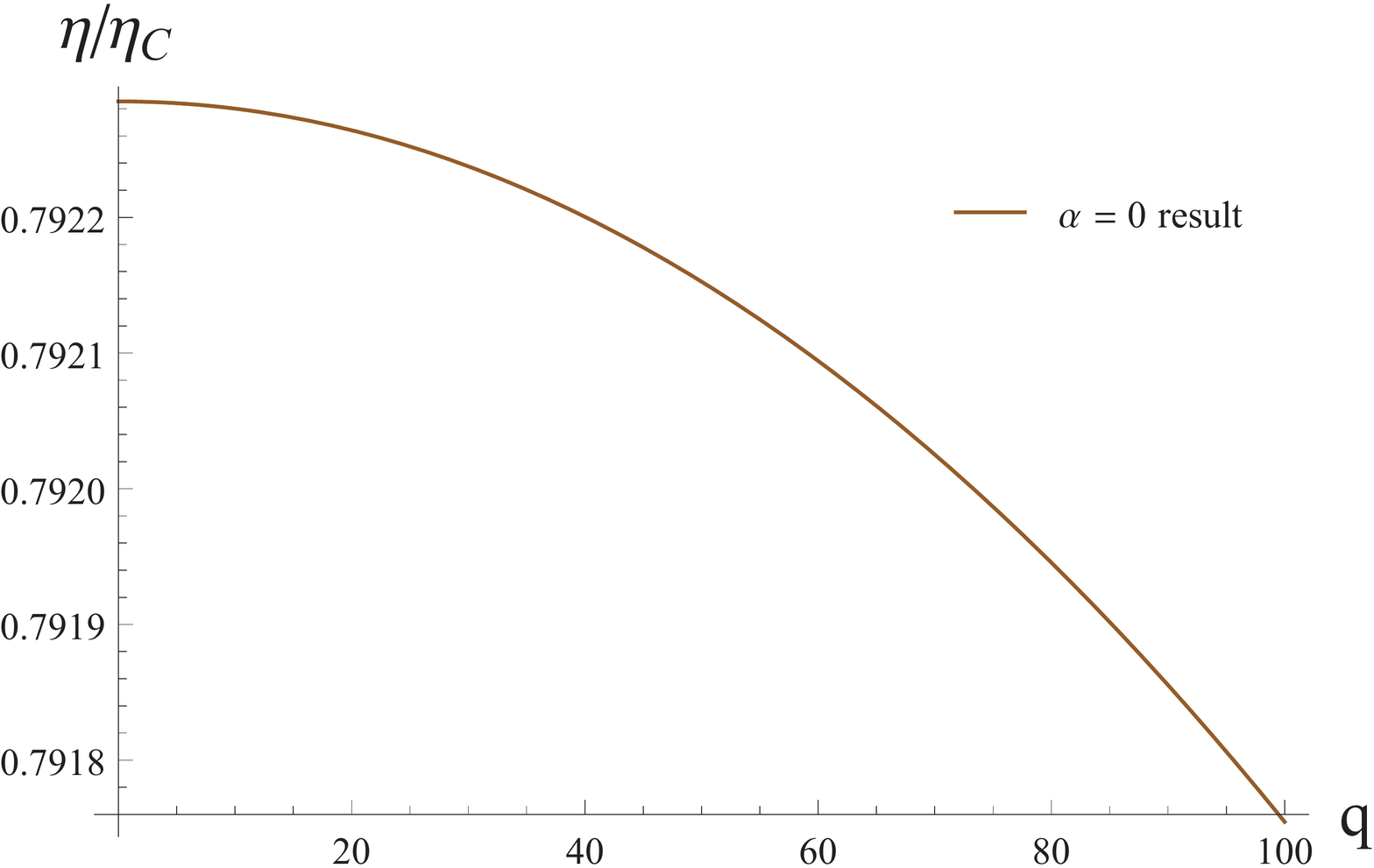} }
			\subfloat[]{	\includegraphics[width=2in]{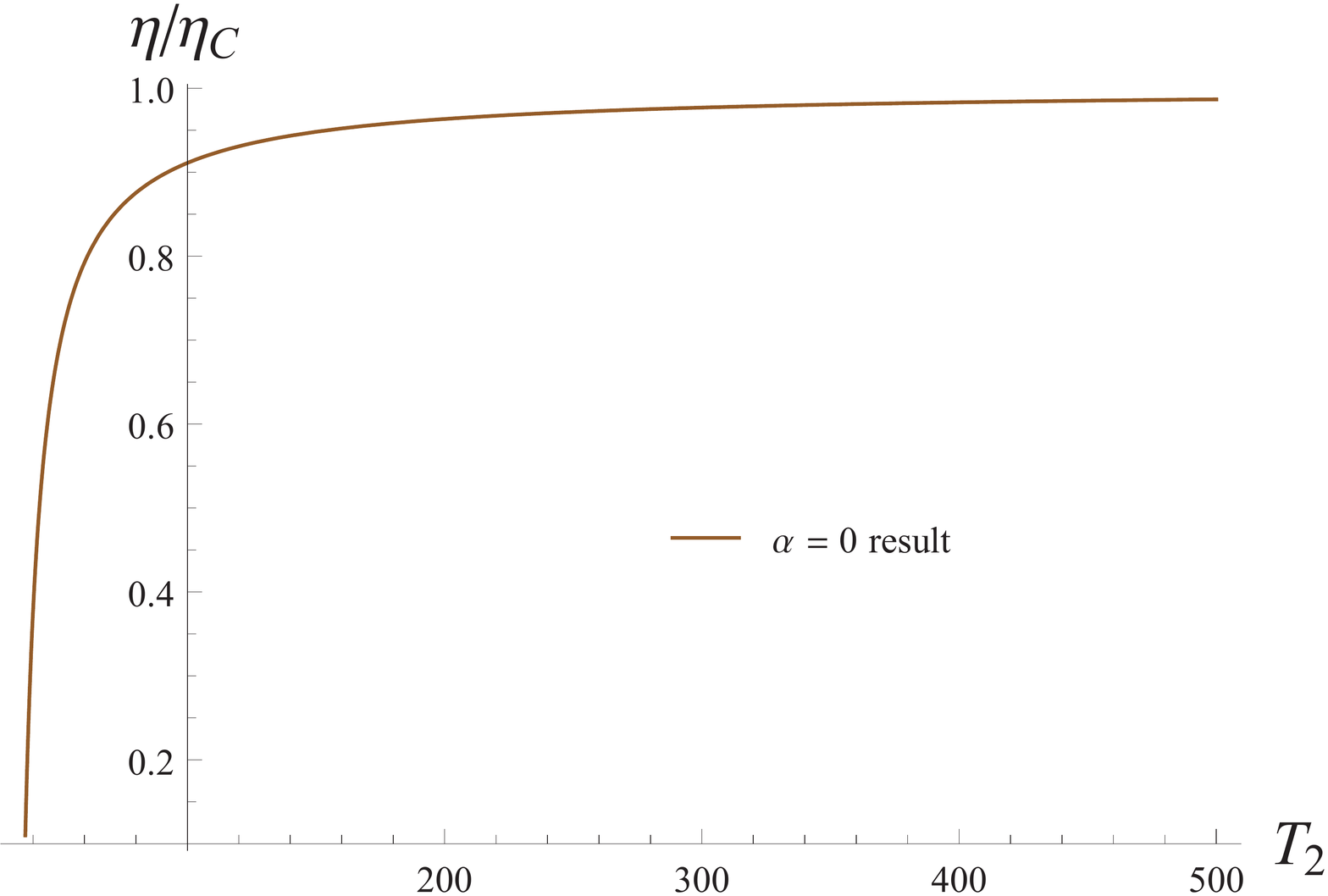} }
			\subfloat[]{	\includegraphics[width=1.9in]{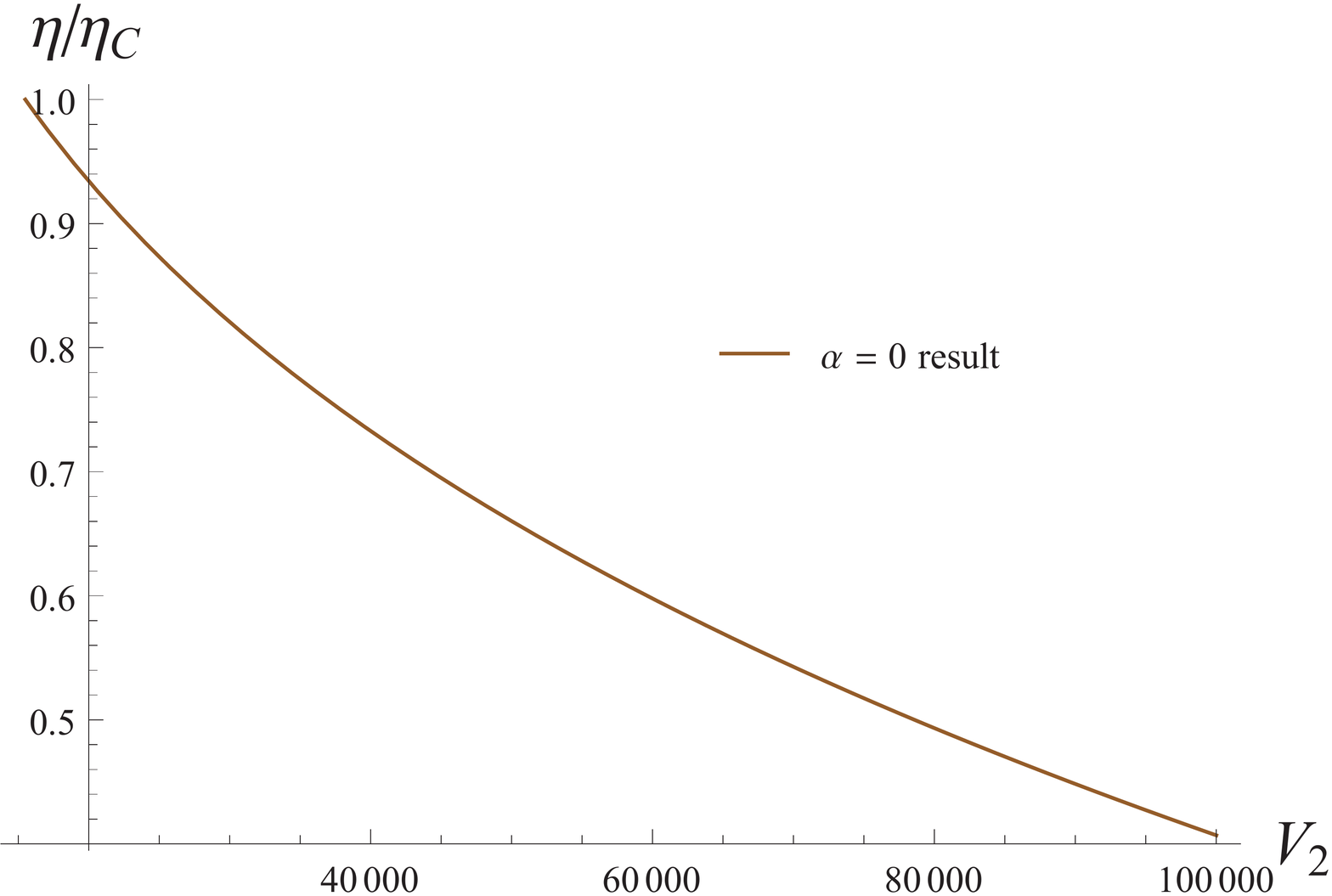} }
			
			\caption{ Effect of parameters on  efficiency,  when other parameters (see the caption of figure 2 for parameter values) are fixed and $ \beta = 5.$} \label{varying parameters in beta}  
			
		}
	\end{center}
\end{figure}

\text{and}
\begin{equation}
\eta \Big|_{(\alpha = 0, \ \beta \rightarrow \infty)} = \Big(1-\frac{p^{\phantom{4}}_{\rm 4}}{p^{\phantom{1}}_{\rm 1}}\Big)\Bigg\{1 - \frac{1}{p_1}\Bigg( \frac{3\sqrt{2}}{8(\sqrt{V_3}+\sqrt{V_4})} \ - \ \frac{\sqrt{2}  \pi^2 q^2}{16   \sqrt{V_4}(V_3+\sqrt{V_3V_4})}\Bigg) + O\Big(\frac{1}{p_1^2}\Big)  \Bigg\}   \end{equation} \text{In fact, for $n=3$, for large volume branch of solutions and neglecting $q$ to leading order, we have} \\ \begin{equation} \eta \Big|_{(\alpha = 0, \ \beta \rightarrow \infty)} = \Big(1-\frac{p^{\phantom{4}}_{\rm 4}}{p^{\phantom{1}}_{\rm 1}}\Big)\Bigg\{1 - \frac{3}{8p_1}\Bigg( \frac{S_2^{1/2}-S_1^{1/2}}{S_2^{3/2}-S_1^{3/2}} \Bigg) + O\Big(\frac{1}{p_1^2}\Big)  \Bigg\}
\end{equation}
 \text{This matches with the equation (20) in \cite{Johnson:2014yja}.} \\ \\ 
From  figure (\ref{fig:BI}), it can be seen that  both the efficiency ratios, i.e., $\eta/\eta_C$  and $\eta/\eta_0$,  grow slowly for a while and then rise, in agreement with the results in~\cite{Johnson:2015fva} for high temperatures. We see from figure (\ref{different q in beta}) that, an increase in q causes significant changes in the efficiency.  In fact, we can see the effect of various parameters \footnote{while varying the parameters one must check the validity of pressures.} on efficiency from figure (\ref{varying parameters in beta}).

\begin{figure}[h]
	\begin{center}
		{\centering
			\includegraphics[width=2.5in]{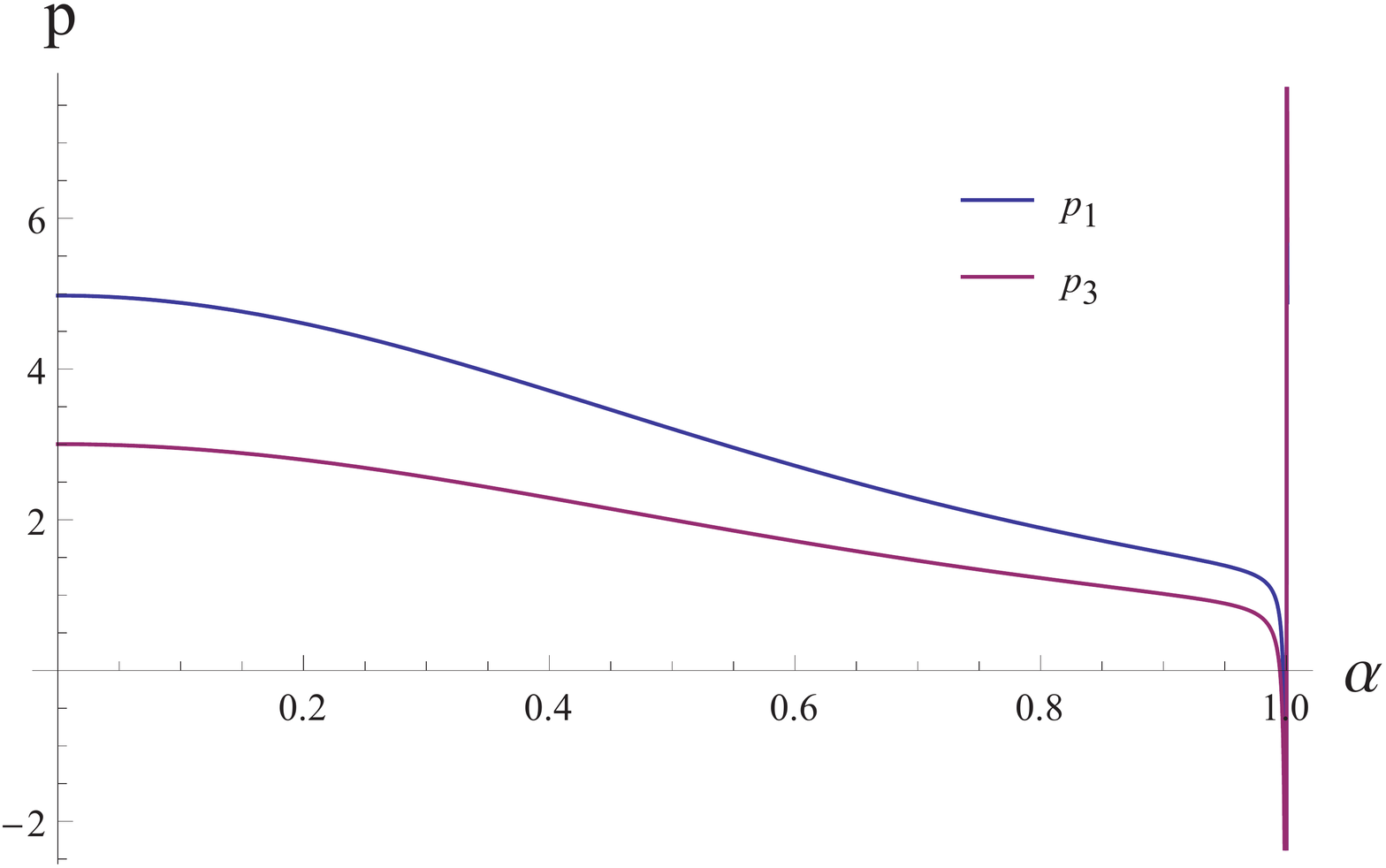} 
			\caption{In the limit $\beta \rightarrow \infty,$ physical range of pressures. (See the caption of figure 2 for parameter values.)}   \label{fig:pressure}
		}
	\end{center}
\end{figure}

 We now keep $\alpha$ non zero and study the resulting efficiency in the limit $\beta \rightarrow \infty$. Now  $\eta$ (equation (\ref{eq:efficiency-prototype})) for $n = 4,$ can be expressed as \begin{equation}
 	\eta \Big|_{ \ \beta \rightarrow \infty} = \frac{\Big(1-\frac{p^{\phantom{4}}_{\rm 4}}{p^{\phantom{1}}_{\rm 1}}\Big) }{\Bigg\{1 \ + \ \frac{(4+\alpha^2)}{8 p^{\phantom{1}}_{\rm 1} \pi (\alpha^2+2)(1-\alpha^2)}\Big[\frac{V_3^{(\frac{2-\gamma}{4-3\gamma} )} \ - \  V_4^{(\frac{2-\gamma}{4-3\gamma} )}}{V_3-V_4}\Big]\Big(\frac{3A_o^{2\gamma-2}}{b^{2\gamma}} \ + \ \frac{q^2A_o^{4\gamma-6}(\alpha^2-1)}{b^{4\gamma}(V_3V_4)^{(\frac{2-\gamma}{4-3\gamma} )}}\Big)\Bigg\}} 
 	\end{equation}
 	 \text{For large  $p_1,$ one obtains}
 	 \begin{equation}
 	 \begin{split}
 	 \eta \Big|_{ \ \beta \rightarrow \infty}  = \ & \Big(1-\frac{p^{\phantom{4}}_{\rm 4}}{p^{\phantom{1}}_{\rm 1}}\Big) \Bigg\{1 \ - \ \frac{(4+\alpha^2)}{8 p^{\phantom{1}}_{\rm 1} \pi (\alpha^2+2)(1-\alpha^2)}\Big[\frac{V_3^{(\frac{2-\gamma}{4-3\gamma} )} \ - \  V_4^{(\frac{2-\gamma}{4-3\gamma} )}}{V_3-V_4}\Big]\Big(\frac{3A_o^{2\gamma-2}}{b^{2\gamma}} \ + \ \frac{q^2A_o^{4\gamma-6}(\alpha^2-1)}{b^{4\gamma}(V_3V_4)^{(\frac{2-\gamma}{4-3\gamma} )}}\Big)  \\ & + \ O \Big(\frac{1}{p_1^2}\Big)\Bigg\} 
 	 \end{split}
 	\end{equation}
 	\\ \text{ which shows the leading behavior of $\eta$ is $(1-p_4/p_1)$, where  $ A_o =\Big(\frac{4-3\gamma}{2\pi^2b^{3\gamma}}\Big)^{1/(4-3\gamma)}$.}  \\	
 	\\
  Efficiency of our engine now depends only on the dilatonic coupling and a  comparison with both $\eta^{\phantom{C}}_{\rm C}$ and $\eta_0$ is again possible. Using the equation of state one can check whether the pressures $(p_1 ,   p_3)$ in the engine remain physical as $\alpha$ changes.  Since, we have fixed $(T_2,V_2)$ and $(T_4,V_4),$  the pressures are now $\alpha$-dependent.
	In fact, pressures become negative as $\alpha$ increases beyond $2$, diverging at $\alpha = 1,2$ since the black hole solution is diverging at these points. If we consider the critical behavior of our black hole~\cite{Dehghani:2016wmw}, the universal ratio $\rho_c$ is positive, provided $\alpha < 1$, so, we restrict ourselves to  the physical range of $\alpha$, i.e., $0< \alpha < 1$ (See figure \ref{fig:pressure}).
\begin{figure}[h]
	\begin{center}
		{\centering
			\subfloat[]{	\includegraphics[width=2.8in]{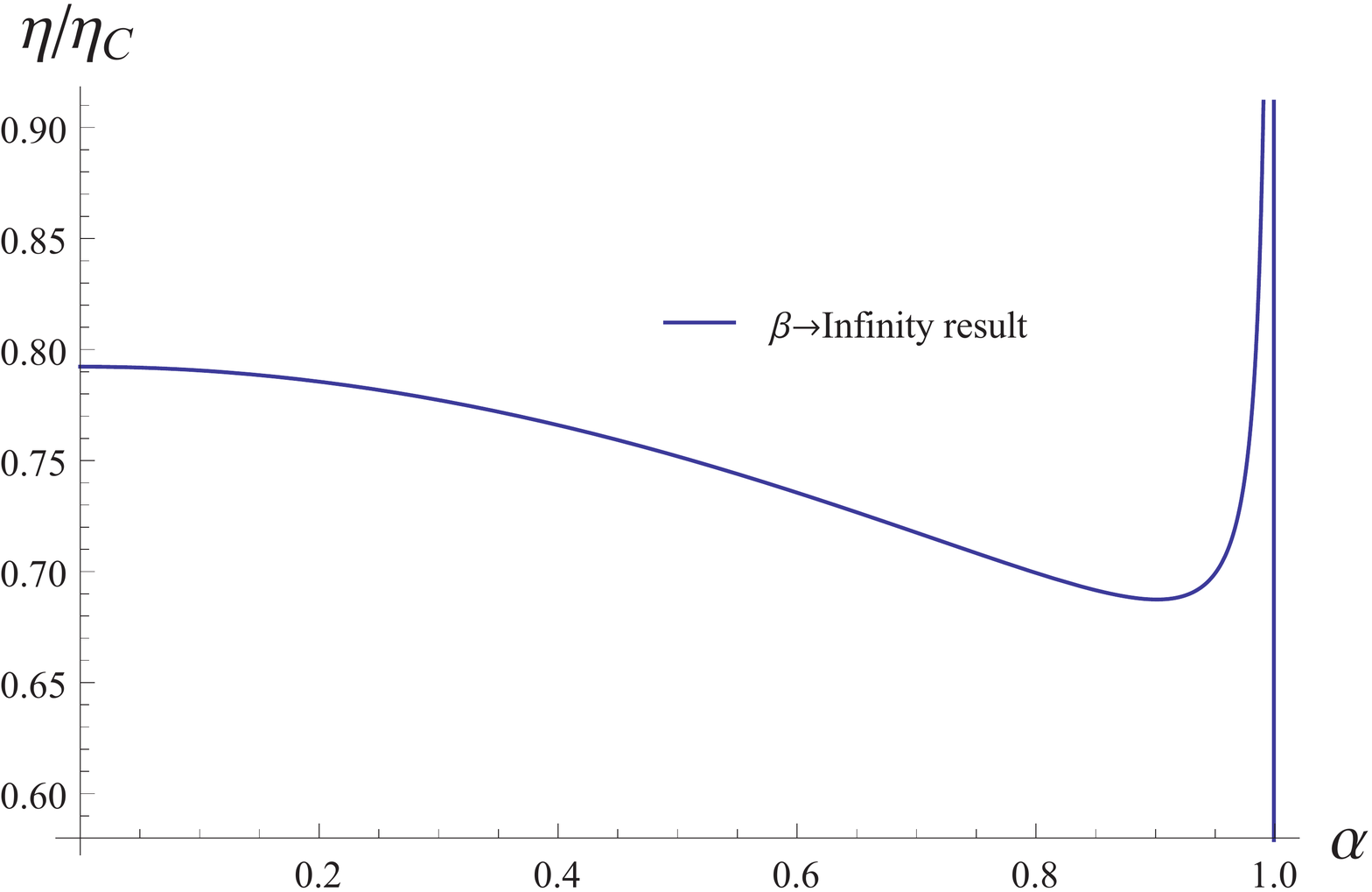} }
			\subfloat[]{	\includegraphics[width=2.8in]{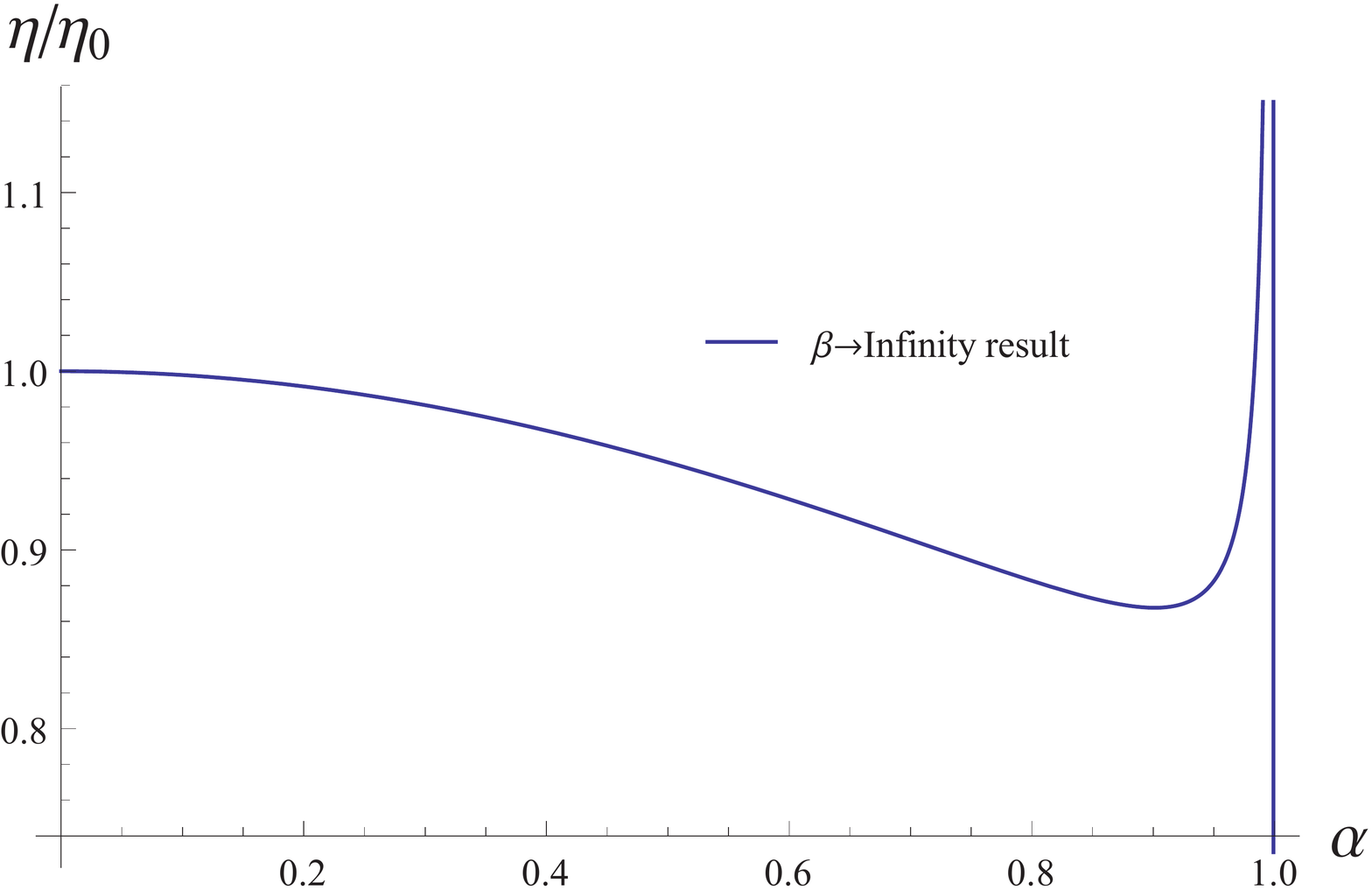} }
			\caption{In the limit $\beta \rightarrow \infty,$ (a) The ratio $\eta/\eta_C$ vs $\alpha.$ (b) The ratio $\eta/\eta_0$ vs $\alpha$.  (See the caption of figure 2 for parameter values.)}   \label{fig:4}
		}
	\end{center}
\end{figure}

 \par A study of how the efficiency $\eta$ varies with respect to  $\alpha$ shows that, initially, it falls as compared to $\eta^{\phantom{C}}_{\rm C}$, but then rises again (See figure \ref{fig:4}). One notes that as $\alpha \rightarrow 1$, results on the efficiency
are less reliable as pressure may no more be positive.  

\begin{figure}[h!]
	\begin{center}
		{\centering
			\subfloat[]{	\includegraphics[width=2.1in]{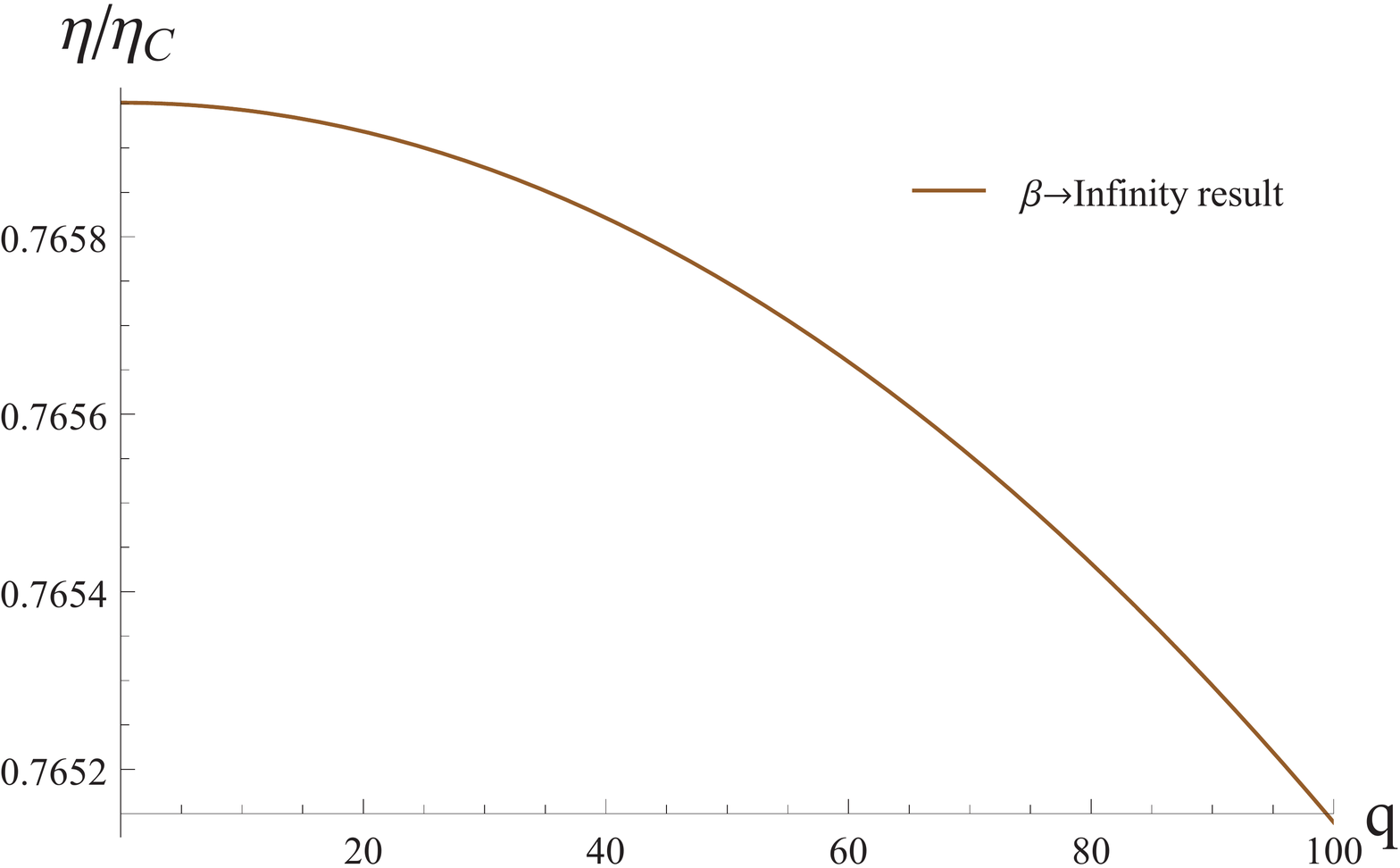} }
			\subfloat[]{	\includegraphics[width=1.9in]{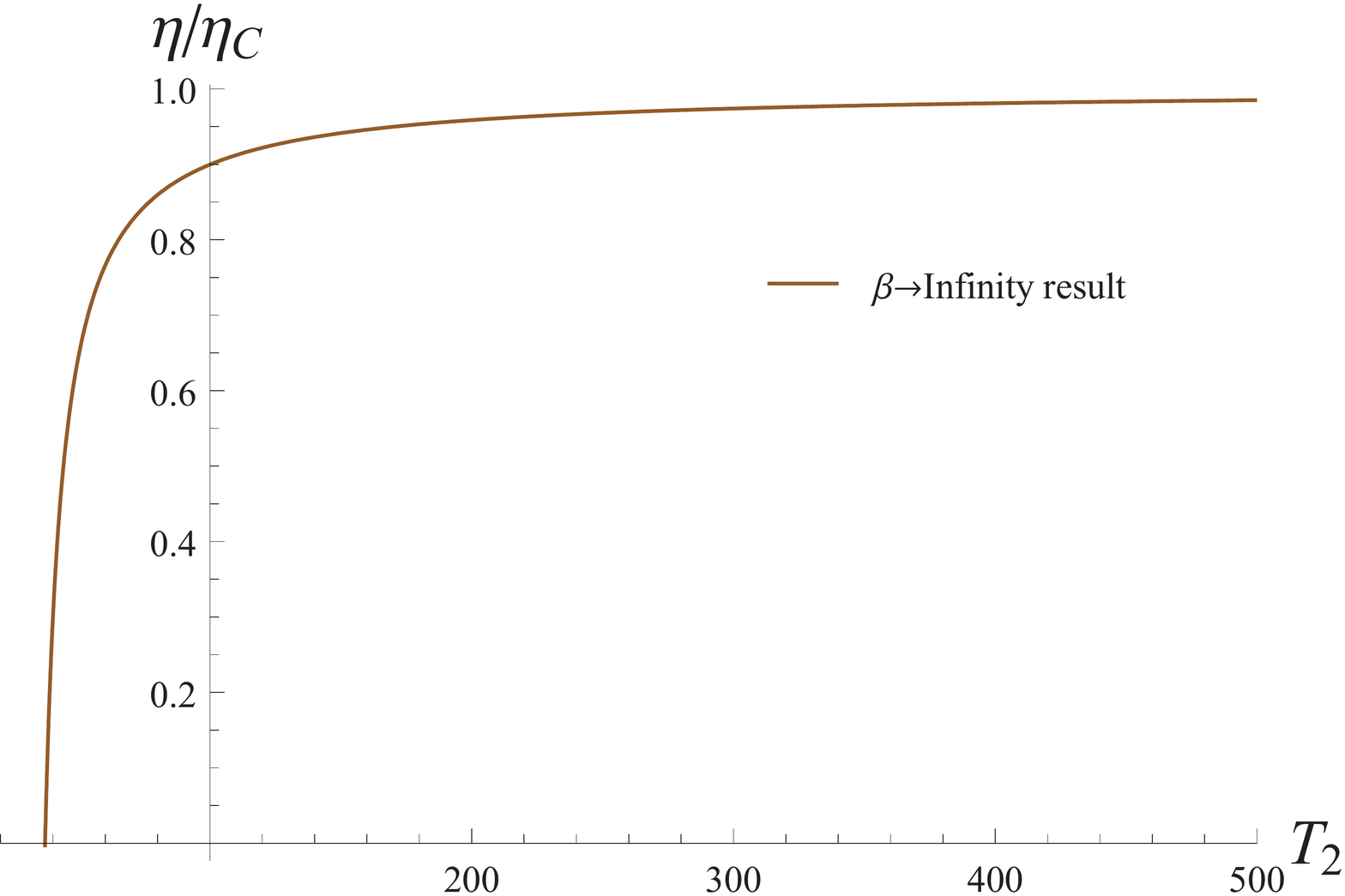} }
			\subfloat[]{	\includegraphics[width=2in]{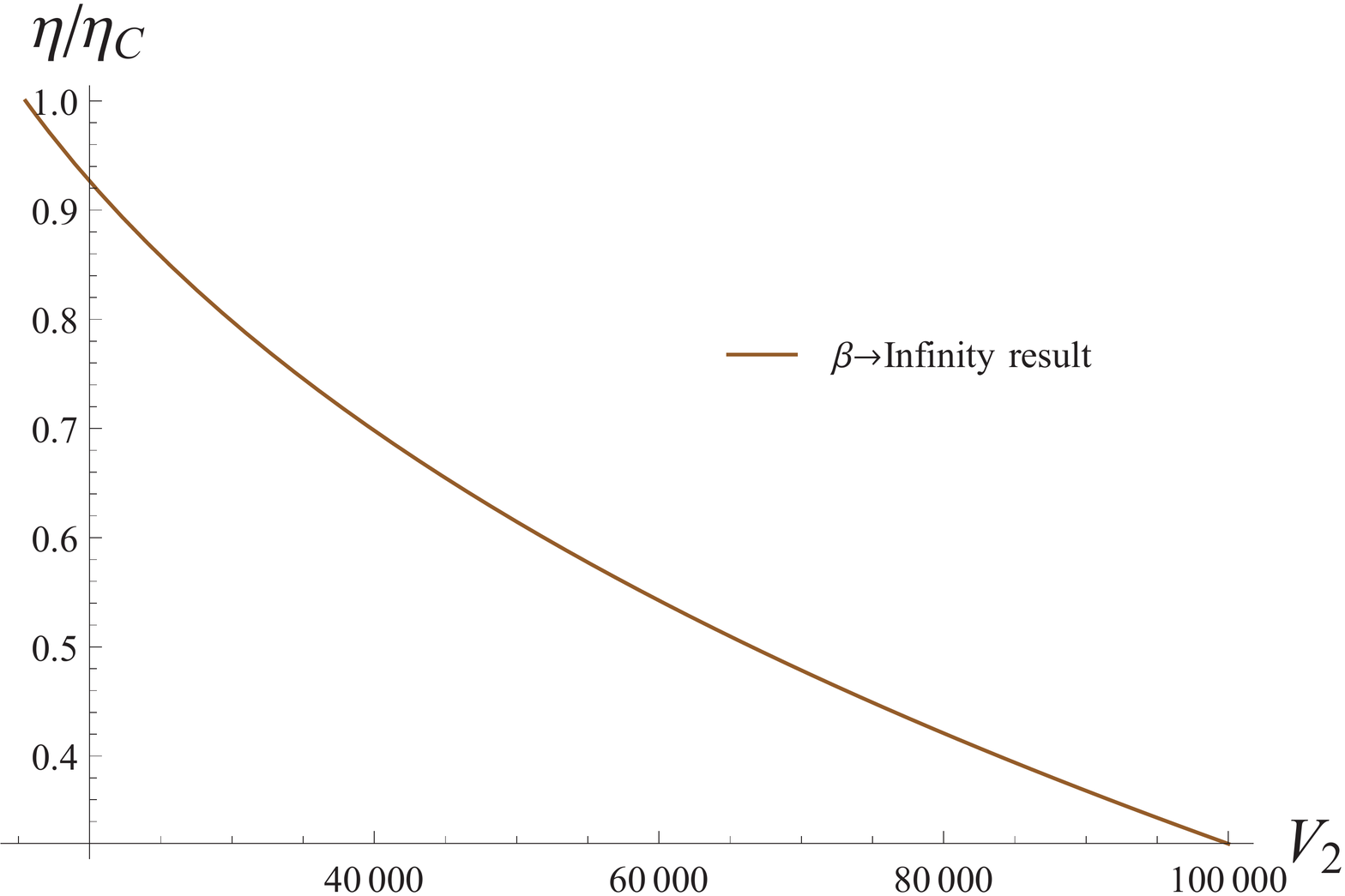} }
			
			\caption{Effect of parameters on  efficiency,  when other parameters (see the caption of figure 2 for parameter values) are fixed and $ \alpha = 0.4.$} \label{varying parameters in alpha}  
			
		}
	\end{center}
\end{figure}

\begin{figure}[h]
	\begin{center}
		{\centering
			\subfloat[]{\includegraphics[width=2.8in]{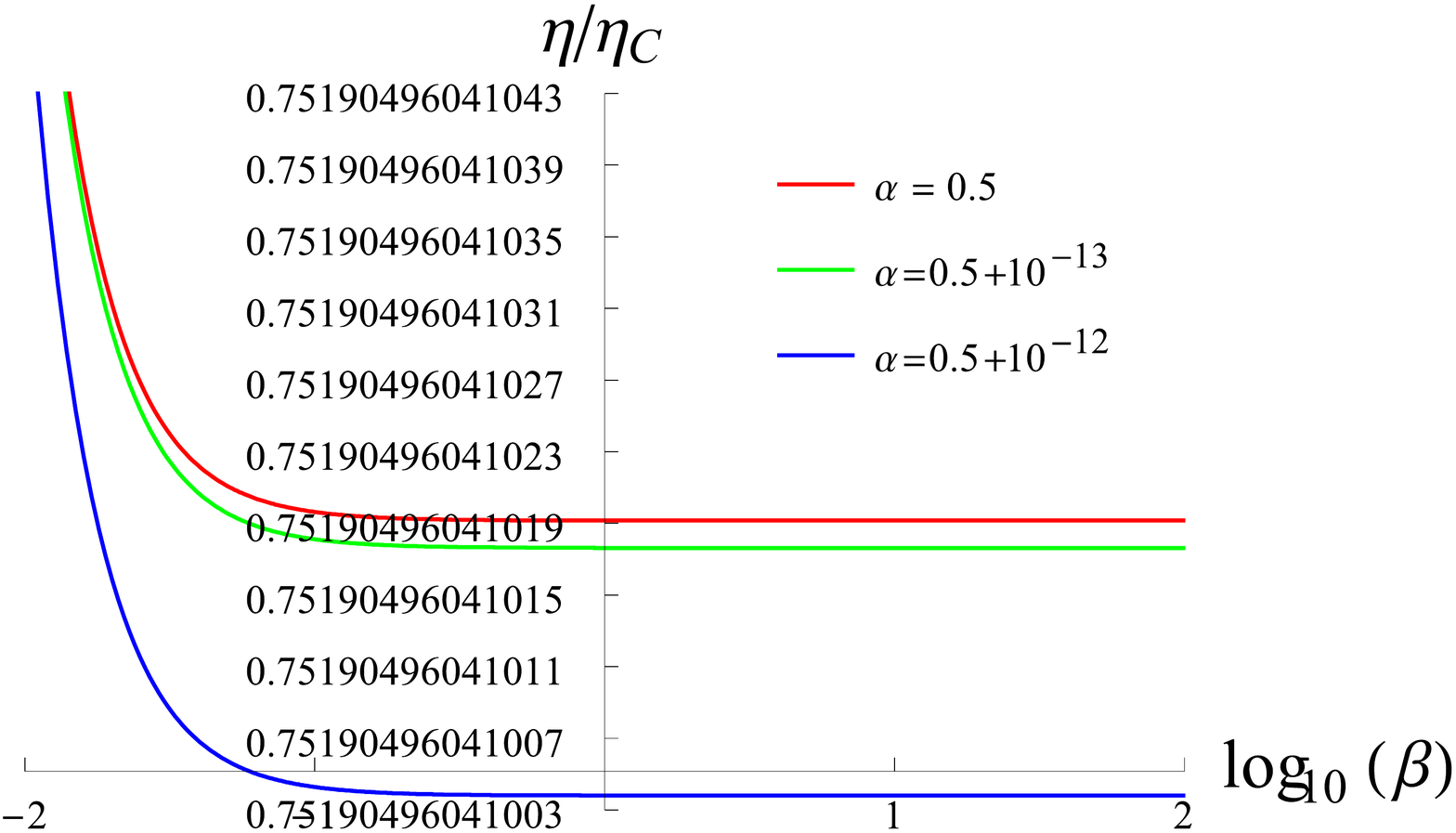} }
			\subfloat[]{\includegraphics[width=2.8in]{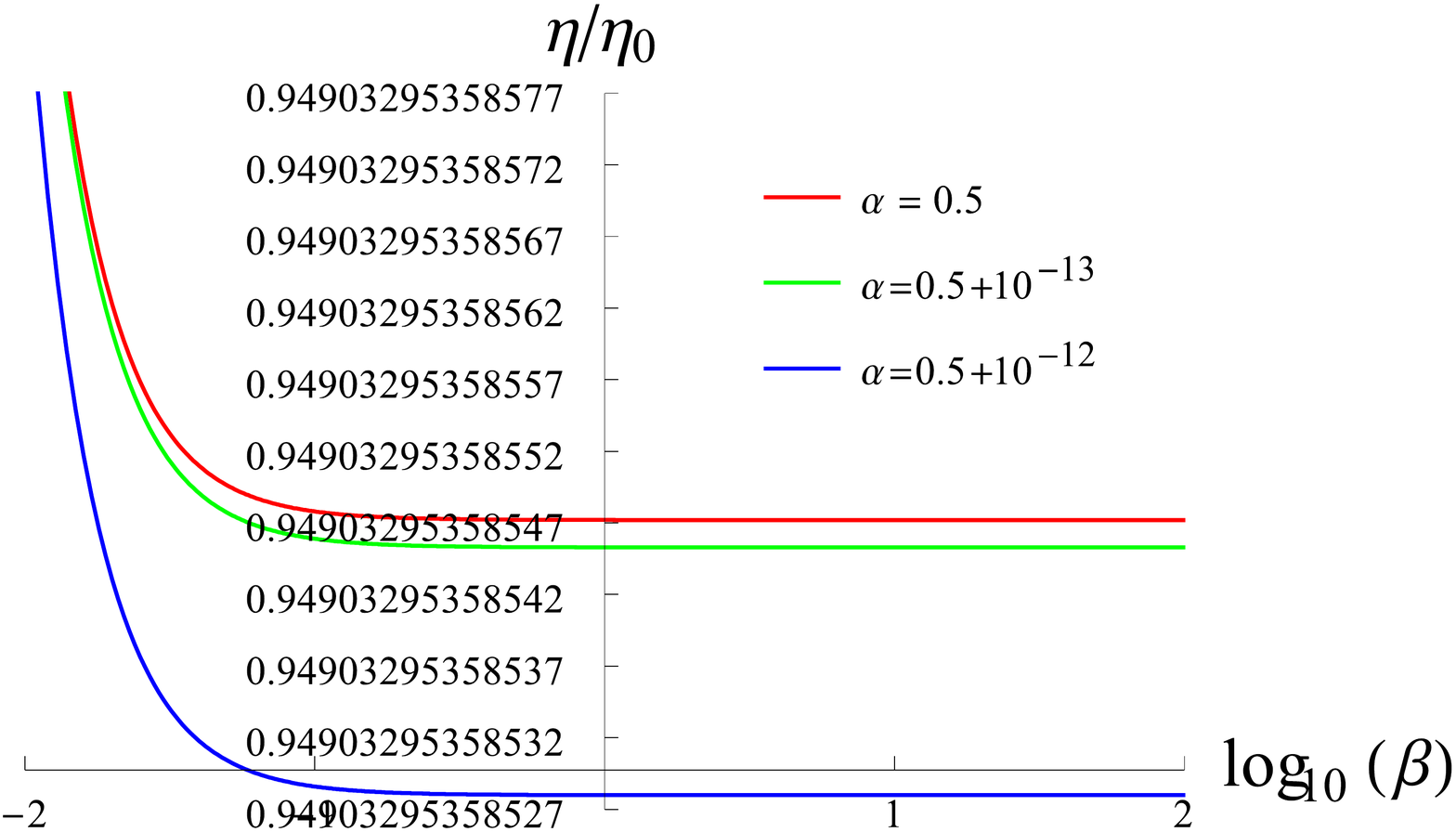} }
			\caption{For the case $\alpha \neq 0$ and $\beta \nrightarrow\infty$, (a) The ratio $\eta/ \eta_C$ vs $\log_{10}(\beta).$ (b) The ratio $\eta/ \eta_0$ vs $\log_{10}(\beta).$ (See the caption of figure 2 for parameter values.) }   \label{fig:5}
		}
	\end{center}
\end{figure}

 When we consider the effect of both the couplings $(\alpha,\beta)$ on the efficiency of our engine (see figure \ref{fig:5}), for a sample range of parameters $10^{-2} < \beta < 10^2$ and $0 < \alpha < 1$ (we checked that the pressures are physical over this sample range of parameters), both the ratios $\eta/\eta^{\phantom{C}}_{\rm C}$ and $\eta/\eta_0$  decrease rapidly in the\textit{ turnaround region} where, roughly, $10^{-2} <\beta <10^{-1}$ and become steady as $\beta$ increases, while as $\alpha$ increases, initially both the ratios  decrease up to $\alpha \approx 0.91$ and then raise again.


\subsection{Brans-Dicke-Born-Infeld Model}
Computation of corrections to efficiency in the Brans-Dicke Born-Infeld theory proceeds by writing down the relevant expression for enthalpy and equation of state~\cite{BDED}(details of black holes and thermodynamic quantities are summarized in Appendix B) as:
\begin{eqnarray} \label{Mass}
	M(r_+, p) &=& \frac{\varpi _{n-1} b^{(n-1)\gamma}(1+\alpha^2)(n-1)}{16\pi}r_+^{(n-1)(1-\gamma)-1}\Bigg\{\frac{(n-2)b^{-2\gamma}}{(1-\alpha^2)(n+\alpha^2-2)}r_+^{2\gamma}-\frac{16\pi p}{(n-1)(\alpha^2-n)}\frac{r_+^{2}}{\Xi} \nonumber \\ 
	&&-\frac{4q^2(\frac{r_+}{b})^{2\gamma(n-2)}}{(n-\alpha^2)r_+^{2(n-2)}}\times \Bigg[\frac{1}{2(n-1)}\digamma_1(\eta_+)-\frac{1}{(\alpha^2+n-2)}\digamma_2(\eta_+) \Bigg]\Bigg\} \, , \\
	&& \nonumber \\
	&&\nonumber \\
p &= & \Bigg[\frac{(n-1)(n-2)}{16\pi(\alpha^2-1)r_+^2}\Big(\frac{r_+}{b}\Big)^{2\gamma}+\frac{(n-1)T}{4(1+\alpha^2)r_+}+\frac{q^2}{8\pi r_+^{2(n-1)}}\Big(\frac{r_+}{b}\Big)^{2\gamma(n-2)}\digamma_1(\eta_+)\Bigg]\Xi.
\end{eqnarray}

Here, $r_+$ is related to the thermodynamic volume $V$ as

\[ 
V = 
\frac{\varpi _{n-1}(1+\alpha^2)}{(n-\alpha^2)}r_+^n\begin{cases}
\Big(\frac{r_+}{b}\Big)^{-\gamma (n-1)},& \text{dilatonic BI,} \\
\Big(\frac{r_+}{b}\Big)^{-\frac{\gamma (n^2-4n-1)}{n-3}},              & \text{BD-BI.}
\end{cases}
\]
where,
\[ 
\Xi  =  
\begin{cases}
1,& \text{dilatonic BI,} \\
\Big(\frac{r_+}{b}\Big)^{-\frac{4\gamma}{n-3}},              & \text{BD-BI.}
\end{cases}
\] 
\begin{eqnarray} 
	\digamma_1(\eta) &=&\text{}_{2}F_{1}\left( \left[ \frac{1}{2},\frac{(n-3) \Upsilon
	}{(\alpha^2+n-2)}\right] ,\left[1+ \frac{(n-3)\Upsilon}{(\alpha^2+n-2)}\right]
	,-\eta\right),  \nonumber \\ 
\digamma_2(\eta) & =& \text{}_{2}F_{1}\left( \left[ \frac{1}{2},\frac{(n-3) \Upsilon
}{2(n-1)}\right] ,\left[1+ \frac{(n-3)\Upsilon}{2(n-1)}\right]
,-\eta\right)\nonumber,\\
\eta & = & \frac{q^2 }{\beta^2r^{2(n-1)}}\Big(\frac{r}{b}\Big)^{2\gamma (n-1)(n-5)/(n-3)} \nonumber,\\
\eta_+ & = & \eta(r) \Big|_{r = r_+}, \nonumber \\
\Upsilon & = & \frac{\alpha^2+n-2}{2\alpha^2+n-3} \nonumber.
\end{eqnarray}
Now, to study  the efficiency of the engine we cast our rectangular cycle in the Einstein frame as well as in the Jordan frame. Hereafter, for simplicity, we  take $\beta \rightarrow \infty $. \\

\begin{figure}[h] 
	\begin{center}
		{\centering
			\subfloat[]{	\includegraphics[width=2.5in]{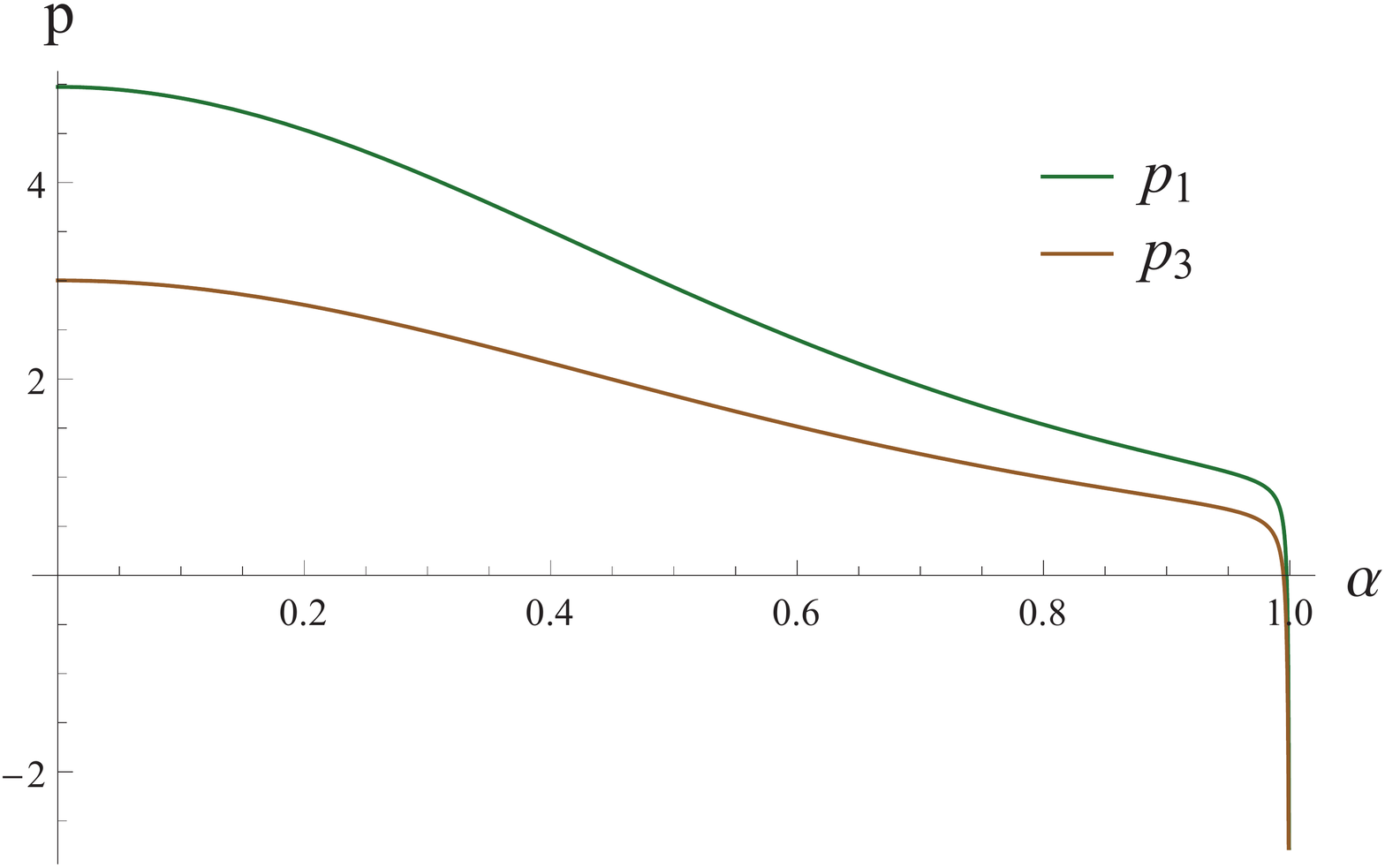} }
			\subfloat[]{	\includegraphics[width=2.5in]{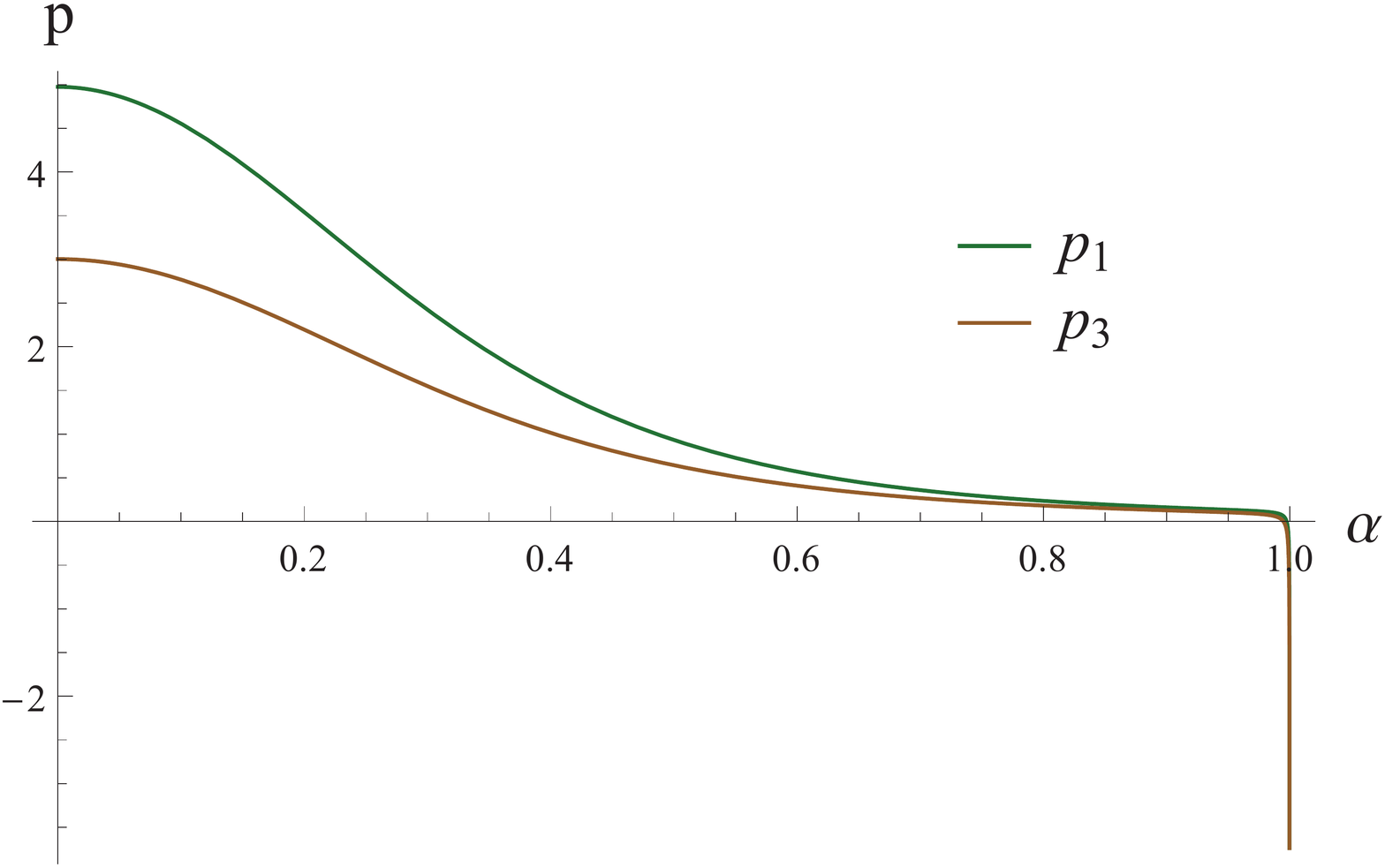} }			
			\caption{  The behavior of the pressures of upper isobar ($p_1$) and  lower isobar ($p_3$) for  engine cycle with respect to $\alpha$ in:   (a) Einstein-BI-dilaton theory and  (b) BD-BI theory.  (See the caption of figure 2 for parameter values.)} \label{pressureEDBD}   
		}	
	\end{center}
		\end{figure}

\par The behavior of pressures in both frames can be seen from figure (\ref{pressureEDBD}), which shows the rapid fall of pressures in Jordan frame, moreover, the pressure of the isobar in the Einstein frame is higher than the pressure of the corresponding isobar in Jordan frame. In fact, the height of the cycle $(p_1-p_3)$ is more for the Einstein frame which leads to more work.\\
 \par
\begin{figure}[h]
	\begin{center}
		{\centering
			\subfloat[]{	\includegraphics[width=3.15in]{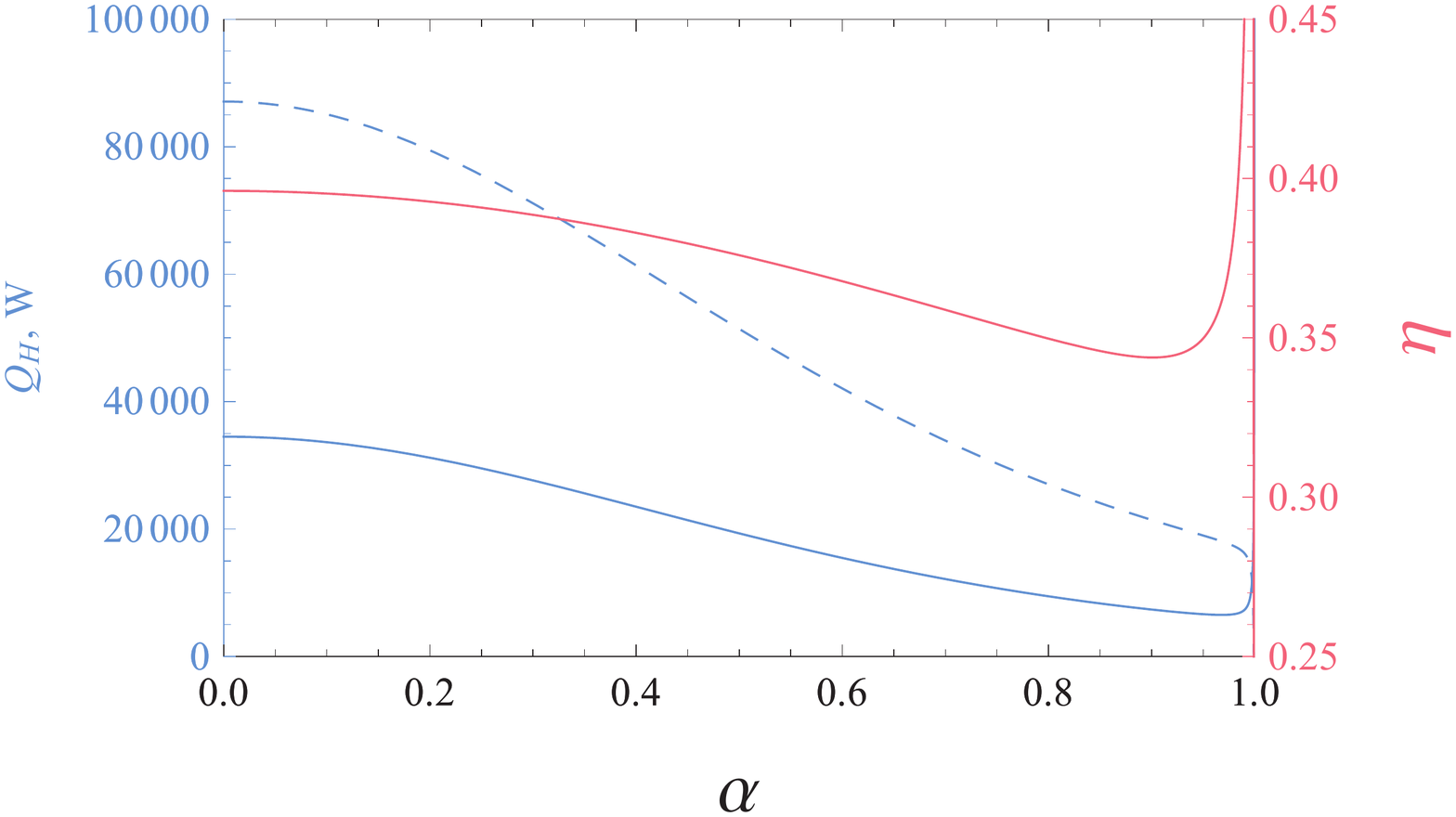} }
			\subfloat[]{	\includegraphics[width=3.15in]{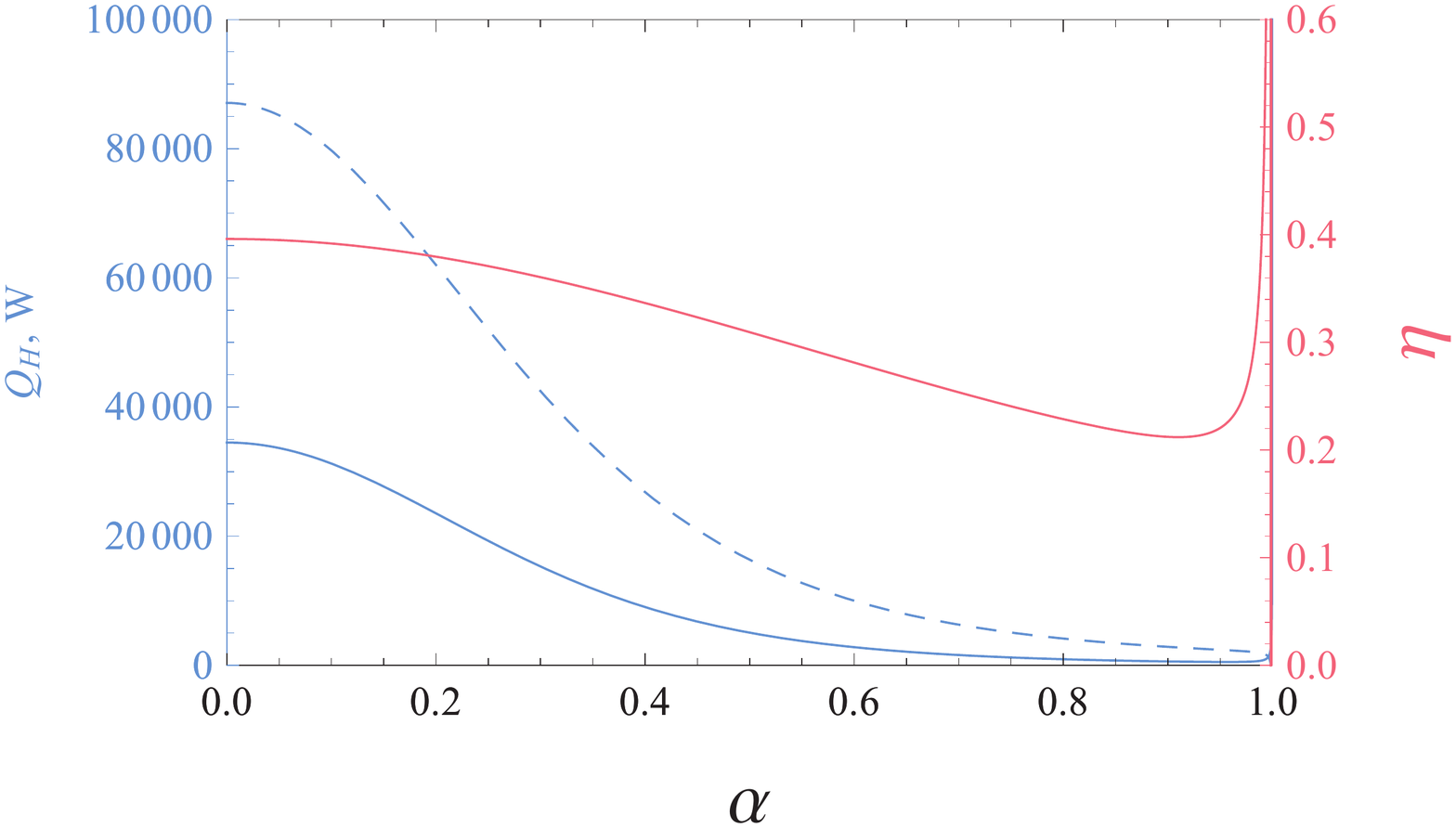} }			
			\caption{ Total work done (solid blue curve)  and a net  inflow of heat (dashed blue curve) for the engine are plotted on left y-axis, efficiency (solid red curve) is plotted on right y-axis for:   (a) Einstein-BI-dilaton theory and  (b) BD-BI theory.  (See the caption of figure 2 for parameter values.)}  \label{QWita}
	}
	\end{center}
\end{figure}

 From figure (\ref{QWita}), we can see that as  $\alpha$ increases, in Einstein frame, the inflow of heat   monotonously decreases  and the work done   decreases up to $\alpha = 0.97$ then raising, while the efficiency is decreasing slowly up to $\alpha= 0.9 $ before a rapid rise.  Whereas, in Jordan frame, $Q_H$ is again  monotonously decreasing and $W$ is decreasing up to $\alpha = 0.96$ then raising while the efficiency shows similar behavior with Einstein frame, however the minimum efficiency occurs at $\alpha = 0.91.$ Regardless of frames, maximum values of $Q_H$ and $W$ occur at $\alpha = 0$, whereas efficiency reaches to higher values when $\alpha \rightarrow 1$. Indeed, ${Q_H}_{Max} = 87089$ and $W_{Max} = 34499.7.$

\begin{figure}[h]
	\begin{center}
		{\centering
			\subfloat[]{	\includegraphics[width=2in]{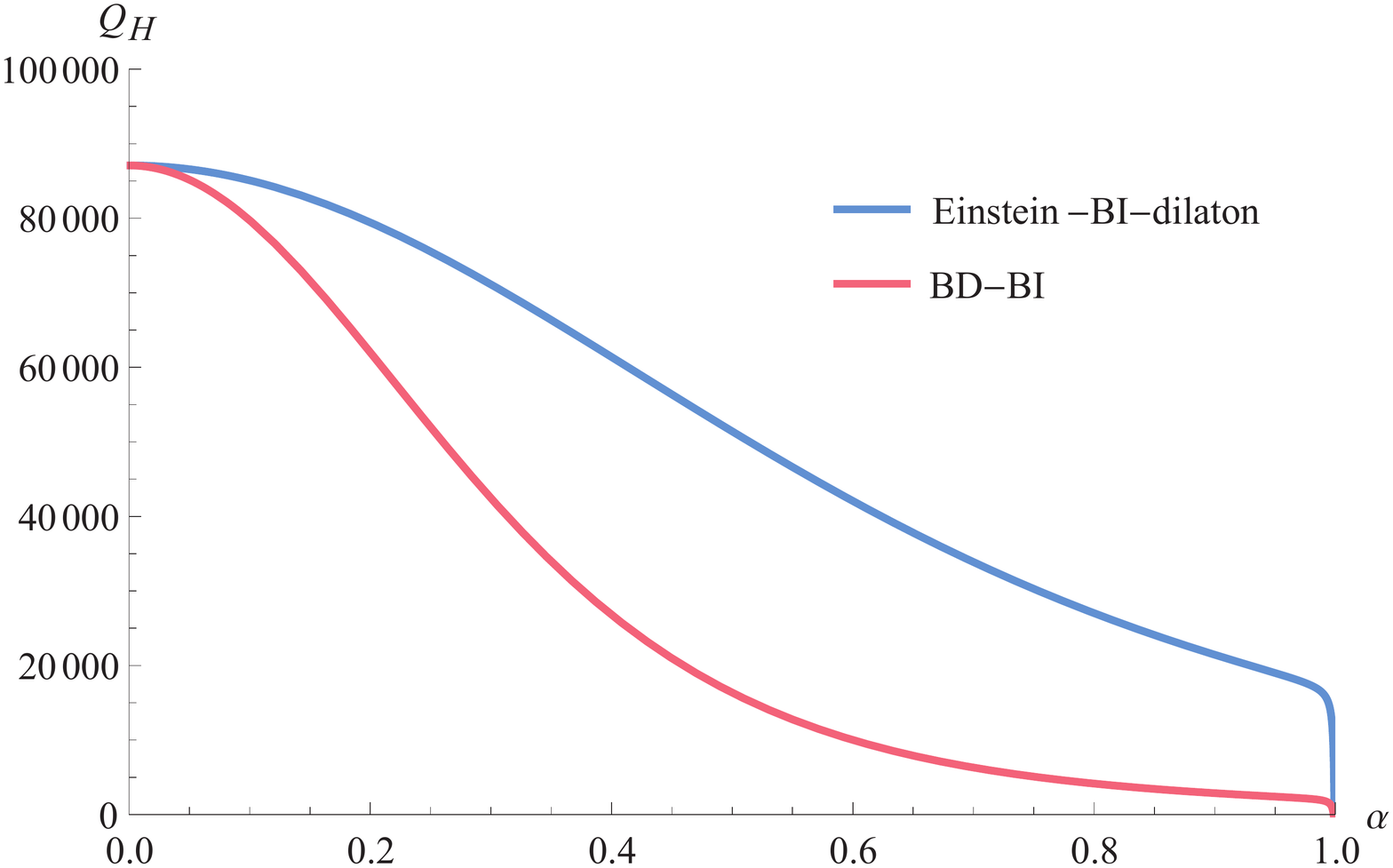} }
			\subfloat[]{	\includegraphics[width=2in]{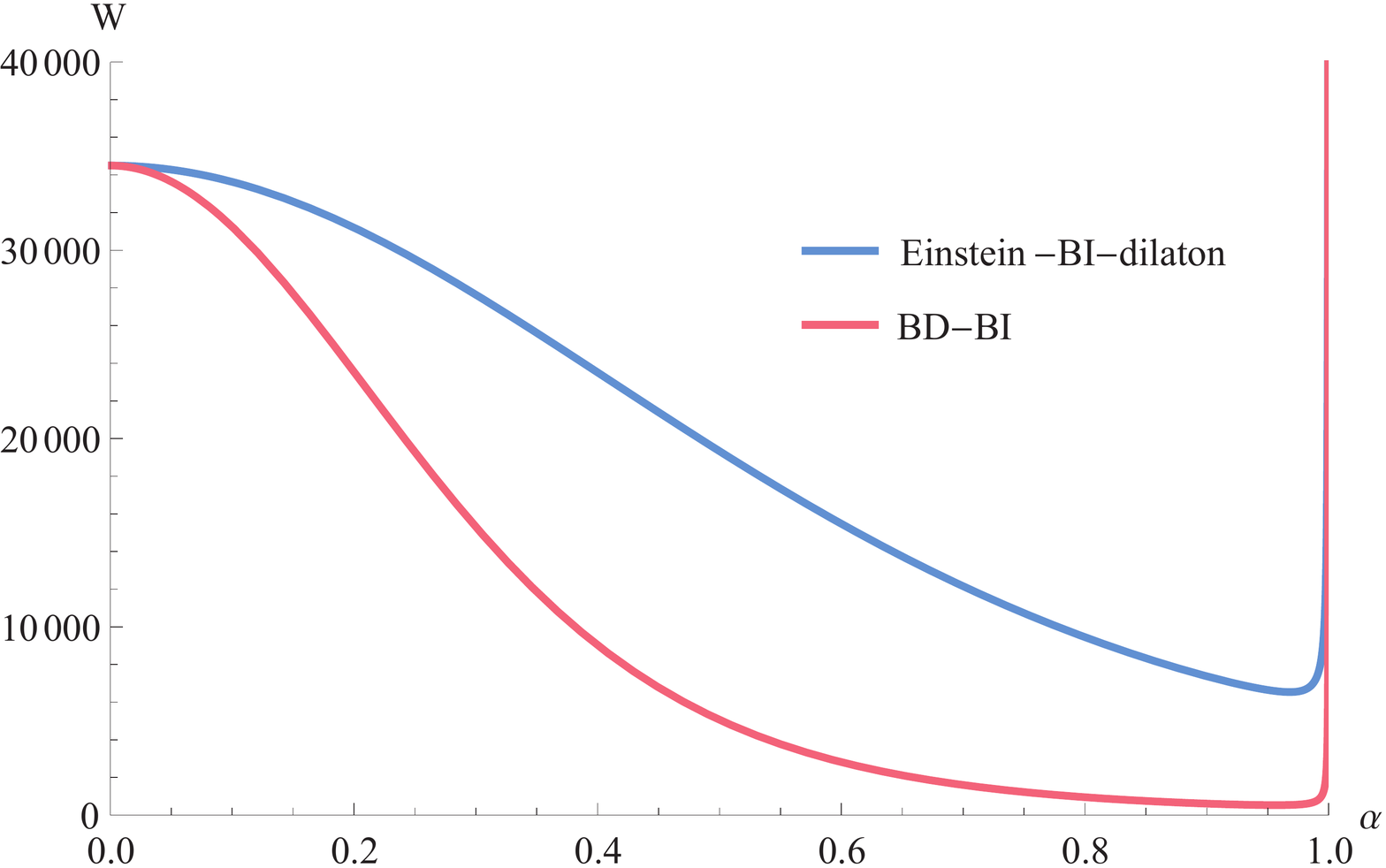} }
				\subfloat[]{	\includegraphics[width=2in]{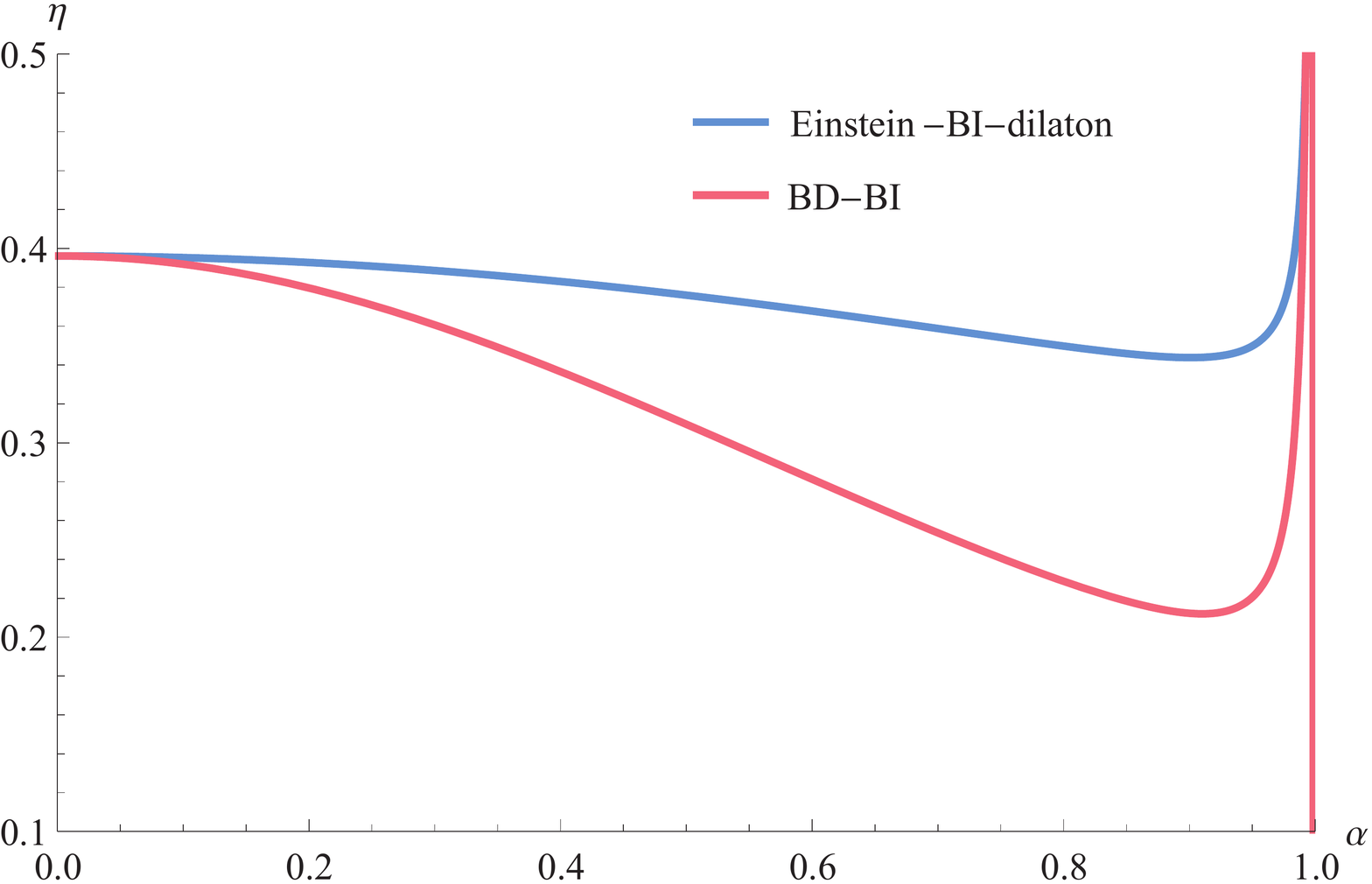} }
			\caption{Plots for (a)  the net inflow of heat (b) total work done and (c) the efficiency vs $\alpha$.  (See the caption of figure 2 for parameter values.)}  \label{comparitiveQWita} 
		}
	\end{center}
\end{figure}

\begin{figure}[h]
	\begin{center}
		{\centering
			\subfloat[]{\label{cnstp}	\includegraphics[width=2.8in]{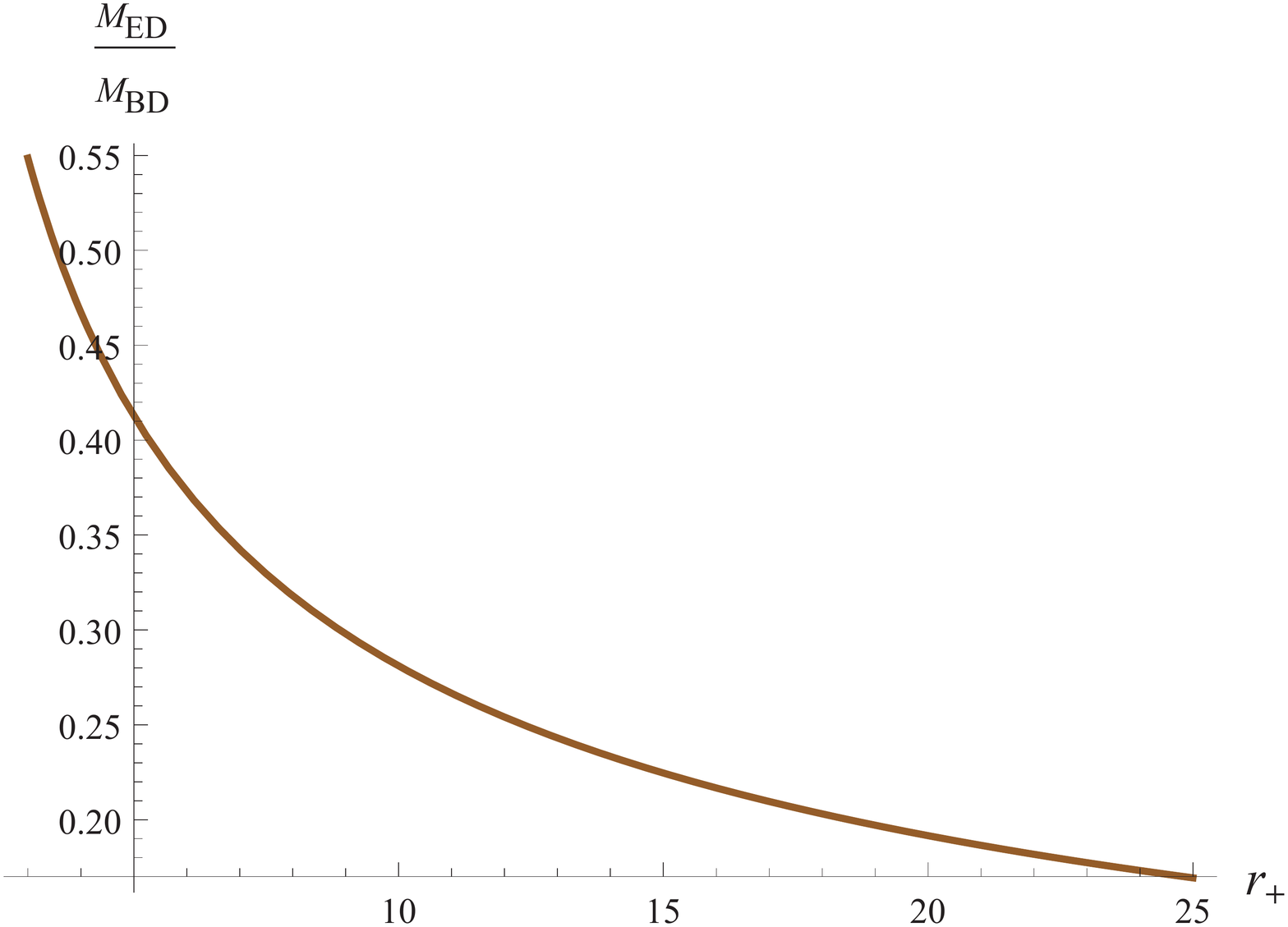} } 
			\subfloat[]{\label{cnstr} 	\includegraphics[width=2.8in]{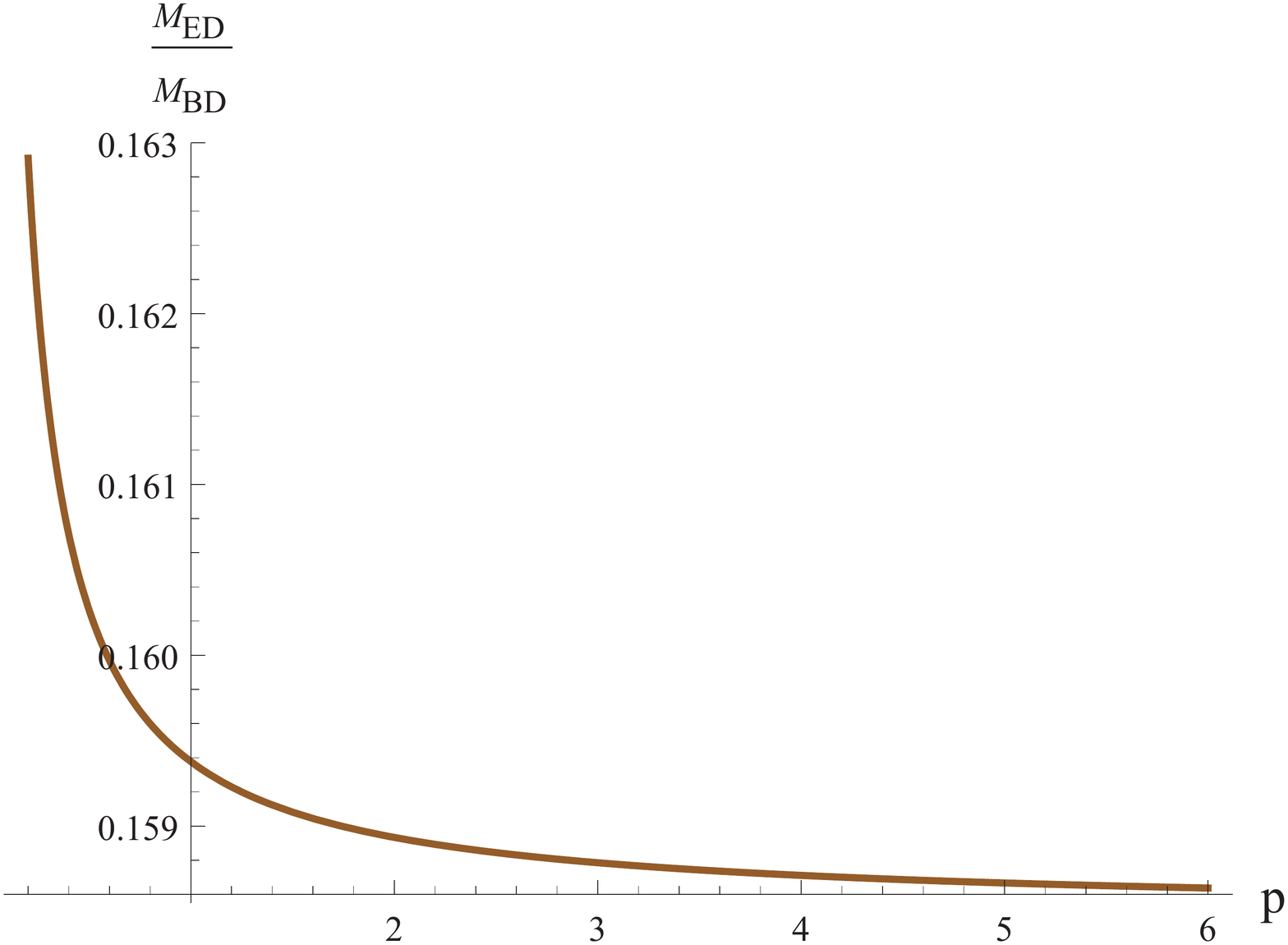} }	
			\caption{Plots for the ratio of enthalpy in Einstein theory to enthalpy in BD theory $(M_{ED}/M_{BD})$   (a) vs $r_+$ at $p = 3, \alpha = 0.4$ (b) vs $p$ at  $r_+ = 10,  \alpha = 0.5$.  ($n=4, b=1, q=0.1.$for both plots)}  
			}
	\end{center}
\end{figure}

For comparison, we plotted $Q_H,  W$ and $\eta$ in figure (\ref{comparitiveQWita}).  It can be seen that for a given value of $\alpha$,  engine running in Einstein frame takes more heat and generates more work and is also more efficient as compared to engine run in Jordan frame.  This is so because  the enthalpies (expressed in $r_+  , p$) are not same for each frame (although the expressions for mass are same). At a given value of $\alpha$, the enthalpy in BD theory dominates over the Einstein theory at the same pressure (figure \ref{cnstp}), as well as at the same horizon radius $r_+$ (figure \ref{cnstr}). This implies that at a given $(r_+, p)$, enthalpy  is more in BD theory than  in the Einstein theory. \\

\par
\begin{figure}[h]
	\begin{center}
		{\centering
		{	\includegraphics[width=2.8in]{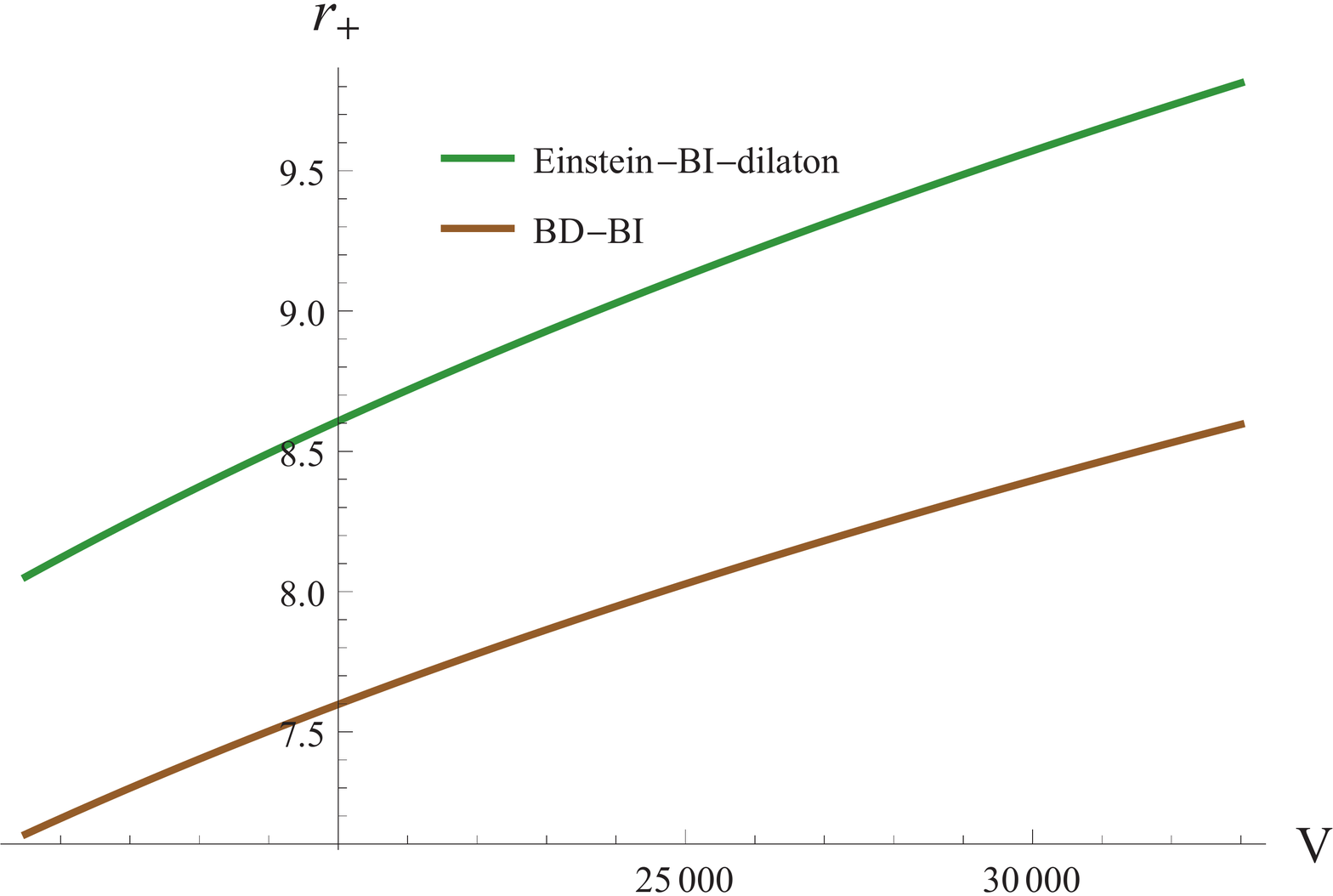} }
						
			\caption{Horizon radius  $r_+$ vs volume $V$  for  $\alpha =0.25,   n=4,   b=1. $}   \label{r+plot}
		}
	\end{center}
\end{figure}

 Although BD theory dominates in enthalpy over Einstein theory, if we evaluate the enthalpies at the corners of the cycle, Einstein theory   dominates over BD theory.   This is because for a given volume, horizon radius $r_+$ of the black hole is large  in Einstein frame than that in the Jordan frame (see figure \ref{r+plot}).  Also, for a given $(V, T)$, the pressure is more in the Einstein frame than that in the Jordan frame. Since, the equation of states (expressed as $p(V, T)$) are not same for both frames (even both  have the same expression for Hawking temperature). 

\begin{figure}[h]
	\begin{center}
		{\centering
			\subfloat[]{	\includegraphics[width=2.1in]{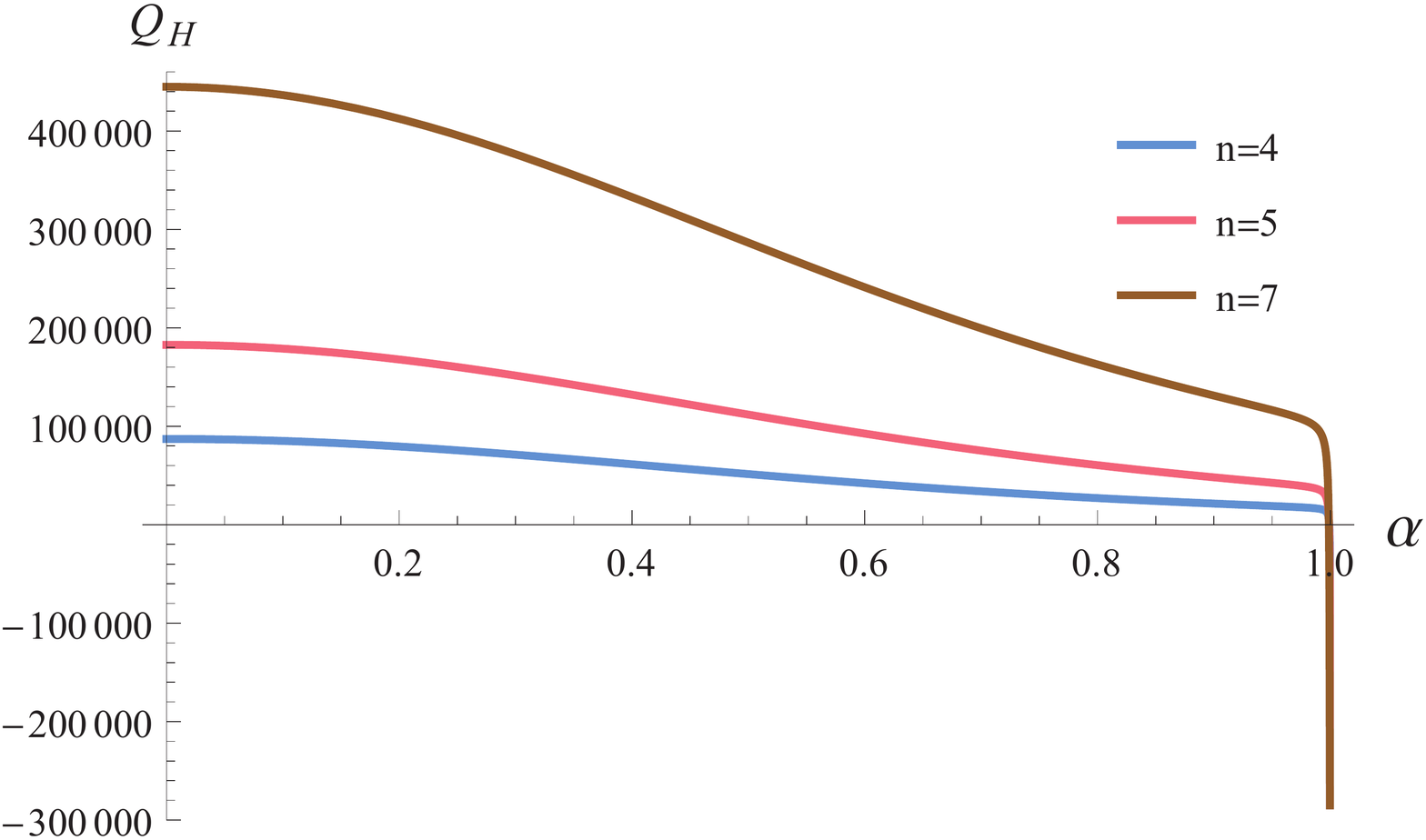} }
			\subfloat[]{	\includegraphics[width=1.9in]{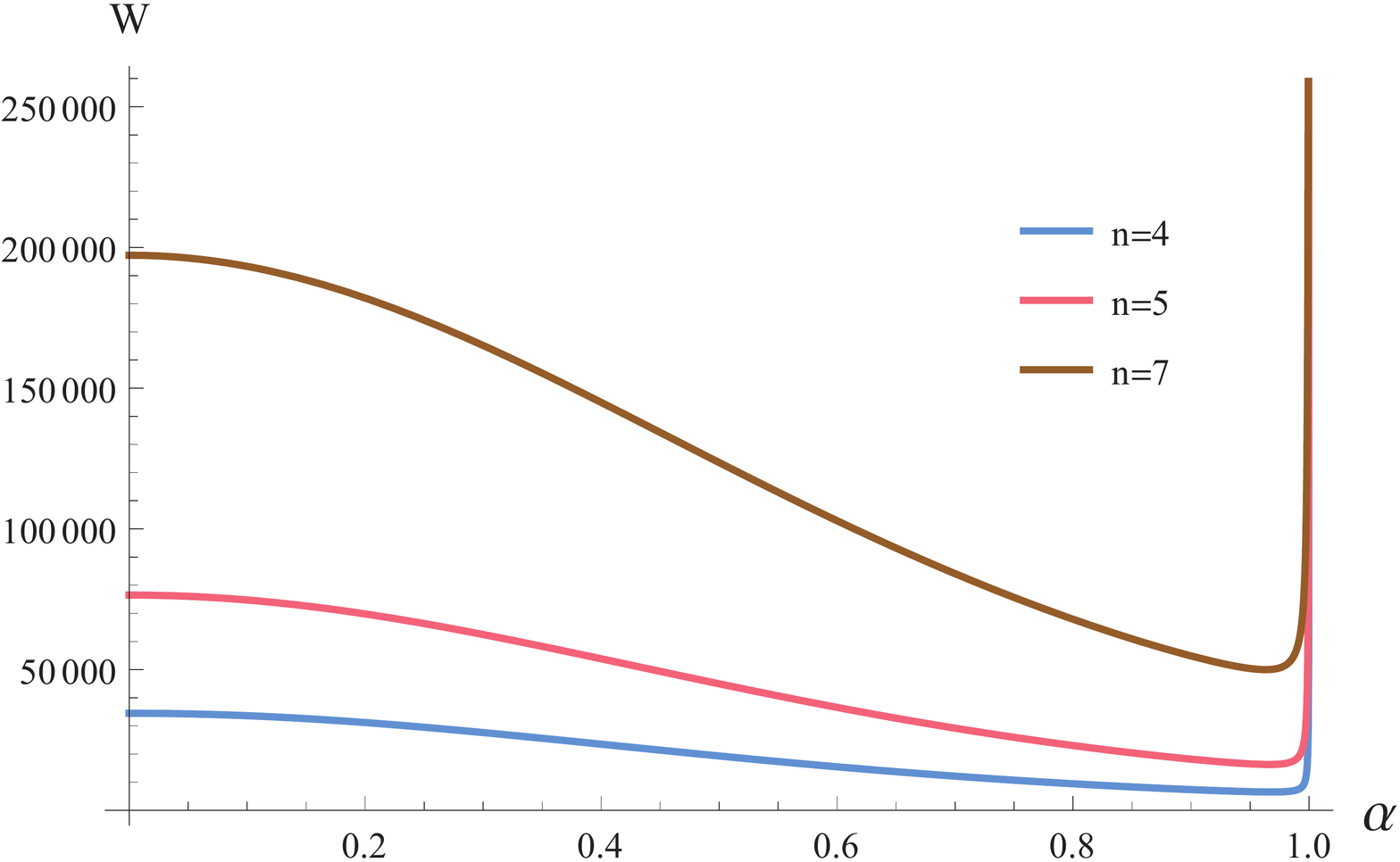} }
			\subfloat[]{	\includegraphics[width=2in]{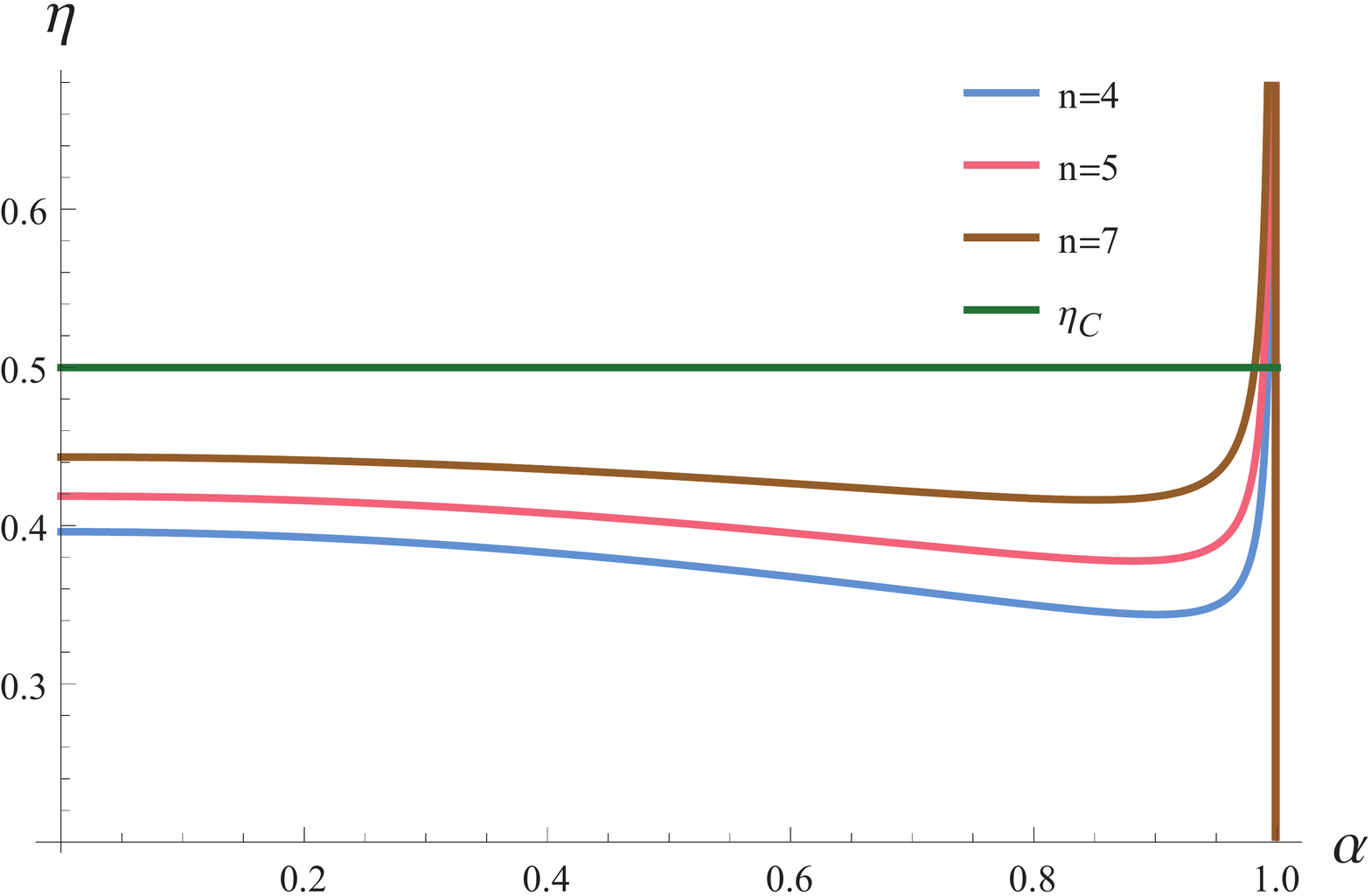} }
			
			\caption{In Einstein frame,  plots for the net inflow of heat $Q_H$, work done  $W$ and the efficiency $\eta$ with respect to $ \alpha $ (See the caption of figure 2 for parameter values.)}  \label{HD} 
		}
	\end{center}
\end{figure}
                     
\begin{figure}[h]
	\begin{center}
		{\centering
			\subfloat[]{	\includegraphics[width=2.1in]{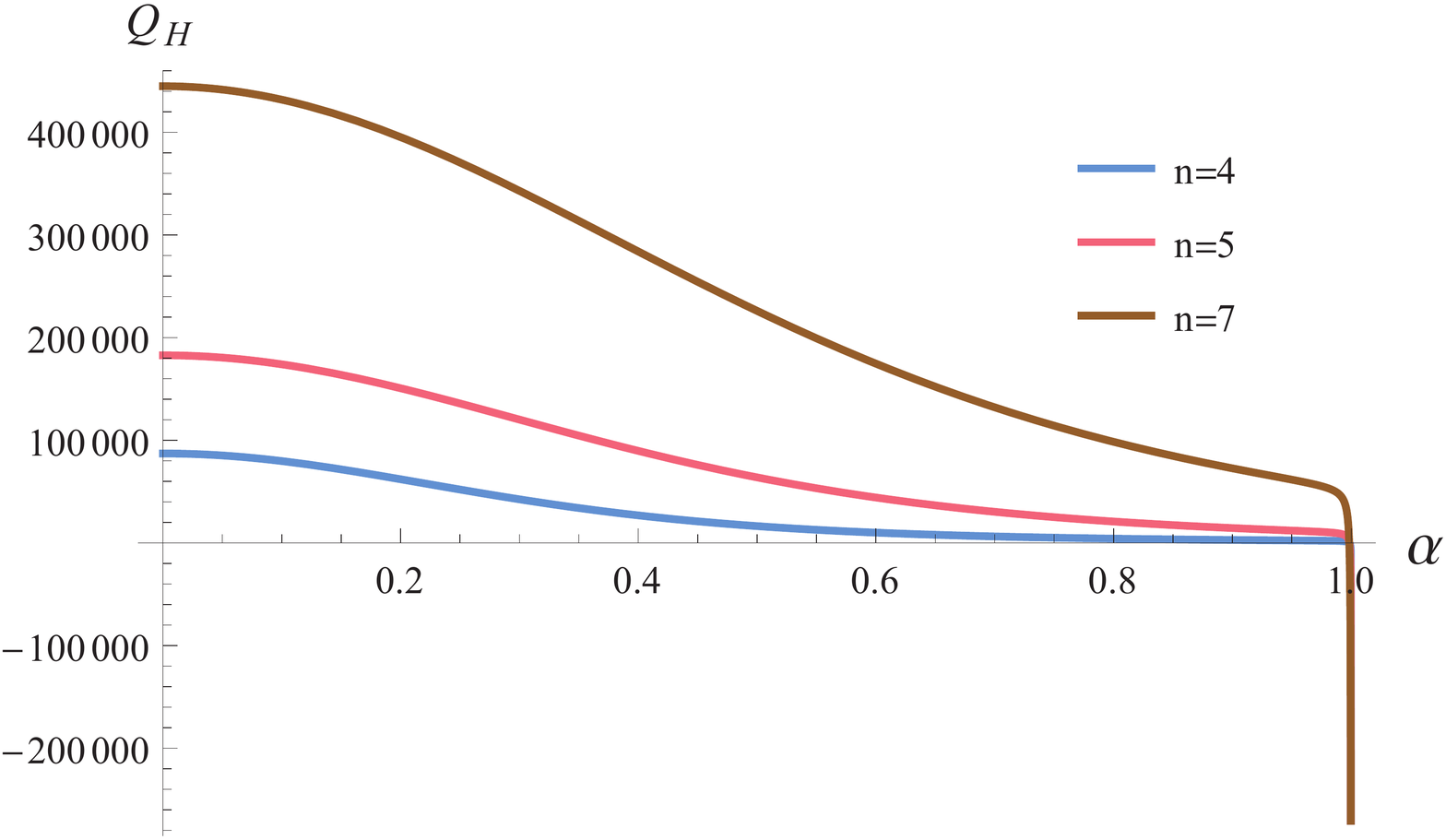} }
			\subfloat[]{	\includegraphics[width=1.9in]{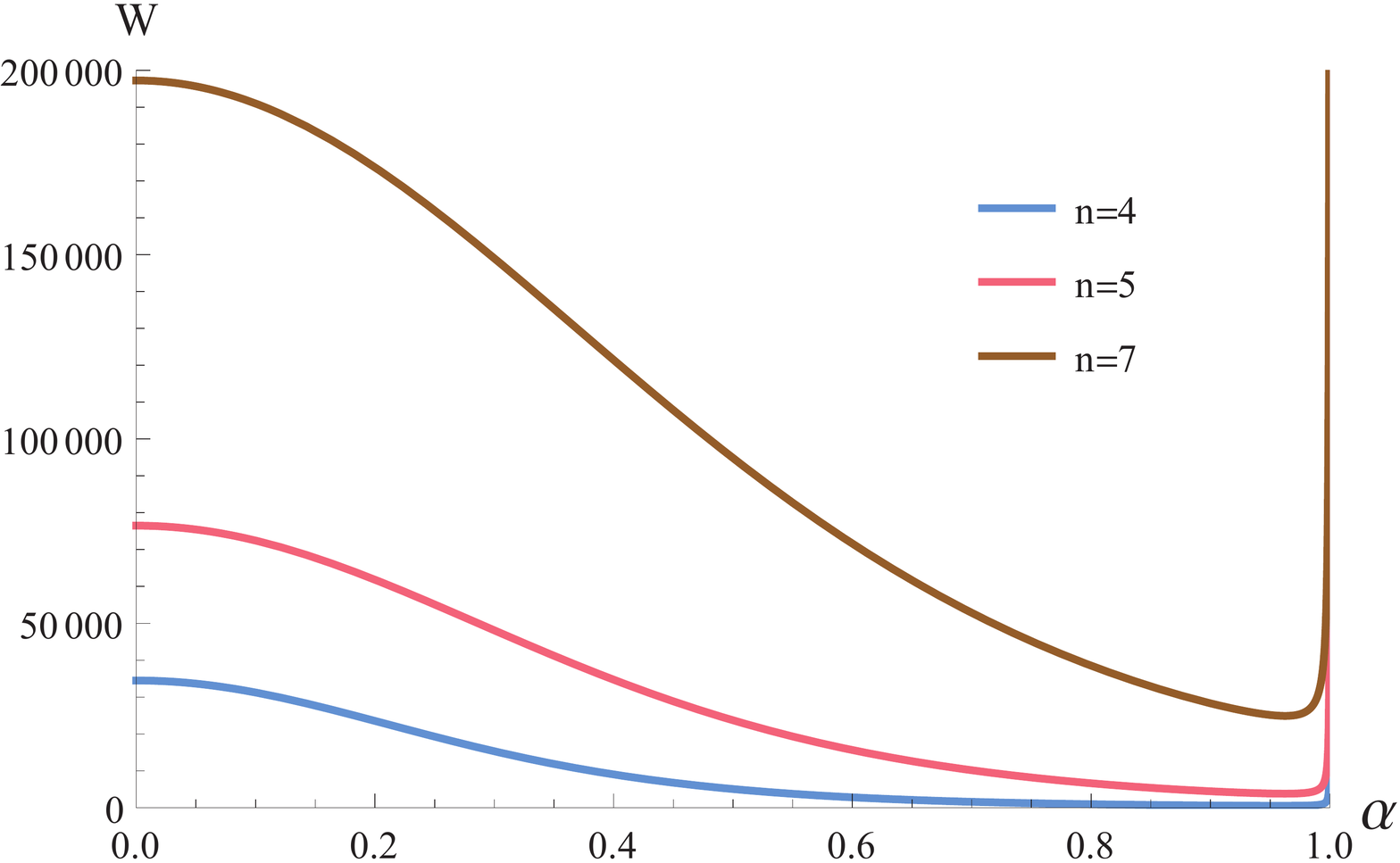} }
			\subfloat[]{	\includegraphics[width=2in]{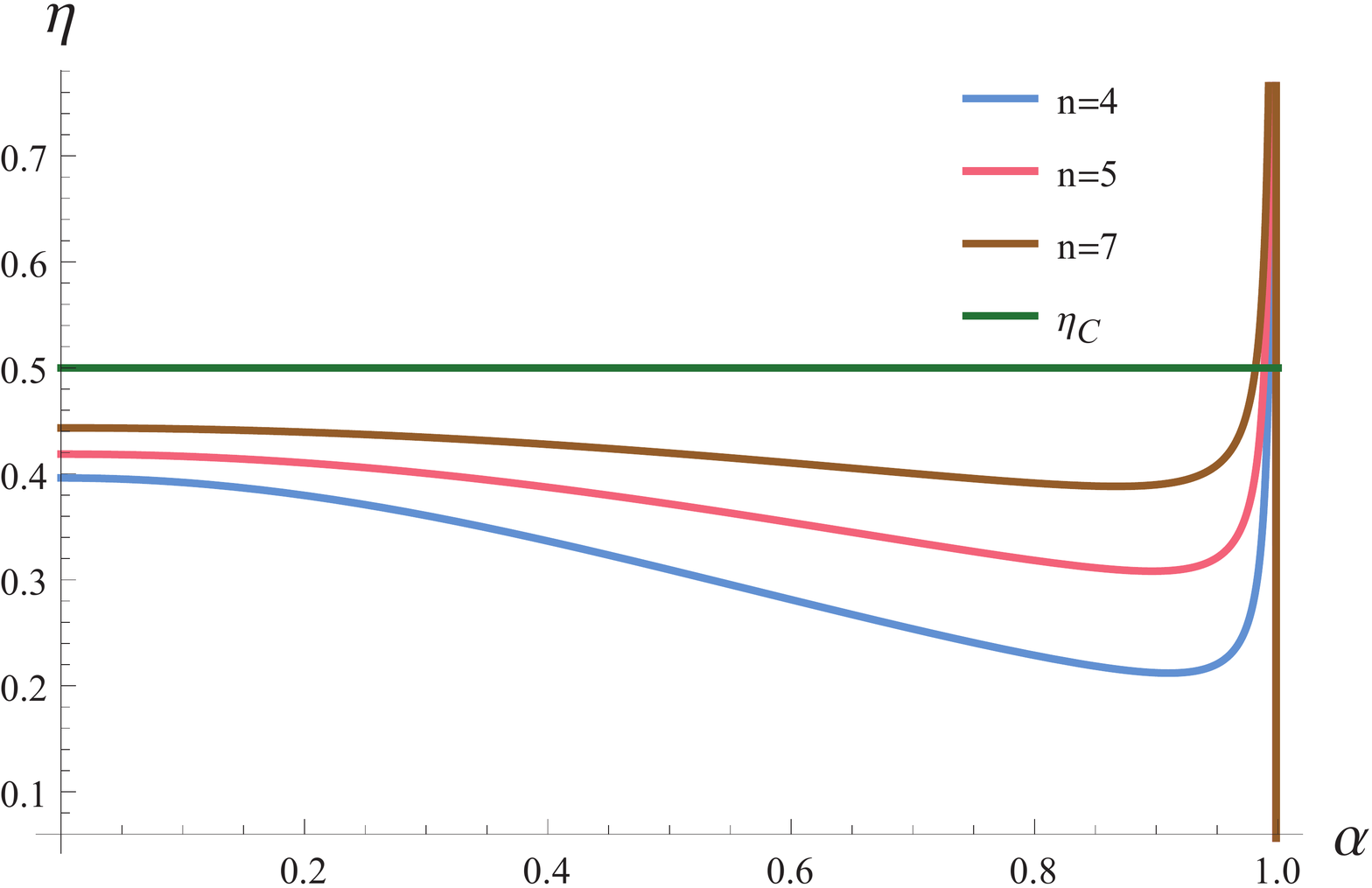} }
			
			\caption{In Jordan frame,  plots for the net inflow of heat $Q_H$, work done  $W$ and the efficiency $\eta$ with respect to $ \alpha $ (See the caption of figure 2 for parameter values.)} \label{HBD}   
		}
	\end{center}
\end{figure}

\par In both frames, we find that  the net inflow of heat $Q_H$, work $W$ and efficiency $\eta$  increase with $n$ (see figures \ref{HD} \& \ref{HBD}), while the allowed range of $\alpha$  decreases from the upper bound when  regulated with Carnot efficiency  $\eta^{\phantom{C}}_{\rm C}  $. In fact the window of the allowed values of $ \alpha $ is wider in Einstein frame  than that in Jordan frame (see Table 1).

	\begin{table*}[h!] 			
\begin{center}
	\begin{tabular}{ ccc } 
		& \textbf{ Table 1:}  Allowed   range of $\alpha$ for the efficiency in figures \ref{HD} and \ref{HBD}.	\\	
		\hline
		n & BI-dilaton & BD-BI \\
	\hline 
		4 &0-0.992536  &0-0.992467  \\ 
		5 &0-0.988880  &0-0.988811  \\ 
		7 &0-0.981655  &0-0.981598  \\ 
		\hline
	
	\end{tabular}
\end{center}
		\end{table*}

\section{Remarks}
We studied the effect of dilaton and Born-Infeld couplings on the efficiency of the holographic heat engines in Einstein Gravity (with  negative cosmological constant), where charged black hole is the working substance, in spite of the dependence of thermodynamic volume on dilaton coupling \cite{sheykhi} and unusual asymptotics \cite{CHM}, $pdV$ terms exist \cite{BDED} and mechanical work is extracted via the $pdV$ terms present in the first law of extended gravitational thermodynamics with a dynamical cosmological constant.  In the case where the dilaton coupling is absent, our exact result agrees with the 
high temperature calculation in~\cite{Johnson:2015fva}. As seen from figure (\ref{fig:5}), this behavior continues to hold even for non zero values of dilaton coupling constant, signifying that the variations in the efficiency with $\beta$ are several orders of magnitude lower than that with $\alpha$. A similar feature was also noticed in the context of heat engines in Gauss-Bonnet black holes~\cite{Johnson:2015fva}. Increasing the parameter $q$ effects the efficiency significantly as seen from figure (\ref{different q in beta}). We noticed  in both Born-Infeld and dilaton cases that the increase  in   charge $q$ and volume $V_2$ lowers the efficiency, where as, the ratio $\frac{\eta}{\eta^{\phantom{C}}_{\rm C}}$ approaches   unity  on the account of  increase in temperature $T_2$. Also  for large $p_1$, leading behavior of the efficiency is $(1-\frac{p_4}{p_1})$. In fact, increase in $q, V_2$ and $T_2$ implicitly changes the height of the cycle $\Delta p \equiv p_1-p_4$ which changes the efficiency accordingly\cite{fr}.
\vskip 0.5cm
\noindent
We also compared the efficiency of engines in dilatonic Born-Infeld theory and Brans-Dicke Born-Infeld theory.
We see that our engine produces more work in the Einstein frame than in the Jordan frame. The Einstein frame provides longer cycle along with high pressures for the corresponding isobars in the Jordan frame. Though, black hole possesses more enthalpy in Brans-Dicke theory, for a fixed volume, the horizon radius is larger for Einstein black hole. Hence, the calculation of $Q_H$ and efficiency $\eta$ as a function of enthalpies evaluated at the corners of cycle yields larger values in the Einstein frame.  We find that irrespective of the frames, the maximum values for $Q_H$ and $W$ occur at $\alpha = 0$, while efficiency $\eta$ reaches to higher values when $\alpha \rightarrow 1$.  We also checked that the qualitative behavior of efficiency (as well as $Q_H, W$) does not alter in higher dimensions, although, the allowed range of $\alpha$
is small.  
\vskip 0.5cm
\noindent
For both the dilatonic Born-Infeld and Brans-Dicke-Born-Infeld, we choose a scheme where highest and lowest temperatures are held fixed to have a get close to the scheme independent answer. In this case, a comparison of our efficiency with two standards, i.e., the case $\eta_0$ (Einstein-Maxwell theory) and the Carnot efficiency $\eta_C$ could be performed. However, it would be nice to study the behavior of efficiency in other possible schemes, as the equation of state still depends on coupling constants of the model and similar behavior is a priori not guaranteed. It would also be nice to have a better holographic understanding of heat engines which have been studied thus far, both with and without dilaton couplings and/or including other higher order gauge/gravity corrections to Einstein Gravity (whether asymptotically AdS/flat or not). An engine operating at the critical point could show further interesting scaling properties\cite{Johnson:varing_q}, especially in the large charge limit \cite{Johnson:2017asf}. We leave these issues for future work.
		
\appendix
\section{Appendix}
The $(n+1)$-dimensional $(n \geq 3)$ action in which
gravity is coupled to dilaton and Born-Infeld fields is~\cite{BID,Dehghani:2016wmw}:
\begin{equation}\label{Act}
S=\frac{1}{16\pi}\int{d^{n+1}x\sqrt{-g}\left(\mathcal{R}\text{
}-\frac{4}{n-1}(\nabla \Phi )^{2}-V(\Phi )+L(F,\Phi)\right)},
\end{equation}
where $\mathcal{R}$ is the Ricci scalar curvature, $\Phi $ is the
dilaton field and $V(\Phi )$ is a potential for $\Phi $:
\begin{equation}\label{v2}
V(\Phi) = 2\Lambda_{0} e^{2\zeta_{0}\Phi} +2 \Lambda e^{2\zeta
\Phi},
\end{equation}
where $\Lambda_{0}$, $\Lambda,$ $\zeta_{0}$ and $ \zeta$ are
constants and the
Born-Infeld $L(F,\Phi)$ part of the action is given by
\begin{equation}
L(F,\Phi )=4\beta ^{2}e^{4\alpha \Phi /(n-1)}\left( 1-\sqrt{1+\frac{%
e^{-8\alpha \Phi /(n-1)}F^{2}}{2\beta ^{2}}}\right).
\end{equation}
Here, $\alpha $ is a constant determining the strength of coupling
of the scalar and electromagnetic fields, $F^2=F_{\mu \nu }F^{\mu
\nu }$, where $F_{\mu \nu }=\partial _{\mu }A_{\nu }-\partial
_{\nu }A_{\mu }$ is the electromagnetic field tensor, and $A_{\mu
}$ is the electromagnetic vector potential. $\beta $ is the
Born-Infeld parameter with the dimension of mass.  A general solution is
\begin{equation}\label{metric}
ds^2=-f(r)dt^2 + {dr^2\over f(r)}+ r^2R^2(r)h_{ij}dx^{i}dx^{j} ,
\end{equation}
where $h_{ij}$ is a function of coordinates $x_{i}$
which spanned an $(n-1)$-dimensional hypersurface with constant
scalar curvature $(n-1)(n-2)k$. Here $k$ is a constant 
characterizing the hypersurface (consider $k>0$). The electromagnetic field (finite at $r=0$) and charge are given respectively as
\begin{equation}\label{Ftr}
F_{tr}=\frac{\beta q e^{4\alpha \Phi/(n-1)}}{\sqrt{\beta^2 \left(
rR\right)^{2n-2}+q^{2}}}, \qquad \qquad {Q}=\frac{q\omega _{n-1}}{4\pi}, 
\end{equation}
where $q$ is an integration constant related to the electric
charge of the black hole and $\omega_{n-1}$ represents the volume of constant curvature
hypersurface described by $h_{ij}dx^idx^j$. $R(r)$ and $\Phi (r)$ are given respectively as:
\begin{equation}
R(r)=e^{2\alpha \Phi /(n-1)},  \qquad \qquad \Phi (r)=\frac{(n-1)\alpha }{2(1+\alpha ^{2})}\ln (\frac{b}{r}),
\end{equation}
where $b$ is an arbitrary constant, $\gamma =\alpha ^{2}/(\alpha
^{2}+1)$.

The ADM (Arnowitt-Deser-Misner) mass $m$ of the
black hole expressed in terms of the
horizon radius $r_h$ is
\begin{eqnarray}\label{mass}
m(r_{h}) &=&-{\frac { k\left(n-2 \right)\left( { \alpha}^{2}+1
\right) ^{2}{b}^{-2\gamma}}{\left( { \alpha}^{2}-1 \right)
\left(n+{\alpha}^{2}-2 \right)
}}{r_{h}}^{n-2+\gamma(3-n)}+\frac{2\Lambda \left( {\alpha}^{2}+1
\right) ^{2}{b}^{2 \gamma}}{(n-1)(\alpha^{2}-n
)}r_{h}^{n(1-\gamma)-\gamma}\nonumber\\
&&-\frac{4\beta^2 (\alpha ^{2}+1)^{2}b^{2\gamma }}{(n-1)(\alpha
^{2}-n)}r_{h}^{n(1-\gamma)-\gamma}\times
\left(1-\text{}_{2}F_{1}\left( \left[ -\frac{1}{2},\frac{\alpha ^{2}-n%
}{2n-2}\right] ,\left[ \frac{\alpha ^{2}+n-2}{2n-2}\right]
,-\eta\right) \right).
\end{eqnarray}
\begin{equation}\label{lam}
\zeta_{0} =\frac{2}{\alpha(n-1)},   \hspace{.8cm}
\zeta=\frac{2\alpha}{n-1}, \hspace{.8cm}    \Lambda_{0} =
\frac{k(n-1)(n-2)\alpha^2 }{2b^2(\alpha^2-1)}, \qquad \eta = \frac{q^{2}b^{2\gamma (1-n)}}{\beta ^{2}r^{2(n-1)(1-\gamma
)}}.
\end{equation}
Finally, $f(r)$ is given as:
\begin{eqnarray}\label{f2}
f(r) &=&-{\frac { k\left(n-2 \right)\left( { \alpha}^{2}+1 \right)
^{2}{b}^{-2\gamma}}{\left( { \alpha}^{2}-1 \right)
\left(n+{\alpha}^{2}-2 \right)
}}{r}^{2\gamma}-\frac{m}{r^{(n-1)(1-\gamma )-1}}+\frac{2\Lambda
\left( {\alpha}^{2}+1 \right) ^{2}{b}^{2
\gamma}}{(n-1)(\alpha^{2}-n )}r^{2(1-\gamma)} \nonumber
\\
&&-\frac{4\beta^2 (\alpha ^{2}+1)^{2}b^{2\gamma }r^{2(1-\gamma
)}}{(n-1)(\alpha ^{2}-n)}\times
\left( 1-\text{{\ }}_{2}F_{1}\left( \left[ -\frac{1}{2},\frac{\alpha ^{2}-n%
}{2n-2}\right] ,\left[ \frac{\alpha ^{2}+n-2}{2n-2}\right]
,-\eta\right) \right).
\end{eqnarray}
Temperature and Entropy of the black hole are given respectively as
\begin{eqnarray}\label{Tem}
T_{+}&=&-\frac{(\alpha ^2+1)b^{2\gamma}r_{+}^{1-2\gamma}}{2\pi
(n-1)}\left(
\frac{k(n-2)(n-1)b^{-4\gamma}}{2(\alpha^2-1)}r_{+}^{4\gamma-2}
+\Lambda -2\beta^2(1-\sqrt{1+\eta_{+}})\right)\nonumber\\
&=&-\frac{k(n-2)(\alpha ^2+1)b^{-2\gamma}}{2\pi(\alpha
^2+n-2)}r_{+}^{2\gamma-1}+ \frac{(n-\alpha ^{2})m}{4\pi(\alpha
^{2}+1)}{r_{+}}^{(n-1)(\gamma -1)}-\frac{q^2(\alpha
^{2}+1)b^{2(2-n)\gamma}}{\pi(\alpha
^2+n-2)}r_{+}^{2(2-n)(1-\gamma)-1}\nonumber\\
&&\times \text{ }_{2}F_{1}\left( %
\left[ {\frac{1}{2},\frac{{n+\alpha }^{2}{-2}}{{2n-2}}}\right] ,\left[ {%
\frac{{3n+\alpha }^{2}{-4}}{{2n-2}}}\right] ,-\eta_{+}\right),
\end{eqnarray}
\begin{equation}
{S}=\frac{b^{(n-1)\gamma}\omega _{n-1}r_{+}^{(n-1)(1-\gamma
)}}{4}.\label{Entropy}
\end{equation}
where $\eta_{+}=\eta(r=r_{+})$.
The gauge $A_{t }$ and electric potential $U$, measured at infinity with respect to
the horizon are
\begin{eqnarray}\label{vectorpot}
A_{t}&=&\frac{qb^{(3-n)\gamma }}{\Upsilon r^{\Upsilon }}\text{ }_{2}F_{1}\left( %
\left[ {\frac{1}{2},\frac{{\alpha }^{2}+n-2}{{2n-2}}}\right] ,\left[ {%
\frac{{\alpha }^{2}+3n-4}{{2n-2}}}\right] ,-\eta\right),
\end{eqnarray}
where $\Upsilon =(n-3)(1-\gamma )+1$, and
\begin{equation}
U=\frac{qb^{(3-n)\gamma }}{ \Upsilon{r_{+}}^{\Upsilon }}\text{ }%
_{2}F_{1}\left( \left[ {\frac{1}{2},\frac{{\alpha}^{2}+n-2}{{2n-2}}}%
\right] ,\left[ {\frac{{\alpha }^{2}+3n-4}{{2n-2}}}\right]
,-\eta_{+}\right). \label{Pot}
\end{equation}
respectively. 
\section{Appendix}
\subsection*{ Einstein–BI-dilaton gravity and its Brans–Dicke counterpart:}

The action of $(n+1)$- dimensional BD theory, in which dilaton field is decoupled from the
matter field (electrodynamics) and  coupled with gravity can be written as~\cite{BDED}

\begin{equation}
	I_{BD-BI}=-\frac{1}{16\pi }\int_{\mathcal{M}}d^{n+1}x\sqrt{-g}\left( \Phi
	\mathcal{R}\text{ }-\frac{\omega }{\Phi }(\nabla \Phi )^{2}-V(\Phi )+%
	\mathcal{L}(\mathcal{F})\right) ,  \label{acBD}
\end{equation}%
where $\mathcal{L}(\mathcal{F})$ is the Lagrangian of BI theory 
\begin{equation}
	\mathcal{L}(\mathcal{F})=4\beta ^{2}\left( 1-\sqrt{1+\frac{\mathcal{F}}{%
			2\beta ^{2}}}\right) ,  
\end{equation}

 $\mathcal{R}$ is the Ricci scalar, $\omega $
is the coupling constant, $\Phi $ denotes the BD scalar field and
$V(\Phi)$ is a self--interaction potential for $\Phi$.

 Indeed,  the BD-BI theory is
	conformally  associated with the Einstein--BI--dilaton gravity. The appropriate
	conformal transformation is as follows
	\begin{equation}
		\bar{g}_{\mu \nu }=\Phi ^{2/(n-1)}g_{\mu \nu },  \label{CT}
	\end{equation}%
	where%
	\begin{eqnarray}
		\bar{\Phi} &=&\frac{n-3}{4\alpha }\ln \Phi ,  \label{Phibar} \\
		\alpha  &=&(n-3)/\sqrt{4(n-1)\omega +4n}.  \label{alpha}
	\end{eqnarray}
	
By means of 	this conformal transformation, one finds that
	the action of BD-BI  transforms to the
	well-known dilatonic-BI gravity as
	\begin{equation}
		\overline{I}_{G}=-\frac{1}{16\pi }\int_{\mathcal{M}}d^{n+1}x\sqrt{-\overline{g}}%
		\left\{ \overline{\mathcal{R}}-\frac{4}{n-1}(\overline{\nabla}\overline{\Phi})^{2}-\overline{V}(%
		\overline{\Phi})+\overline{L}\left(
		\overline{\mathcal{F}},\overline{\Phi }\right) \right\} ,
		\label{acED}
	\end{equation}%
 where, the
			potential $\overline{V}\left( \overline{\Phi }\right) $ and
			the BI-dilaton coupling Lagrangian $\overline{L}\left( \overline{F},%
			\overline{\Phi }\right) $ are, respectively,
			\begin{equation}
				\overline{V}(\overline{\Phi})=\Phi ^{-(n+1)/(n-1)}V(\Phi ),
				\label{vrelation}
			\end{equation}%
			and
			\begin{equation}
				\overline{L}\left( \overline{\mathcal{F}},\overline{\Phi }\right)
				=4\beta ^{2}e^{-4\alpha \left( n+1\right) \overline{\Phi }/\left[
					\left( n-1\right) \left( n-3\right) \right] }\left(
				1-\sqrt{1+\frac{e^{16\alpha \overline{\Phi
							}/\left[ \left( n-1\right) \left( n-3\right) \right] }\overline{\mathcal{F}}}{%
						2\beta ^{2}}}\right) .  \label{LFP}
			\end{equation}

			\subsubsection*{\textbf{Black holes in Einstein frame (Einstein-dilaton-BI Theory):}}
			
		 For the black hole solution, we assume the  metric
		
			\begin{equation}
				d\overline{s}^{2}=-Z(r)dt^{2}+\frac{dr^{2}}{Z(r)}+r^{2}R^{2}(r)d\Omega
				_{k}^{2}, \label{metricED}
			\end{equation}%
			 and the potential $ \mathbf{\overline{V}}(\overline{\Phi})$ as
			 
			 \begin{equation}
			 \mathbf{\overline{V}}(\overline{\Phi})=2\Lambda \exp \left( \frac{4\alpha \overline{\Phi}}{%
			 	n-1}\right) +\frac{k(n-1)(n-2)\alpha ^{2}}{b^{2}\left( \alpha ^{2}-1\right) }%
			 \exp \left( \frac{4\overline{\Phi}}{(n-1)\alpha }\right)
			 +\frac{W(r)}{\beta ^{2}}, \label{potenED}
			 \end{equation}%
			 \\
			where $d\Omega _{k}^{2}$  betokens the Euclidean metric of an $(n-1)$-dimensional
			hypersurface with constant curvature $(n-1)(n-2)k$ and volume $\varpi_{n-1}$ (hereafter we optate $k = 1$ ).\\
			\\		
			Now,  the metric (\ref{metricED}) with Equations of motion for the action (\ref{acED}) admit the following solution.
			\begin{eqnarray}
				F_{tr} &=&E(r)=\frac{qe^{\left( \frac{4\alpha \overline{\Phi }(r)}{n-1}%
						\right) }}{(rR(r))^{(n-1)}\sqrt{1+\frac{e^{(\frac{8\alpha \overline{\Phi }(r)%
								}{n-3})}q^{2}(rR(r))^{-2(n-1)}}{\beta ^{2}}}},  \label{E} \\
				\overline{\Phi} &=&\frac{(n-1)\alpha }{2(1+\alpha ^{2})}\ln \left(
				\frac{b}{r}\right) \label{phi}
			\end{eqnarray}%
			\begin{equation}
				W(r)=\frac{4q(n-1)\beta ^{2}R(r)}{\left( 1+\alpha ^{2}\right)
					r^{\gamma
					}b^{n\gamma }}\int \frac{E(r)}{r^{n(1-\gamma )-\gamma }}dr+\frac{4\beta ^{4}%
				}{R(r)^{\frac{2(n+1)}{n-3}}}\left( 1-\frac{E(r)R(r)^{(n-3)}}{qr^{1-n}}%
				\right) -\frac{4q\beta ^{2}E(r)}{r^{n-1}}(\frac{r}{b})^{\gamma
					(n-1)}, \label{W}
			\end{equation}%
			\begin{eqnarray}
				Z(r) &=&-\frac{k\left( n-2\right) \left( \alpha ^{2}+1\right)
					^{2}b^{-2\gamma }r^{2\gamma }}{\left( \alpha ^{2}+n-2\right)
					\left( \alpha
					^{2}-1\right) }+\left( \frac{(1+\alpha ^{2})^{2}r^{2}}{(n-1)}\right) \frac{%
					2\Lambda \left( \frac{r}{b}\right) ^{-2\gamma }}{(\alpha ^{2}-n)}-\frac{m}{%
					r^{(n-1)(1-\gamma )-1}}  \notag \\
				&&-\frac{4(1+\alpha ^{2})^{2}q^{2}(\frac{r}{b})^{2\gamma
						(n-2)}}{(n-\alpha ^{2})r^{2(n-2)}}\left( \frac{1}{2(n-1)}\digamma
				_{1}(\eta )-\frac{1}{\alpha ^{2}+n-2}\digamma _{2}(\eta )\right) ,
				\label{f}
			\end{eqnarray}%
			where $m$ and $b$ are integration constants related to the mass
			and scalar field, respectively, and
			\begin{eqnarray*}
				\digamma _{1}(\eta ) &=&\text{ }_{2}F_{1}\left( \left[ \frac{1}{2},\frac{%
					(n-3)\Upsilon }{\alpha ^{2}+n-2}\right] ,\left[ 1+\frac{(n-3)\Upsilon }{%
					\alpha ^{2}+n-2}\right] ,-\eta \right) , \\
				\digamma _{2}(\eta ) &=&\text{ }_{2}F_{1}\left( \left[ \frac{1}{2},\frac{%
					(n-3)\Upsilon }{2(n-1)}\right] ,\left[ 1+\frac{(n-3)\Upsilon
				}{2(n-1)}\right]
				,-\eta \right) , \\
				\Upsilon  &=&\frac{\alpha ^{2}+n-2}{2\alpha ^{2}+n-3}, \\
				\eta  &=&\frac{q^{2}(\frac{r}{b})^{2\gamma
						(n-1)(n-5)/(n-3)}}{\beta
					^{2}r^{2(n-1)}}, \\
				R(r) &=&\exp \left( \frac{2\alpha \overline{\Phi }}{n-1}\right)
				=\left( \frac{r}{b}\right) ^{-\gamma }.
			\end{eqnarray*}

			\subsubsection*{\textbf{Black holes in Jordan frame (BD-BI Theory):}}
			
			We invoke the conformal transformation to obtain  black hole solutions of the BD-BI
			theory. The
			potential $\mathbf{V}(\Phi )$ in Jordan frame using the relation (\ref{vrelation}) is
			\begin{equation}
				\mathbf{V}(\Phi )=2\Lambda \Phi ^{2}+\frac{k(n-1)(n-2)\alpha ^{2}}{%
					b^{2}\left( \alpha ^{2}-1\right) }\Phi ^{\lbrack (n+1)(1+\alpha
					^{2})-4]/[(n-1)\alpha ^{2}]}+\Phi ^{(n+1)/(n-1)}\frac{W(r)}{\beta
					^{2}}. \label{potenBD}
			\end{equation}
			
			Taking into account the solutions in an Einstein frame with the
			mentioned conformal transformation, we are able to acquire the solutions of
			field equations for the BD-BI action (\ref{acBD}).  Considering the following
			$(n+1)-$dimensional metric
			\begin{equation}
				ds^{2}=-A(r)dt^{2}+\frac{dr^{2}}{B(r)}+r^{2}H^{2}(r)d\Omega
				_{k}^{2}, \label{metric1}
			\end{equation}%
			we find that the functions $A(r)$ and $B(r)$ are
			\begin{eqnarray}
				A(r) &=&\left( \frac{r}{b}\right) ^{4\gamma /\left( n-3\right)
				}Z\left(
				r\right) ,  \label{A(r)} \\
				B(r) &=&\left( \frac{r}{b}\right) ^{-4\gamma /\left( n-3\right)
				}Z\left(
				r\right) ,  \label{B(r)} \\
				H(r) &=&\left( \frac{r}{b}\right) ^{-\gamma (\frac{n-5}{n-3})},
				\label{H(r)}
				\\
				\Phi \left( r\right)  &=&\left( \frac{r}{b}\right)
				^{-\frac{2\gamma \left( n-1\right) }{n-3}}.  \label{Phi}
			\end{eqnarray}

			\subsubsection*{\textbf{Thermodynamic quantities:}}
			
		 In both the frames,  Hawking temperature, 	mass,  entropy and the electric charge of the black hole  are taking the subsequent forms:

			\begin{eqnarray}
				T & = & \frac{\left( \alpha ^{2}+1\right) }{2\pi \left( n-1\right) }\left[ -\frac{%
					\left( n-2\right) (n-1)}{2\left( \alpha ^{2}-1\right) r_{+}}\left( \frac{%
					r_{+}}{b}\right) ^{2\gamma }-\Lambda r_{+}\left(
				\frac{r_{+}}{b}\right) ^{-2\gamma }+\Gamma _{+}\right] ,  \label{Td2} \\
				M &=&\frac{\varpi _{n-1}b^{(n-1)\gamma }}{16\pi }\left(
				\frac{n-1}{1+\alpha
					^{2}}\right) m,  \label{Md} \\
				S &=&\frac{\varpi _{n-1}b^{(n-1)\gamma }}{4}r_{+}^{(n-1)\left(
					1-\gamma \right) }.  \label{Sd} \\
					Q & = & \frac{q}{4\pi },  \label{Qd}
			\end{eqnarray}
				where
				\begin{eqnarray}
				\Gamma _{+} &=&-\frac{\left( \alpha ^{2}+1\right) ^{2}q^{2}}{2\pi (n-1)}%
				\left( \frac{r_{+}}{b}\right) ^{2\gamma \left( n-2\right)
				}r_{+}^{3-2n}\digamma _{1}(\eta _{+}),  \label{GAMMA} \\
				\eta _{+} &=&\eta \Big|_{r=r_{+}}, \\
					m &=& (1+\alpha^2)^2 r_+^{(n-1)(1-\gamma)-1}\Bigg\{\frac{(n-2)b^{-2\gamma}}{(1-\alpha^2)(n+\alpha^2-2)}r_+^{2\gamma}+\frac{2\Lambda r_+^2}{(n-1)(\alpha^2-n)}\Big(\frac{r_+}{b}\Big)^{-2\gamma} \nonumber \\ 
				&&-\frac{4q^2(\frac{r_+}{b})^{2\gamma(n-2)}}{(n-\alpha^2)r_+^{2(n-2)}}\times \Bigg[\frac{1}{2(n-1)}\digamma_1(\eta_+)-\frac{1}{(\alpha^2+n-2)}\digamma_2(\eta_+) \Bigg]\Bigg\} . 
				\end{eqnarray}

			In addition, in the extended  phase space, thermodynamical
			pressure  and  volume are given by
			\begin{equation}
			p = -\frac{\Lambda }{8\pi }\times \left\{
			\begin{array}{cc}
			\left( \frac{r_{+}}{b}\right) ^{-2\gamma }, & \text{dilatonic BI} \\
			\left( \frac{r_{+}}{b}\right) ^{-\frac{2\gamma \left( n-1\right)
				}{n-3}}, &
			\text{BD-BI}%
			\end{array}%
			\right.   \label{Pd}
		\end{equation}
			
			\begin{equation}		
			V=\frac{\varpi _{n-1}\left( 1+\alpha ^{2}\right) r_{+}^{n}}{n-\alpha ^{2}}%
			\left\{
			\begin{array}{cc}
			\left( \frac{r_{+}}{b}\right) ^{-\gamma \left( n-1\right) }, & \text{%
				dilatonic BI} \\
			\left( \frac{r_{+}}{b}\right) ^{-\frac{\gamma (n^{2}-4n-1)}{n-3}}, & \text{%
				BD-BI}%
			\end{array}%
			\right. .  \label{Vd}	
			\end{equation}


\begin{thebibliography}{10}




\bibitem{Caldarelli:1999xj}
M.~M. Caldarelli, G.~Cognola, and D.~Klemm, ``{Thermodynamics of
  Kerr-Newman-AdS black holes and conformal field theories},''
  \href{http://dx.doi.org/10.1088/0264-9381/17/2/310}{{\em Class.Quant.Grav.}
  {\bf 17} (2000)  399--420},
\href{http://arxiv.org/abs/hep-th/9908022}{{\tt arXiv:hep-th/9908022
  [hep-th]}}.

\bibitem{Wang:2006eb}
S.~Wang, S.-Q. Wu, F.~Xie, and L.~Dan, ``{The First laws of thermodynamics of
  the (2+1)-dimensional BTZ black holes and Kerr-de Sitter spacetimes},''
  \href{http://dx.doi.org/10.1088/0256-307X/23/5/009}{{\em Chin.Phys.Lett.}
  {\bf 23} (2006)  1096--1098},
\href{http://arxiv.org/abs/hep-th/0601147}{{\tt arXiv:hep-th/0601147
  [hep-th]}}.

\bibitem{Sekiwa:2006qj}
Y.~Sekiwa, ``{Thermodynamics of de Sitter black holes: Thermal cosmological
  constant},'' \href{http://dx.doi.org/10.1103/PhysRevD.73.084009}{{\em
  Phys.Rev.} {\bf D73} (2006)  084009},
\href{http://arxiv.org/abs/hep-th/0602269}{{\tt arXiv:hep-th/0602269
  [hep-th]}}.

\bibitem{LarranagaRubio:2007ut}
E.~A. Larranaga~Rubio, ``{Stringy Generalization of the First Law of
  Thermodynamics for Rotating BTZ Black Hole with a Cosmological Constant as
  State Parameter},''
\href{http://arxiv.org/abs/0711.0012}{{\tt arXiv:0711.0012 [gr-qc]}}.

\bibitem{Kastor:2009wy}
D.~Kastor, S.~Ray, and J.~Traschen, ``{Enthalpy and the Mechanics of AdS Black
  Holes},'' \href{http://dx.doi.org/10.1088/0264-9381/26/19/195011}{{\em
  Class.Quant.Grav.} {\bf 26} (2009)  195011},
\href{http://arxiv.org/abs/0904.2765}{{\tt arXiv:0904.2765 [hep-th]}}.

\bibitem{Dolan:2010ha}
B.~P. Dolan, ``{The cosmological constant and the black hole equation of
  state},'' \href{http://dx.doi.org/10.1088/0264-9381/28/12/125020}{{\em
  Class.Quant.Grav.} {\bf 28} (2011)  125020},
\href{http://arxiv.org/abs/1008.5023}{{\tt arXiv:1008.5023 [gr-qc]}}.

\bibitem{Cvetic:2010jb}
M.~Cvetic, G.~Gibbons, D.~Kubiznak, and C.~Pope, ``{Black Hole Enthalpy and an
  Entropy Inequality for the Thermodynamic Volume},''
  \href{http://dx.doi.org/10.1103/PhysRevD.84.024037}{{\em Phys.Rev.} {\bf D84}
  (2011)  024037},
\href{http://arxiv.org/abs/1012.2888}{{\tt arXiv:1012.2888 [hep-th]}}.

\bibitem{Dolan:2011jm}
B.~P. Dolan, ``{Compressibility of rotating black holes},''
  \href{http://dx.doi.org/10.1103/PhysRevD.84.127503}{{\em Phys.Rev.} {\bf D84}
  (2011)  127503},
\href{http://arxiv.org/abs/1109.0198}{{\tt arXiv:1109.0198 [gr-qc]}}.

\bibitem{Dolan:2011xt}
B.~P. Dolan, ``{Pressure and volume in the first law of black hole
  thermodynamics},''
  \href{http://dx.doi.org/10.1088/0264-9381/28/23/235017}{{\em
  Class.Quant.Grav.} {\bf 28} (2011)  235017},
\href{http://arxiv.org/abs/1106.6260}{{\tt arXiv:1106.6260 [gr-qc]}}.

\bibitem{Dolan:2012jh}
B.~P. Dolan, ``{Where is the PdV term in the fist law of black hole
  thermodynamics?},''
\href{http://arxiv.org/abs/1209.1272}{{\tt arXiv:1209.1272 [gr-qc]}}.

\bibitem{Altamirano:2014tva}
N.~Altamirano, D.~Kubiznak, R.~B. Mann, and Z.~Sherkatghanad, ``{Thermodynamics
  of rotating black holes and black rings: phase transitions and thermodynamic
  volume},'' \href{http://dx.doi.org/10.3390/galaxies2010089}{{\em Galaxies}
  {\bf 2} (2014)  89--159},
\href{http://arxiv.org/abs/1401.2586}{{\tt arXiv:1401.2586 [hep-th]}}.

\bibitem{Kubiznak:2016qmn}
D.~Kubiznak, R.~B. Mann, and M.~Teo, ``{Black hole chemistry: thermodynamics
  with Lambda},'' \href{http://iopscience.iop.org/article/10.1088/1361-6382/aa5c69/meta;jsessionid=B56C3CA7D56EAB9E5307647D0EC19E6A.c3.iopscience.cld.iop.org#}{{\tt Class. Quantum Grav. 34 (2017) 063001}},  
\href{http://arxiv.org/abs/1608.06147}{{\tt arXiv:1608.06147 [hep-th]}}.


\bibitem{Henneaux:1984ji}
M.~Henneaux and C.~Teitelboim, ``{The Cosmological Constant as a Canonical
  Variable},''
\href{http://dx.doi.org/10.1016/0370-2693(84)91493-X}{{\em Phys.Lett.} {\bf
  B143} (1984)  415--420}.

\bibitem{Teitelboim:1985dp}
C.~Teitelboim, ``{The Cosmological Constant as a Thermodynamic Black Hole
  Parameter},''
\href{http://dx.doi.org/10.1016/0370-2693(85)91186-4}{{\em Phys.Lett.} {\bf
  B158} (1985)  293--297}.

\bibitem{Henneaux:1989zc}
M.~Henneaux and C.~Teitelboim, ``{The Cosmological Constant and General
  Covariance},''
\href{http://dx.doi.org/10.1016/0370-2693(89)91251-3}{{\em Phys.Lett.} {\bf
  B222} (1989)  195--199}.




\bibitem{Bekenstein:1973ur}
J.~D. Bekenstein, ``{Black holes and entropy},''
\href{http://dx.doi.org/10.1103/PhysRevD.7.2333}{{\em Phys.Rev.} {\bf D7}
  (1973)  2333--2346}.

\bibitem{Bekenstein:1974ax}
J.~D. Bekenstein, ``{Generalized second law of thermodynamics in black hole
  physics},''
\href{http://dx.doi.org/10.1103/PhysRevD.9.3292}{{\em Phys.Rev.} {\bf D9}
  (1974)  3292--3300}.

\bibitem{Hawking:1974sw}
S.~Hawking, ``{Particle Creation by Black Holes},''
\href{http://dx.doi.org/10.1007/BF02345020}{{\em Commun.Math.Phys.} {\bf 43}
  (1975)  199--220}.

\bibitem{Hawking:1976de}
S.~Hawking, ``{Black Holes and Thermodynamics},''
\href{http://dx.doi.org/10.1103/PhysRevD.13.191}{{\em Phys.Rev.} {\bf D13}
  (1976)  191--197}.



\bibitem{Parikh:2005qs}
M.~K. Parikh, ``{The Volume of black holes},''
  \href{http://dx.doi.org/10.1103/PhysRevD.73.124021}{{\em Phys.Rev.} {\bf D73}
  (2006)  124021},
\href{http://arxiv.org/abs/hep-th/0508108}{{\tt arXiv:hep-th/0508108
  [hep-th]}}.


 


\bibitem{Johnson:2014xza} 
  C.~V.~Johnson, ``Thermodynamic Volumes for AdS-Taub-NUT and AdS-Taub-Bolt," \href{http://iopscience.iop.org/article/10.1088/0264-9381/31/23/235003/meta}{{\tt Class.\ Quant.\ Grav.\  {\bf 31}, no. 23, 235003 (2014),}}     \href{https://arxiv.org/abs/1405.5941}{{\tt arXiv:1405.5941 [hep-th]}}.
  

\bibitem{Hawking:1998jf} 
  S.~W.~Hawking and C.~J.~Hunter,
   ``Gravitational entropy and global structure,"  \href{https://journals.aps.org/prd/abstract/10.1103/PhysRevD.59.044025}{{\tt  Phys.\ Rev.\ D {\bf 59}, 044025 (1999),}}  
    \href{https://arxiv.org/abs/hep-th/9808085}{{\tt  arXiv:hep-th/9808085 [hep-th]}}.
 
\bibitem{BID}
M. H. Dehghani, S. Kamrani, and A. Sheykhi, ``P-V criticality
of charged dilatonic black holes," \href{https://journals.aps.org/prd/abstract/10.1103/PhysRevD.90.104020}{{\tt  Phys. Rev. D 90,	104020 (2014)}};
A. Sheykhi, ``Thermodynamical properties of topological
Born-Infeld-dilaton black holes," \href{http://www.worldscientific.com/doi/abs/10.1142/S021827180901425X}{{\tt Int. J. Mod. Phys. D, 18, 25 (2009)}}; A. Sheykhi and N. Riazi, ``Thermodynamics of black
holes in (n + 1)-dimensional Einstein-Born-Infeld dilaton
gravity," \href{https://journals.aps.org/prd/abstract/10.1103/PhysRevD.75.024021}{{\tt  Phys. Rev. D 75, 024021 (2007)}}; K. C. K. Chan, J. H. Horne, and R. B. Mann, ``Charged
dilaton black holes with unusual asymptotics," \href{http://www.sciencedirect.com/science/article/pii/0550321395002057?via%3Dihub}{{\tt  Nucl. Phys.	B447, 441 (1995).}}

\bibitem{6} A. Chamblin, R. Emparan, C. V. Johnson, and R. C. Myers, ``Charged AdS black holes and catastrophic holography," \href{https://journals.aps.org/prd/abstract/10.1103/PhysRevD.60.064018}{{\tt  Phys. Rev. D \textbf{60}, 064018 (1999)}};  A. Chamblin, R. Emparan, C. V. Johnson, and R. C. Myers, ``Holography, thermodynamics, and fluctuations of charged AdS black holes," \href{https://journals.aps.org/prd/abstract/10.1103/PhysRevD.60.104026}{{\tt Phys. Rev. D \textbf{60}, 104026 (1999).}}  

\bibitem{8} D. Kubiznak, R. B. Maan, ``P-V criticality of charged AdS black holes," \href{https://link.springer.com/article/10.1007%2FJHEP07%282012%29033}{{\tt JHEP \textbf{1207}, 033 (2012)}}, \href{https://arXiv.org/abs/1205.0559}{{\tt arxiv:1205.0559 [hep-th]}}. 

\bibitem{Johnson:2014yja} 
  C.~V.~Johnson,
  ``Holographic Heat Engines,'' \href{http://iopscience.iop.org/article/10.1088/0264-9381/31/20/205002/meta;jsessionid=8FA26BD5DD7F585CB14439156F38AF57.c2.iopscience.cld.iop.org}{{\tt  Class.\ Quant.\ Grav.\  {\bf 31}, 205002 (2014)}}, \href{https://arxiv.org/abs/1404.5982}{{\tt    arXiv:1404.5982 [hep-th]}}.
 
  


\bibitem{Belhaj:2015hha}
A.~Belhaj, M.~Chabab, H.~El~Moumni, K.~Masmar, M.~B. Sedra, and A.~Segui, ``{On
  Heat Properties of AdS Black Holes in Higher Dimensions},''
  \href{http://dx.doi.org/10.1007/JHEP05(2015)149}{{\em JHEP} {\bf 05} (2015)
  149},
\href{http://arxiv.org/abs/1503.07308}{{\tt arXiv:1503.07308 [hep-th]}}.

\bibitem{Sadeghi:2015ksa}
J.~Sadeghi and K.~Jafarzade, ``{Heat Engine of black holes},''
\href{http://arxiv.org/abs/1504.07744}{{\tt arXiv:1504.07744 [hep-th]}}.

\bibitem{Caceres:2015vsa}
E.~Caceres, P.~H. Nguyen, and J.~F. Pedraza, ``{Holographic entanglement
  entropy and the extended phase structure of STU black holes},''
  \href{http://dx.doi.org/10.1007/JHEP09(2015)184}{{\em JHEP} {\bf 09} (2015)
  184},
\href{http://arxiv.org/abs/1507.06069}{{\tt arXiv:1507.06069 [hep-th]}}.

\bibitem{Setare:2015yra}
M.~R. Setare and H.~Adami, ``{Polytropic black hole as a heat engine},''
\href{http://dx.doi.org/10.1007/s10714-015-1979-0}{{\em Gen. Rel. Grav.} {\bf
  47} (2015) no.~11, 133}.

  \bibitem{Johnson:2015ekr}
C.~V. Johnson, ``{Gauss-Bonnet Black Holes and Holographic Heat Engines Beyond
  Large N},'' \href{http://iopscience.iop.org/article/10.1088/0264-9381/33/21/215009/meta}{{\tt Class.Quant.Grav. 33 (2016) no.21, 215009}}, 
\href{http://arxiv.org/abs/1511.08782}{{\tt arXiv:1511.08782 [hep-th]}}.

\bibitem{Johnson:2015fva}
C.~V. Johnson, ``{Born-Infeld AdS Black Holes as Heat Engines},'' \href{http://iopscience.iop.org/article/10.1088/0264-9381/33/13/135001/meta}{{\tt Class. Quantum Grav. 33 (2016) 135001}}, 
\href{http://arxiv.org/abs/1512.01746}{{\tt arXiv:1512.01746 [hep-th]}}.

\bibitem{Johnson:2016pfa} 
  C.~V.~Johnson,
  ``An Exact Efficiency Formula for Holographic Heat Engines,'' \href{http://www.mdpi.com/1099-4300/18/4/120}{{\tt Entropy 2016, 18(4), 120}},  \href{https://arxiv.org/abs/1602.02838}{{\tt  arXiv:1602.02838 [hep-th]}}.
 



\bibitem{Maldacena:1997re}
J.~M. Maldacena, ``The large N limit of superconformal field theories and
  supergravity,'' \href{https://link.springer.com/article/10.1023%2FA%3A1026654312961}{{\tt {\em Adv. Theor. Math. Phys.} {\bf 2} (1998)  231--252}},
\href{http://arxiv.org/abs/hep-th/9711200}{{\tt arXiv:hep-th/9711200 [hep-th]}}.

\bibitem{Witten:1998qj}
E.~Witten, ``Anti-de sitter space and holography,''  {\em Adv. Theor. Math.
  Phys.} {\bf 2} (1998)  253--291,
\href{http://arxiv.org/abs/hep-th/9802150}{{\tt arXiv:hep-th/9802150 [hep-th]}}.

\bibitem{Gubser:1998bc}
S.~S. Gubser, I.~R. Klebanov, and A.~M. Polyakov, ``Gauge theory correlators
  from non-critical string theory,'' \href{http://www.sciencedirect.com/science/article/pii/S0370269398003773}{{\tt {\em Phys. Lett.} {\bf B428} (1998)
  105--114}},
\href{http://arxiv.org/abs/hep-th/9802109}{{\tt arXiv:hep-th/9802109 [hep-th]}}.

\bibitem{Witten:1998zw}
E.~Witten, ``Anti-de sitter space, thermal phase transition, and confinement in
  gauge theories,'' {\em Adv. Theor. Math. Phys.} {\bf 2} (1998)  505--532,
\href{http://arxiv.org/abs/hep-th/9803131}{{\tt arXiv:hep-th/9803131 [hep-th]}}.

\bibitem{Aharony:1999ti}
O.~Aharony, S.~S. Gubser, J.~M. Maldacena, H.~Ooguri, and Y.~Oz, ``{Large N
  field theories, string theory and gravity},''
  \href{http://dx.doi.org/10.1016/S0370-1573(99)00083-6}{{\em Phys. Rept.} {\bf
  323} (2000)  183--386},
\href{http://arxiv.org/abs/hep-th/9905111}{{\tt arXiv:hep-th/9905111
  [hep-th]}}.

\bibitem{lowstring}
	M. B. Green, J. H. Schwarz and E. Witten, ``Superstring
	Theory" (Cambridge University Press, Cambridge England,
	1987).


\bibitem{Dehghani:2016wmw} 
M.~H.~Dehghani, A.~Sheykhi and Z.~Dayyani,
``Critical behavior of Born-Infeld dilaton black holes,'' \href{https://journals.aps.org/prd/abstract/10.1103/PhysRevD.93.024022}{{\tt Phys.\ Rev.\ D {\bf 93}, no. 2, 024022 (2016)}}, \href{https://arxiv.org/abs/1611.08978v1}{{\tt arXiv:1611.08978v1 [hep-th]}}.

\bibitem{BDED} Hendi, S.~H., Tad, R.~M., Armanfard, Z. et al, ``Extended phase space thermodynamics and P–V criticality: Brans–Dicke–Born–Infeld vs. Einstein–Born–Infeld-dilaton black holes,"  \href{https://link.springer.com/article/10.1140%2Fepjc%2Fs10052-016-4106-9}{{\tt Eur. Phys. J. C (2016) 76: 263}}, \href{https://arxiv.org/abs/1511.02761}{{\tt arXiv:1511.02761 [gr-qc]}}.
	
	\bibitem{hyp_inflation} P.~J. Steinhardt, F.~S. Accetta, ``Hyperextended inflation," \href{https://journals.aps.org/prl/abstract/10.1103/PhysRevLett.64.2740}{{\tt  Phys. Rev. Lett. 64, 2740 (1990).}}
	
	\bibitem{weinberg} P. A. M. Dirac, ``A New Basis for Cosmology," \href{http://rspa.royalsocietypublishing.org/content/165/921/199}{{\tt Proc. R. Soc. Lond. A 165, 199 (1938)}}; S. Weinberg: Gravitation and Cosmology (Wiley, 1972).
	
	\bibitem{BDth1}
	A. Sheykhi, M.~M. Yazdanpanah, ``Thermodynamics of charged Brans–Dicke AdS black holes,"
	\href{http://www.sciencedirect.com/science/article/pii/S0370269309008934}{{\tt Physics Letters B 679 (2009) 311-316}}, \href{https://arxiv.org/abs/0904.1777}{{\tt arXiv:0904.1777 [hep-th]}}.
	
	\bibitem{BDth2}
	S.~H. Hendi, Z. Armanfard, ``Extended phase space thermodynamics and P-V criticality of charged black holes in Brans-Dicke theory," \href{https://link.springer.com/article/10.1007%2Fs10714-015-1965-6}{{\tt  Gen. Relativ. Gravity 47, 125 (2015)}}, 
	 \href{https://arxiv.org/abs/1503.07070}{\tt arXiv:1503.07070 [gr-qc]}.
	\bibitem{BDth3}
	M. Kord Zangeneh, M. H. Dehghani, And A. Sheykhi, ``Thermodynamics of topological black holes in Brans-Dicke gravity with a power-law Maxwell field," \href{https://journals.aps.org/prd/abstract/10.1103/PhysRevD.92.104035}{{\tt  Phys. Rev. D 92, 104035 (2015)}}, \href{https://arxiv.org/abs/1509.05990}{{\tt arXiv:1509.05990 [gr-qc]}}.
	
	\bibitem{BDth4}
	Hendi, S.~H., Panahiyan, S., Panah, B.~E. et al, ``Phase transition of charged Black Holes in Brans–Dicke theory through geometrical thermodynamics," \href{https://link.springer.com/article/10.1140/epjc/s10052-016-4235-1}{{\tt  Eur. Phys. J. C (2016) 76: 396}}, \href{https://arxiv.org/abs/1511.00598}{{\tt arXiv:1511.00598 [gr-qc]}}.
	
	\bibitem{solar} A.~D. Felice, et al, ``Relaxing nucleosynthesis constraints on Brans-Dicke theories,"
	\href{https://journals.aps.org/prd/abstract/10.1103/PhysRevD.74.103005}{{\tt Phys. Rev. D 74, 103005 (2006)}}, \href{https://arxiv.org/abs/astro-ph/0510359}{{\tt arXiv:astro-ph/0510359}}.
	
	\bibitem{BI}
	M. Born and L. Infeld, ``Foundations of the New Field Theory," \href{http://rspa.royalsocietypublishing.org/content/144/852/425}{{\tt Proc.
			R. Soc. A 144, 425 (1934).}} 
	
	\bibitem{All}
	R. G. Cai, D. W. Pang and A. Wang, ``Born-Infeld black holes in (A)dS spaces",  \href{http://arxiv.org/abs/hep-th/0410158}{{\tt arXiv:hep-th/0410158 [hep-th]}}; D. Kubiznak and R. B. Mann, ``P-V criticality of charged Ads
	black holes," \href{https://link.springer.com/article/10.1007%2FJHEP07%282012%29033}{{\tt J. High Energy Phys. 07 (2012) 033}};
		Sh. Gunasekaran, D. Kubiznak and R. B. Mann, ``Extended
		phase space thermodynamics for charged and rotating black
		holes and Born-Infeld vacuum polarization," \href{https://link.springer.com/article/10.1007/JHEP11(2012)110}{{\tt J. High Energy Phys. 11 (2012) 110}}; De. Ch. Zou, Sh.-J. Zhang and B. Wang, ``Critical behavior
		of Born-Infeld AdS black holes in the extended phase space
		thermodynamics," \href{https://journals.aps.org/prd/abstract/10.1103/PhysRevD.89.044002}{{\tt  Phys. Rev. D 89, 044002 (2014)}};
		R. Banerjee and D. R Roychowdhury, ``Critical phenomena
		in Born-Infeld AdS black holes," \href{https://journals.aps.org/prd/abstract/10.1103/PhysRevD.85.044040}{ Phys. Rev. D 85, 044040 (2012)};  
		R. Banerjee and D. R Roychowdhury, ``Critical behavior of Born-Infeld AdS black holes in
		higher dimensions," \href{https://journals.aps.org/prd/abstract/10.1103/PhysRevD.85.104043}{{\tt Phys. Rev. D 85, 104043 (2012)}}; 
		J. X. Mo and W. B. Liu, ``P-V criticality of topological black
		holes in Lovelock-Born-Infeld gravity," \href{https://link.springer.com/article/10.1140/epjc/s10052-014-2836-0}{{\tt Eur. Phys. J. C 74,	2836 (2014)}}; S. H. Hendi, S. Panahiyan and B. Eslam Panah, ``P–V criticality and geometrical thermodynamics of black holes with Born–Infeld type nonlinear electrodynamics," \href{http://www.worldscientific.com/doi/abs/10.1142/S0218271816500103}{{\tt Int. J. Mod. Phys. D 25, 1650010 (2016)}}; 
		S. H. Hendi and M. H. Vahidinia, ``Extended phase space
		thermodynamics and P-V criticality of black holes with
		a nonlinear source," \href{https://journals.aps.org/prd/abstract/10.1103/PhysRevD.88.084045}{{\tt Phys. Rev. D 88, 084045 (2013)}}; 
		De. Zou, Y. Liu and B. Wang, ``Critical behavior of charged
		Gauss-Bonnet AdS black holes in the grand canonical
		ensemble," \href{https://journals.aps.org/prd/abstract/10.1103/PhysRevD.90.044063}{{\tt Phys. Rev. D 90, 044063 (2014)}}; 
		M. B. Jahani Poshteh, B. Mirza and Z. Sherkatghanad,
		``Phase transition, critical behavior, and critical exponents
		of Myers-Perry black holes," \href{https://journals.aps.org/prd/abstract/10.1103/PhysRevD.88.024005}{{\tt Phys. Rev. D 88, 024005 (2013)}};
		Z. Sherkatghanad, B. Mirza, Z. Mirzaeyan and S. A.
		Hosseini Mansoori, ``Critical behaviors and phase transitions
		of black holes in higher order gravities and extended phase
		spaces," \href{http://www.worldscientific.com/doi/abs/10.1142/S0218271817500171}{{\tt Int. J. Mod. Phys. D 26, 1750017 (2017)}}, \href{https://arxiv.org/abs/1412.5028}{{\tt  arXiv:1412.5028 [gr-qc]}}.
		
		\bibitem{Zhao:2013oza} 
		R.~Zhao, H.~H.~Zhao, M.~S.~Ma and L.~C.~Zhang,
		``On the critical phenomena and thermodynamics of charged topological dilaton AdS black holes,'' \href{https://link.springer.com/article/10.1140%2Fepjc%2Fs10052-013-2645-x}{{\tt  Eur.\ Phys.\ J.\ C {\bf 73}, 2645 (2013)}},  \href{https://arxiv.org/abs/1305.3725}{{\tt  arXiv:1305.3725 [gr-qc]}}.
			
			
			
			
			\bibitem{sheykhi}
			A. Sheykhi, ``Thermodynamics of charged topological
			dilaton black holes," \href{https://journals.aps.org/prd/abstract/10.1103/PhysRevD.76.124025}{{\tt  Phys. Rev. D 76, 124025 (2007)}}, \href{https://arxiv.org/abs/0709.3619}{{\tt arXiv:0709.3619 [hep-th]}}.
			
			
			\bibitem{CDB1}  G. W. Gibbons and K. Maeda, ``Black holes and membranes in higher-dimensional theories with dilaton fields," \href{http://www.sciencedirect.com/science/article/pii/0550321388900065?via%3Dihub#}{{\tt  Nucl. Phys. {\bf B298}, 741 (1988)}}; T. Koikawa and M. Yoshimura, ``Dilaton fields and event horizon," \href{http://www.sciencedirect.com/science/article/pii/0370269387912640?via%3Dihub}{{\tt Phys. Lett. {\bf B189}, 29 (1987)}}; D. Brill and G. Horowitz, ``Negative energy in string theory," \href{http://www.sciencedirect.com/science/article/pii/037026939190618Z#}{{\tt Phys. Lett. {\bf B262}, 437	(1991).}}
				
				\bibitem{CDB2}  D. Garfinkle, G. T. Horowitz and A. Strominger, ``Charged black holes in string theory," \href{https://journals.aps.org/prd/abstract/10.1103/PhysRevD.43.3140}{{\tt Phys. Rev. D {\bf 43}, 3140 (1991)}}; R. Gregory and J. A. Harvey, ``Black holes with a massive dilaton," \href{https://journals.aps.org/prd/abstract/10.1103/PhysRevD.47.2411}{{\tt Phys. Rev. D {\bf 47}, 2411 (1993)}}; M. Rakhmanov, ``Dilaton black holes with electric charge," \href{https://journals.aps.org/prd/abstract/10.1103/PhysRevD.50.5155}{{\tt  Phys. Rev. D {\bf 50}, 5155 (1994).}}
				
					\bibitem{nonsusy}	
					L. J. Dixon and J. A. Harvey, ``String Theories In Ten-Dimensions Without Space-Time Supersymmetry," \href{http://www.sciencedirect.com/science/article/pii/055032138690619X}{{\tt Nucl. Phys. B	274, 93 (1986)}}; L. Alvarez-Gaume, P. H. Ginsparg, G. W. Moore and C. Vafa, ``An O(16) X O(16) Heterotic String," \href{http://www.sciencedirect.com/science/article/pii/0370269386915248}{{\tt Phys. Lett. B 171, 155 (1986)}}; A. Sagnotti, ``Some properties of open string theories," \href{https://arxiv.org/abs/hep-th/9509080}{{\tt arXiv:hep-th/9509080 [hep-th]}}; A. Sagnotti,``Surprises in open-string perturbation theory," \href{http://www.sciencedirect.com/science/article/pii/S0920563297003447?via%3Dihub}{{\tt Nucl.Phys.Proc.Suppl.56B:332-343,1997}}, \href{https://arxiv.org/abs/hep-th/9702093}{{\tt arXiv:hep-th/9702093 [hep-th]}}; S. Sugimoto,
						``Anomaly cancellations in type I D9-D9-bar system and the USp(32) string,"  \href{https://academic.oup.com/ptp/article-lookup/doi/10.1143/PTP.102.685}{{\tt Prog.Theor.Phys. 102 (1999) 685-699}}, \href{https://arxiv.org/pdf/hep-th/9905159.pdf}{{\tt arXiv:hepth/9905159 [hep-th]}}.
						
						\bibitem{moudas} 
						E. Dudas and J. Mourad, ``Brane solutions in strings with broken supersymmetry and dilaton tadpoles," \href{http://www.sciencedirect.com/science/article/pii/S0370269300007346}{{\tt Phys.Lett. B486 (2000) 172-178}},   \href{https://arxiv.org/pdf/hep-th/0004165.pdf}{{\tt arXiv:hep-th/0004165 [hep-th]}}. 
						
						\bibitem{CLSZ} 
						C. Charmousis, D. Langlois, D. Steer, R. Zegers, ``Rotating Spacetimes with a Cosmological Constant," \href{http://iopscience.iop.org/article/10.1088/1126-6708/2007/02/064/meta}{{\tt JHEP 0702:064, 2007}},  \href{https://arxiv.org/abs/gr-qc/0610091}{{\tt arXiv:gr-qc/0610091 [gr-qc]}}. 
						
						\bibitem{d-reduction} 
						Blaise.~G, Jelena Smolic, et al, ``Holography for Einstein-Maxwell-dilaton theories from
						generalized dimensional reduction," \href{https://link.springer.com/article/10.1007%2FJHEP01%282012%29089}{{\tt JHEP 1201:089, 2012}},  \href{https://arxiv.org/abs/1110.2320}{{\tt arXiv:1110.2320v2 [hep-th]}}.  	
							
				
	
	
\bibitem{Ahar}  O. Aharony, M. Berkooz, D. Kutasov and N. Seiberg, ``Linear dilatons, NS5-branes and holography," \href{http://iopscience.iop.org/article/10.1088/1126-6708/1998/10/004/meta}{{\tt J. High Energy Phys. \textbf{10},
		004 (1998)}}, \href{https://arxiv.org/abs/hep-th/9808149}{{\tt arXiv:hep-th/9808149 [hep-th]}}.






\bibitem{MW}  S. Mignemi and D. Wiltshire, ``Spherically symmetric solutions in dimensionally reduced spacetimes," \href{http://iopscience.iop.org/article/10.1088/0264-9381/6/7/006/meta#}{{\tt Class. Quant. Gravit. {\bf 6}, 987 (1989)}}; D. L. Wiltshire, ``Spherically symmetric solutions in dimensionally reduced spacetimes with a higher-dimensional cosmological constant," \href{https://journals.aps.org/prd/abstract/10.1103/PhysRevD.44.1100}{{\tt {Phys. Rev}. D {\bf 44}, 1100 (1991)}}; S. Mignemi and D. L. Wiltshire, ``Black holes in higher-derivative gravity theories," \href{https://journals.aps.org/prd/abstract/10.1103/PhysRevD.46.1475}{{\tt {Phys. Rev}. D {\bf 46}, 1475 (1992).}}

\bibitem{PW}  S. J. Poletti and D. L.   Wiltshire, ``Global properties of static spherically symmetric charged dilaton spacetimes with a Liouville potential," \href{https://journals.aps.org/prd/abstract/10.1103/PhysRevD.50.7260}{{\tt  Phys. Rev. D {\bf 50}, 7260 (1994)}}, \href{https://arxiv.org/abs/gr-qc/9407021}{{\tt arXiv:gr-qc/9407021 [gr-qc]}}.

\bibitem{CHM}  K. C. K. Chan, J. H. Horne and R. B. Mann, ``Charged dilaton black holes with unusual asymptotics,"   \href{http://www.sciencedirect.com/science/article/pii/0550321395002057?via%3Dihub}{{\tt Nucl. Phys. {\bf %
	B447}, 441 (1995)}}, \href{https://arxiv.org/abs/gr-qc/9502042v1}{{\tt arXiv:gr-qc/9502042v1 [gr-qc]}}.

\bibitem{Cai}  R. G. Cai, J. Y. Ji and K. S. Soh, ``Topological dilaton black holes," \href{https://journals.aps.org/prd/abstract/10.1103/PhysRevD.57.6547}{{\tt Phys. Rev D {\bf 57}, 6547 (1998)}}; R. G. Cai and Y. Z. Zhang, ``Holography and brane cosmology in domain wall backgrounds," \href{https://journals.aps.org/prd/abstract/10.1103/PhysRevD.64.104015}{{\tt Phys. Rev D {\bf 64}, 104015 (2001)}}; R. G. Cai and Y. Z. Zhang, ``Black plane solutions in four-dimensional spacetimes," \href{https://journals.aps.org/prd/abstract/10.1103/PhysRevD.54.4891}{{\tt Phys. Rev D {\bf 54}, 4891 (1996).}}

\bibitem{Clem}  G. Clement, D. Gal'tsov and C. Leygnac, ``Linear dilaton black holes," \href{https://journals.aps.org/prd/abstract/10.1103/PhysRevD.67.024012}{{\tt Phys. Rev. D {\bf67}, 024012 (2003)}}, \href{https://arxiv.org/abs/hep-th/0208225}{{\tt arXiv:hep-th/0208225 [hep-th]}}; G. Clement and C. Leygnac, ``Non-asymptotically flat, non-AdS dilaton black holes," \href{https://journals.aps.org/prd/abstract/10.1103/PhysRevD.70.084018}{{\tt  Phys. Rev. D {\bf 70}, 084018 (2004)}}, \href{https://arxiv.org/abs/gr-qc/0405034}{{\tt arXiv:gr-qc/0405034 [gr-qc]}}.  

\bibitem{Mitra} T. Ghosh and P. Mitra, ``Asymptotically non-flat rotating dilaton black holes,"  \href{http://iopscience.iop.org/article/10.1088/0264-9381/20/7/311?pageTitle=IOPscience}{{\tt  Class. Quant. Gravit. \textbf{20}, 1403
		(2003)}}, \href{https://arxiv.org/abs/gr-qc/0212057}{{\tt arXiv:gr-qc/0212057 [gr-qc]}}.
\bibitem{SDRP} A. Sheykhi, M. H. Dehghani, N. Riazi and J. Pakravan, ``Thermodynamics of rotating solutions in (n+1)-dimensional Einstein-Maxwell-dilaton gravity," 
 \href{https://journals.aps.org/prd/abstract/10.1103/PhysRevD.74.084016}{{\tt Phys. Rev. D {\bf74},  084016 (2006).}}

\bibitem{yaz} S. S. Yazadjiev, ``Einstein-Born-Infeld-dilaton black holes in nonasymptotically flat spacetimes,"   \href{https://journals.aps.org/prd/abstract/10.1103/PhysRevD.72.044006}{\tt {Phys. Rev. D {\bf 72}, 044006 (2005)}}, \href{https://arxiv.org/abs/hep-th/0504152}{{\tt arXiv:hep-th/0504152 [hep-th]}}.

\bibitem{DHSR} M. H Dehghani, S. H. Hendi, A. Sheykhi and H. Rastegar Sedehi, ``Thermodynamics of rotating black branes in Einstein–Born–Infeld-dilaton gravity," \href{http://iopscience.iop.org/article/10.1088/1475-7516/2007/02/020/fulltext/}{{\tt JCAP {\bf0702} (2007)
		020}}, \href{https://arxiv.org/abs/hep-th/0611288}{{\tt arXiv:hep-th/0611288 [hep-th]}}.

	\bibitem{mass_in_lifshtz} 
	W.~G. Brenna, Robert B. Mann and Miok Park, ``{Mass and thermodynamic volume in Lifshitz spacetimes}," \href{https://journals.aps.org/prd/abstract/10.1103/PhysRevD.92.044015}{{\tt Phys. Rev. D 92, 044015 (2015)}}, \href{https://arxiv.org/abs/1505.06331}{{\tt arXiv:1505.06331 [hep-th]}}. 
	
	\bibitem{ref42_of_mann}
	Shamit Kachru, Xiao Liu and Michael Mulligan, ``{Gravity duals of Lifshitz-like fixed points},''
	\href{https://journals.aps.org/prd/abstract/10.1103/PhysRevD.78.106005}{{\em Phys. Rev. D 78, 106005 (2008)}}, \href{https://arxiv.org/abs/0808.1725}{{\tt arXiv:0808.1725 [hep-th]}}.
	
	\bibitem{Erdmenger:2016wyp} 
	J.~Erdmenger, D.~Fernandez, P.~Goulart and P.~Witkowski,
	``Conductivities from attractors,''
\href{https://link.springer.com/article/10.1007%2FJHEP03%282017%29147}{{\tt 	JHEP {\bf 1703}, 147 (2017)}}, \href{https://arxiv.org/abs/1611.09381}{{\tt arXiv:1611.09381 [hep-th]}}.	
		
	
\bibitem{fr}
Ming Zhang and Wen-Biao Liu, ``{f(R) Black Holes as Heat Engines},''
\href{http://download.springer.com/static/pdf/449/art%253A10.1007%252Fs10773-016-3134-4.pdf?originUrl=http%3A%2F%2Flink.springer.com%2Farticle%2F10.1007%2Fs10773-016-3134-4&token2=exp=1492344171~acl=%2Fstatic%2Fpdf%2F449%2Fart%25253A10.1007%25252Fs10773-016-3134-4.pdf%3ForiginUrl%3Dhttp%253A%252F%252Flink.springer.com%252Farticle%252F10.1007%252Fs10773-016-3134-4*~hmac=257e9475958e0b2449e0f60af2902d866fc9cb13f243c7d32e17e3f0bf14b89c}{{\em  Int J Theor Phys (2016) 55: 5136}}.
	
	\bibitem{Johnson:varing_q} 
	C.~V.~Johnson,
	``Approaching the Carnot Limit at Finite Power: An Exact Solution,''  \href{https://arxiv.org/abs/1703.06119}{{\tt arXiv:1703.06119v1 [hep-th]}}.	
	
\bibitem{Johnson:2017asf} 
C.~V.~Johnson,
``Critical Black Holes in a Large Charge Limit,'' \href{https://arxiv.org/abs/1705.01154}{{\tt arXiv:1705.01154 [hep-th].}}

\end{thebibliography}
\end{document}